%% file: pumptrack_benchmark_v1.tex
\documentclass[letterpaper, 10 pt, conference]{ieeeconf}
\pdfoutput=1

\IEEEoverridecommandlockouts                              
\usepackage[utf8]{inputenc}
\usepackage{microtype}
\usepackage{pgfplots}
\pgfplotsset{compat=1.8}
\usepackage{bm}
\usepackage{mathtools}
\usepackage{breqn}
\usepackage{graphicx,import}
\usepackage[justification=centering]{caption}
\usepackage{subcaption}
\usepackage{float}
\usepackage{soul}  
\usepackage{amssymb}
\usepackage{url}
\usepackage{booktabs}
\usepackage{tabularx}
\usepackage{multirow}
\usepackage[hidelinks]{hyperref}
\usepackage{comment}
\usepackage{xcolor}

\title{\LARGE \bf Tutorial Problems for Nonsmooth Dynamics and Optimal Control:\\ Ski Jumping and Accelerating a Bike Without Pedaling
}

\author{Julian Golembiewski and Timm Faulwasser
\thanks{\textbf{Julian Golembiewski} and \textbf{Timm Faulwasser} are with the Institute of Control Systems, Hamburg University of Technology, 21079 Hamburg, Germany. E-mail: 
        {\tt\small julian.golembiewski@tuhh.de, timm.faulwasser@ieee.org}}%
}

\begin{document}
\maketitle
\thispagestyle{empty}
\pagestyle{empty}


\begin{abstract}
Nonsmooth phenomena, such as abrupt changes, impacts, and switching behaviors, frequently arise in real-world systems and present significant challenges for traditional optimal control methods, which typically assume smoothness and differentiability. These phenomena introduce numerical challenges in both simulation and optimization, highlighting the need for specialized solution methods. Although various applications and test problems have been documented in the literature, many are either overly simplified, excessively complex, or narrowly focused on specific domains. On this canvas, this paper proposes two novel tutorial problems that are both conceptually accessible and allow for further scaling of problem difficulty. The first problem features a simple ski jump model, characterized by state-dependent jumps and sliding motion on impact surfaces. This system does not involve control inputs and serves as a testbed for simulating nonsmooth dynamics. The second problem considers optimal control of a special type of bicycle model. This problem is inspired by practical techniques observed in BMX riding and mountain biking, where riders accelerate their bike without pedaling by strategically shifting their center of mass in response to the track's slope.
\end{abstract}


\input{Content/1_intro.tex}
\input{Content/2_setting.tex}
\input{Content/3_num_oc.tex}
\input{Content/4_benchmarks.tex}
\input{Content/5_results.tex}
\input{Content/6_conclusion.tex}

\bibliographystyle{unsrt}
\bibliography{Literature/Literatur}

\end{document}

%% file: Content/1_intro.tex
\section{INTRODUCTION}
\label{cha:intro}
Many real-world control problems are characterized by nonsmooth phenomena, such as abrupt changes, impacts, switching behaviors, and nondifferentiable dynamics. These systems can generally be classified into two categories: those with internal switches, where nonsmoothness arises from state-dependent discontinuities, and those with external switches, where it is due to control-dependent discontinuities~\cite{van2007introduction}. Additionally, nonsmooth systems can be further categorized based on the type of nonsmoothness they exhibit, such as continuous systems with nondifferentiable right-hand sides, systems with discontinuous dynamics, or systems with state discontinuities~\cite{leine2007stability}.

Established optimal control methods, which rely on smoothness and differentiability assumptions, are often insufficient for handling these complexities. This necessitates the development of specialized numerical methods to solve optimal control problems involving nonsmooth dynamics~\cite{acary2008numerical}. To evaluate and compare these methods, various applications and benchmark problems have been proposed in the literature.

For systems with controlled switches, a benchmark library of mixed-integer optimal control problems is provided in~\cite{sager2011benchmark}. Other examples include a supermarket refrigeration system in~\cite{larsen2007supermarket} and a simulated moving bed chromatography system in~\cite{alamir2006benchmark}. Switched affine systems have been studied using an abstract system model in~\cite{seatzu2006optimal}. Systems with autonomous state jumps, such as a bipedal system mimicking a walking-like motion, a bouncing ball, and a moon landing problem have been explored in~\cite{kirches2018numerical,nurkanovic2022nosnoc,ross2002direct}. Additionally, a benchmark library for Mathematical Programs with Complementarity Constraints (MPCCs) is given in~\cite{nurkanovic2023solving}.

While numerous test problems exist, many are either too abstract to be engaging for educational purposes, overly complex for testing numerical methods and prototyping, or narrowly tailored to specific application domains.
In this paper, we present two novel tutorial problems in nonsmooth optimal control, combining physical intuitiveness with numerical challenges arising from their nonsmooth dynamics. These problems are introduced using basic models, with potential extensions suggested to scale complexity. First, we propose a ski jump model as an introductory problem for discretizing and simulating nonsmooth dynamical systems. Moreover, inspired by~\cite{lunze_fahrrad,astrom_bicycle_dynamics}, which use bicycle dynamics as educational example for control theory, we present a slightly different bicycle model to explore nonsmooth optimal control. The proposed Optimal Control Problem (OCP) involves nonsmooth dynamics with internal state discontinuities and sliding motion on an impact surface, a common feature in mechanical impact systems. While it seems counter-intuitive to accelerate a bike without pedaling, it is well known in mountain biking and BMX riding, that skilled riders can accelerate by shifting their body weight according to the slope of the track. On so called \textit{pump tracks} or, e.g., in BMX Cycling at the Olympics, this is the dominant way to accelerate the bicycle.

The contribution of this paper is twofold: First, we introduce two novel tutorial problems -- one serves as an entry-level example for simulating nonsmooth dynamical systems, and the other with the potential to serve as a scalable benchmark for nonsmooth optimal control. Both problems are accessible and engaging, making them well-suited for teaching purposes, while also enabling the prototyping and testing of numerical methods. We provide simulation results based on the time-freezing approach from~\cite{nurkanovic2023time} to numerically address the occurring state jumps.
Second, we extend the initial investigations into modeling and optimal control of a bicycle on a pump track from~\cite{golembiewski2024dynamics}, by explicitly incorporating jumps that result in mechanical impacts.

The remainder of this paper is organized as follows: Section~\ref{sec:setting} outlines the class of nonsmooth systems under consideration and presents a generic formulation of the OCP. Section~\ref{sec:NumOC} discusses numerical methods for discretizing nonsmooth dynamics and solving nonsmooth OCPs. Section~
\ref{sec:benchmark} introduces two tutorial problems: ski jumping and accelerating a bike without pedaling. Section~\ref{sec:results} presents simulation results for these problems, and, finally, Section~\ref{sec:conclusion} concludes the paper and suggests potential directions for future research.

\textit{Notation:} The concatenation of two column vectors $x\in\mathbb{R}^{n}$ and $y\in\mathbb{R}^{m}$ is denoted by $(x,y):=\left[ x^\top,y^\top \right]^\top$.

%% file: Content/2_setting.tex
\section{PROBLEM SETTING AND PRELIMINARIES}
\label{sec:setting}

\subsection{Class of nonsmooth dynamical systems}
\label{subsec:nonsmooth_dyn_systems}
The problems addressed in this work involve mechanical impact systems with unilateral impact surfaces. Modeling frameworks for this class of nonsmooth dynamical systems include, but are not limited to, piecewise-smooth systems, differential inclusions, Filippov systems, dynamic complementarity systems, and Moreau's sweeping processes~\cite{bernardo2008piecewise, liberzon2003switching, filippov2013differential, heemels2003complementarity, brogliato2020dynamical}. Studies on the optimal control of such nonsmooth systems are available, e.g.,~\cite{Westenbroek2019piecewiseOC, colombo2016optimal, vieira2020optimal}, while \cite{nurkanovic2023numerical} provides a comprehensive overview with a focus on the numerical solution of nonsmooth OCPs.

In this work, we consider Nonlinear Complementarity Systems (NLCS). In this framework, the nonsmooth behavior of the dynamics is expressed as a complementarity condition. The NLCS for a mechanical impact system with unilateral impact surfaces is formulated as: \begin{subequations}
    \label{eq:nlcs}
    \begin{align}
         \dot{x} = f(x,\lambda_n, u, t) \label{eq:nlcs1}\\
        0\leq f_c(q) \perp \lambda_n\geq 0 \label{eq:nlcs2}\\
        \textrm{impact laws},\label{eq:nlcs_impact_law}
    \end{align}
\end{subequations} where the state is $x=(q,\, \dot{q}) \in \mathbb{R}^{n_x}$, and $q, \dot{q}\in \mathbb{R}^{n_q}$ are the generalized coordinates and velocities, respectively. The input is $u \in \mathbb{R}^{n_u}$. The impact surface is denoted by $f_c(q)$, and $\lambda_n$ represents the normal contact force. The complementarity condition ensures that either (i) there is no contact force while maintaining a positive distance from the impact surface ($\lambda_n=0,\:f_c(q)>0$), or (ii) a contact force exists when the distance to the impact surface is zero ($\lambda_n>0,\:f_c(q)=0$).

In cases where the system undergoes state jumps due to impacts, the \textit{impact laws} \eqref{eq:nlcs_impact_law} govern the associated state transitions. When the impact laws are neglected, the complementarity system defined by\eqref{eq:nlcs1}--\eqref{eq:nlcs2} can be efficiently solved using tailored methods~\cite{kim2020mpec}. However, in this work, we explicitly model the complexities introduced by impact laws and address them numerically using the time-freezing method outlined in~\cite{nurkanovic2023time}.

\subsection{Optimal control formulation}
\label{subsec:nonsmooth_ocp}
Consider the continuous-time OCP 
\begin{subequations}
    \label{eq:nonsmooth_ocp}
    \begin{align}
        \min_{u(\cdot),t_{\mathrm{f}}}& \int_{0}^{t_{\mathrm{f}}} \ell(t,x(t),u(t)) \,\mathrm{d}t + V_{\mathrm{f}}(x(t_{\mathrm{f}})) \\
        \text{subject to } & \text{for almost all } t \in[0,t_{\mathrm{f}}] \notag \\
        \eqref{eq:nlcs},& \quad x(0) = x_0 \label{eq:nonsmooth_ocp_dynamics}\\
        x(t) &\in \mathbb{X} \subset \mathbb{R}^{n_x} \\
        u(t) &\in \mathbb{U} \subset \mathbb{R}^{n_u}\\
        x(t_{\mathrm{f}}) &\in \mathbb{X}_{\mathrm{f}} \subset \mathbb{R}^{n_x},
    \end{align}
\end{subequations} where~\eqref{eq:nonsmooth_ocp_dynamics} describes a nonsmooth dynamical system with states and inputs  $x(t)$ and $u(t)$, respectively. This implies that the nonsmoothness may propagate state limitations, for example, through impact surfaces, leading to jump discontinuities in the velocities. Moreover, the OCP is subject to state, input, and terminal constraints. These constraints are given as compact subsets of $\mathbb{R}^{n_x}$ and $\mathbb{R}^{n_u}$, respectively. The objective is to minimize the running cost $\ell:\mathbb{R}\times\mathbb{R}^{n_x}\times\mathbb{R}^{n_u}\rightarrow\mathbb{R}_0^+$ and the terminal cost $V_{\mathrm{f}}:\mathbb{R}^{n_x}\rightarrow\mathbb{R}_0^+$ over the time horizon $t_{\mathrm{f}}\in\mathbb{R}^+$. The terminal time $t_{\mathrm{f}}$ can also be treated as a decision variable within the OCP, enabling the formulation of free end-time problems.

%% file: Content/3_num_oc.tex
\section{NONSMOOTH OPTIMAL CONTROL}
\label{sec:NumOC} 
Solution methods for OCPs can be broadly categorized as either indirect or direct~\cite{diehl2006fast}. In indirect methods, the OCP is first optimized analytically by deriving the necessary conditions of optimality using Pontryagin's Maximum Principle. These conditions are then discretized and solved numerically. In contrast, direct methods begin by discretizing the OCP itself, transforming it into a finite-dimensional nonlinear program, which is then solved numerically.

In this work, we focus on applying direct methods, which require the explicit discretization of the system dynamics. Consequently, we must address the discretization of ODEs with discontinuous right-hand sides.
However, standard integration methods fail to detect switches accurately and therefore are limited to at most first-order accuracy. Additionally, the derivative information at points of discontinuity is incorrect, leading to integration errors that can accumulate over time. This can also cause the numerical solution to oscillate near discontinuities. Thus, tailored methods are required to discretize the resulting nonsmooth ODEs efficiently. Due to space limitations, we provide only a brief overview of those methods. For a more comprehensive discussion, we refer to~\cite{acary2008numerical,nurkanovic2023numerical,brogliato1999nonsmooth}. For detailed analysis and convergence results, see \cite{acary2008numerical,stewart2010optimal}.

\subsection{Discretization of nonsmooth dynamical systems}
\label{subsec:overview_num_oc}
Numerical methods for discretizing and simulating nonsmooth ODEs can be broadly categorized into event-driven schemes, time-stepping schemes, and smoothing techniques~\cite{acary2008numerical}.
Event-driven methods focus on accurately detecting and handling discrete events where the system dynamics exhibit discontinuities. These methods define and detect events, solve the nonsmooth dynamics using reinitialization rules at the events, and perform smooth integration between events, achieving high accuracy during smooth periods. However, event-driven schemes can become inefficient when the number of events is large and are sensitive to the tolerances set for event detection.

In contrast, time-stepping methods discretize both the smooth parts of the dynamics and the complementarity conditions associated with the nonsmooth dynamics. By considering events concurrently with the integration of smooth dynamics, time-stepping schemes offer greater efficiency in scenarios with numerous events. However, they generally provide lower-order accuracy than event-driven schemes, as they do not detect events exactly and produce incorrect gradient information at discontinuities.

An alternative approach involves smoothing techniques, which approximate the discontinuous right-hand side of the dynamics by using a smooth function. This method allows for simpler implementation and can be particularly useful for generating initial guesses in optimal control. Nonetheless, achieving accurate results with smoothing techniques requires a sufficiently small step size relative to the smoothing parameter which proves to be difficult in practice \cite{nurkanovic2023numerical}.

\subsection{Time-freezing approach for systems with state jumps}
\label{subsec:time_freezing}
In the numerical results presented in this paper, we utilize the time-freezing approach for systems with state jumps, as introduced by~\cite{nurkanovic2023time}. This approach converts a system with state jumps into a Filippov differential inclusion by emulating the state jump within the infeasible region of the system. The resulting formulation ensures that discontinuities are present only in the right-hand side of the dynamics, while the state trajectories remain smooth. To this end, an auxiliary dynamical system $f_{\text{aux}}(x)$ is defined in the infeasible region, where the trajectory endpoints adhere to the impact law while the progression of time is effectively frozen. We outline the main steps of the time-freezing approach in the following.

Consider system~\eqref{eq:nlcs} with state vector $x=(q,\, \dot{q})$, impact surface $f_c(q)$, and input $u$. For $f_c(q)>0$ the reduced dynamics are given as $\dot{x}=f_{\text{free}}(x,u)$. One introduces a parameter for the numerical time $\tau$ to define a clock state $t(\tau)$ that stops evolving whenever $f_{\text{aux}}(x)$ is active. This allows to differentiate between the numerical time $\tau$ and the physical time $t$ in the time-freezing system. To incorporate time evolution into our dynamics, one extends the state vector to $\hat{x}(\tau):=(x(\tau),\, t(\tau))\in\mathbb{R}^{n_x+1}$. Time derivatives of the extended state vector with respect to the numerical time are compactly denoted as $\hat{x}^\prime(\tau):=\frac{d}{d\tau}\hat{x}(\tau)$.

Next, one extends the impact surface using its time derivative to define the switching function $c(\hat{x})= (c_1(\hat{x}),\, c_2(\hat{x})) = (f_c(\hat{x}),\, \frac{d}{dt} f_c(\hat{x}))$ and specify the following regions
\begin{equation}
    \label{eq:tf_regions}
    \begin{aligned}
    R_1 &= \left\{ \hat{x}\in\mathbb{R}^{n_x+1} \: | \: c_1(\hat{x})>0\right\} \cup \left\{ \hat{x} \: | \: c_1(\hat{x})<0, c_2(\hat{x})>0\right\},\\
    R_2 &= \left\{ \hat{x} \: | \: c_1(\hat{x})<0, c_2(\hat{x})<0\right\}.
    \end{aligned} 
\end{equation} 
Considering these regions and the extended state vector, the dynamics in region $R_1$ are described by $\hat{x}^\prime=\hat{f}_{\text{free}}(\hat{x},u)=(f_{\text{free}}(x,u),\, 1)$, while the auxiliary dynamics in region $R_2$ are given by $\hat{x}^\prime=\hat{f}_{\text{aux}}(\hat{x})=(f_{\text{aux}}(x),\, 0)$. Note that the control input $u$ does not influence the system in $f_{\text{aux}}(x)$. The state and input trajectory for the original problem can be recovered by considering only the segments where $\hat{x}\in R_1$.

Finally, one can state the time-freezing system as \begin{equation}
    \label{eq:tf_system_di}
    \begin{split}
        \hat{x}^\prime \in F_{\text{TF}}(\hat{x},u) = \left\{ \theta_1\hat{f}_{\text{free}}(\hat{x},u)+\theta_2\hat{f}_{\text{aux}}(\hat{x}) \: \middle| \right.\\
        \left. \phantom{\hat{f}} \theta_1+\theta_2=1, \, \theta_i\geq0,\, \theta_i=0\: \text{if} \: \hat{x}\notin \overline{R}_i, \forall i\in\{1,2\} \right\}.&
    \end{split}
\end{equation} For the simulation results in Section~\ref{sec:results} we define such systems using the benchmarks introduced in Section~\ref{sec:benchmark}.

%% file: Content/4_benchmarks.tex
\section{TWO TUTORIAL PROBLEMS}
\label{sec:benchmark}

In the following, we present two tutorial problems that may also serve as benchmarks for nonsmooth dynamics and optimal control. The first problem concerns a ski jump scenario where the system model does not include any control inputs. Thus, we only focus on the numerical integration of the dynamics subject to state jumps and sliding modes. The second example is an OCP for bicycle dynamics navigating an uneven mountain bike track, with the objective of reaching a target location as quickly as possible without pedaling. The control input in this scenario is the reciprocal motion between the rider and the bicycle, referred to as \textit{pumping}, which is the sole means of acceleration~\cite{golembiewski2024dynamics}. As discussed  in the previous section, we employ the time-freezing approach for our simulations to model the impact behavior of the systems. To this end, we derive $f_{\text{free}}(x,u)$ as well as the impact surfaces $f_c(q)$ for both systems.

\subsection{Ski jump}
\label{subsec:bench_ski}
We consider a ski jump as an example of a system exhibiting impacts and sliding motion. The system comprises a skier who jumps off a \textit{take-off table} and lands on a \textit{landing slope} with no control inputs. 
Simulations of this system allow us to examine how the starting height affects the point of impact, i.e., the distance covered by the jump.

The skier is modeled as a point mass with position $q=(y,\, z)=(q_1,\, q_2)\in\mathbb{R}^2$ and velocity vector $\dot{q}$. The state of the system is $x=(q,\, \dot{q}) \in \mathbb{R}^4$. A sketch of the system is shown in Figure~\ref{fig:ski_model}. Consequently, the system dynamics during free flight are given by \begin{equation}
    \label{eq:ski_free_dynamics}
    \dot{x}=f_{\text{free}}(x)=\begin{bmatrix}
       \dot{q}^\top,\:0,\:-g
    \end{bmatrix}^\top
\end{equation} where $g$ is the gravitational constant. We model the ski jump by considering the take-off table and the landing slope as $h_{\text{jump}}(q_1)$ and $h_{\text{land}}(q_1)$. The impact surfaces from \eqref{eq:nlcs2}, where the height of the skier $q_2$ impacts the ground, are given by 
\begin{equation}
    \label{eq:ski_impact_surfaces}
    \begin{aligned}
        f_{c1}(q)&=q_2-h_{\text{jump}}(q_1),\quad f_{c2}(q)&=q_2-h_{\text{land}}(q_1).
    \end{aligned}
\end{equation} 
All impacts are assumed to be perfectly inelastic. Using these impact surfaces we define the following three regions: 
\begin{equation*}
\label{eq:ski_regions}
\begin{aligned}
    R_1&=\left\{ x\in\mathbb{R}^4 \: \middle| \: f_{c1}(q_1)>0 \wedge f_{c2}(q_1)>0 \right\},\\
    R_2&=\left\{ x\in\mathbb{R}^4 \: \middle| \: f_{c1}(q_1)<0 \right\},\\
    R_3&=\left\{ x\in\mathbb{R}^4 \: \middle| \: f_{c2}(q_1)<0 \right\}.
\end{aligned}
\end{equation*} 
\begin{figure}
    \centering
    \input{Graphics/tikz/ski_model.tex}  
    \caption{Model of a ski jumping venue.}
    \label{fig:ski_model}
\end{figure}
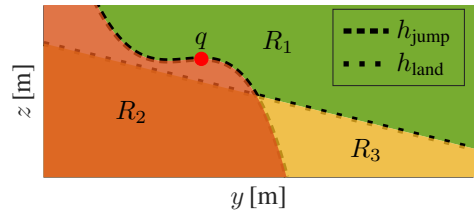 One can identify two operating modes: the free flight within the interior of $R_1$ and the sliding motion on the boundary defined by \begin{align*}
    \partial(R_1, R_2\cup R_3)=\left\{ x\in\mathbb{R}^4 \: \middle| \: x\in\partial \overline{R_1} \cap x\in \partial\overline{R_2\cup R_3}  \right\}.
\end{align*} This boundary lies on the intersection of $\partial \overline{R_1}$ and the closure of $\partial \overline{R_2\cup R_2}$, encompassing their boundaries and limit points. Note that the infeasible region is in the interior of $R_2 \cup R_3$, and the system should not be initialized in this region.

\subsection{Accelerating a bike without pedaling}
\label{subsec:bench_pumptrack}
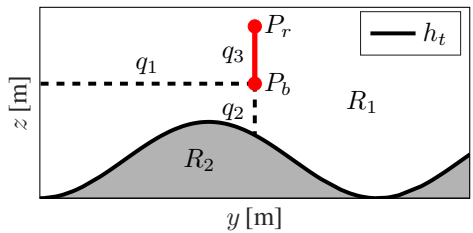
\begin{figure}
   \centering
   \input{Graphics/tikz/pump_model.tex}  
   \caption{Bicycle model on a pump track.}
   \label{fig:pump_model}\vspace*{-3mm}
\end{figure}

In this section, we formulate an OCP where a bicycle rider aims to reach a goal as quickly as possible without pedaling. Consider a bicycle and its rider on an uneven mountain bike track, also known as a pump track. As it is common knowledge in competitive cycling (in particular in BMX riding and mountain biking), riders can accelerate by strategically shifting their center of mass in response to the track’s slope. Initial investigations as well as experimental results of this problem can be found in~\cite{golembiewski2024dynamics}. In this work, we extend these investigations by incorporating jumps that introduce nonsmoothness into the system. Similar to our previous work, we use a two-mass model connected by a mass-less link as bicycle-rider model. Figure~\ref{fig:pump_model} shows a sketch of the system.
The generalized coordinates $q~=~(q_1,\, q_2,\, q_3)$ are sufficient to describe the center of masses of bike and rider with masses $m_b$ and $m_r$, respectively. The positions in the $y$-$z$-plane are \begin{align*}
   P_b=\begin{bmatrix}
       q_1\\
       h_t(q_1)+q_2
   \end{bmatrix}, \quad P_r=\begin{bmatrix}
       q_1\\
       h_t(q_1)+q_2+q_3
   \end{bmatrix},
\end{align*} 
where $q_1$ corresponds to their $y$-coordinate, $q_2$ is the slack above the ground, and the length of their connection is given by $q_3$. The height of the track is given as a nonlinear, differentiable function $h_t(q_1)$. For compact notation we denote its derivatives with respect to $q_1$ as \begin{align*}
   h_t^{\prime}=\frac{d h_t}{d q_1},\quad h_t^{\prime\prime}=\frac{d^2 h_t}{d^2 q_1}.
\end{align*} The state vector is given by $x=(q,\, \dot{q})\in \mathbb{R}^{6}$. We define our control input $u$ as the exogenous force acting on $q_3$, which causes the reciprocal motion between both masses.

Using the Lagrangian formalism with generalized coordinates $q(t)$, we derive the reduced dynamics as \begin{equation}
    \label{eq:pump_free_dyn}
    f_{\text{free}}(x,u)=\begin{bmatrix}
        \dot{q}\\
        0\\
        -(m_1\dot{q}_1^2h_t^{\prime\prime}+u+gm_1)/(m_1)\\
        u(m_1+m_2)/(m_1m_2)
    \end{bmatrix}\in\mathbb{R}^{6}.
\end{equation}

Similar to the ski jump, we consider the impacts to be perfectly inelastic. In this case, since we use a slack variable describing the height above the track, the impact surface from \eqref{eq:nlcs2} is trivially given by $f_c=q_2$. We define the regions \begin{equation*}
    \label{eq:pump_regions}
    \begin{aligned}
        R_1=\left\{ x\in\mathbb{R}^6 \: \middle| \: q_2>0 \right\},\:R_2=\left\{ x\in\mathbb{R}^6 \: \middle| \: q_2<0 \right\}
    \end{aligned}
\end{equation*} 
to distinguish the operating modes of the system. In particular, the free motion in the interior of $R_1$ and the sliding motion on the boundary between $R_1$ and $R_2$, which will be addressed by extending to a time-freezing system~\eqref{eq:tf_system_di} in Section~\ref{sec:results}. The infeasible region is in the interior of $R_2$.

The goal is to reach a target location as quickly as possible using only pumping motion, leading to the following nonsmooth time-optimal control problem: \begin{subequations}
    \label{eq:pump_time_optimal_ocp}
    \begin{align}
        \min_{u(\cdot), t_{\mathrm{f}}}& \: t_{\mathrm{f}} \\
        \text{subject to }  &\text{for almost all } t \in[0,t_{\mathrm{f}}] \notag \\
        &\eqref{eq:nlcs},\quad x(0) = x_0 \\
        &l_{\text{min}} \leq q_3 \leq l_{\text{max}},\:u_{\text{min}} \leq u \leq u_{\text{max}}  \label{eq:pump_time_optimal_ocp_constraints} \\
        &q_1(t_{\mathrm{f}}) = q_{\text{goal}}, \label{eq:pump_time_optimal_ocp_terminal_constraint}
    \end{align}
\end{subequations} where, $t_{\mathrm{f}}$ is the free terminal time, $x_0$ is the initial condition, and \eqref{eq:pump_time_optimal_ocp_terminal_constraint} is the terminal constraint. To impose realistic state and input constraints~\eqref{eq:pump_time_optimal_ocp_constraints}, experimental results were reported in~\cite{golembiewski2024dynamics}. Specifically, during test rides on a real pump track, the centers of mass of the bicycle and the rider were captured separately using a 3D camera system. This data is utilized to analyze the distance \( l = q_3 \) between the two masses, as well as the corresponding acceleration \( \ddot{q}_3 \). This led to the boundaries of the state constraint \( \mathbb{X} = \{ x \mid l_{\text{min}} \leq q_3 \leq l_{\text{max}} \} \). Using the relation \( \ddot{q}_3 = u(m_1 + m_2)/(m_1 m_2) \) from~\eqref{eq:pump_free_dyn}, we can determine the input constraint \( \mathbb{U} = \{ u \mid u_{\text{min}} \leq u \leq u_{\text{max}} \} \) based on the measurements of \( \ddot{q}_3 \).

\subsection{Scalability of the proposed problems}
In its current form, the ski jump model serves as a testbed solely for simulating nonsmooth dynamics. However, it can be easily extended to be used for optimal control. One straightforward extension, requiring no additional modeling effort from the user, is to include the skier's initial velocity as a decision variable in the OCP. This can be interpreted as the impulse generated when the athlete pushes off the starting bench to accelerate. Alternatively, or in addition, the model can be extended to include a control action at the take-off table. Given the significance of air drag in ski jumping, incorporating it into the model would provide valuable insights into its impact on performance.

Similarly, the complexity of the pump track problem can be scaled to challenge numerical methods. For instance, the bicycle model can be extended to include two ground-contact points, resulting in simultaneous impacts that must be handled by the numerical approach. The total number of impacts can be varied by adjusting the initial velocity, simulation time, or characteristics of the impact surface. Additionally, more complex impact surfaces could be considered, such as a discontinuous surface to model the bicycle jumping over a gap, or a series of impact surfaces to represent a complete track. Finally, modeling ground contact using Coulomb friction would impose a second type of nonsmoothness.

%% file: Graphics/tikz/ski_model.tex
%
%
\definecolor{mycolor1}{rgb}{0.92900,0.69400,0.12500}%
\definecolor{mycolor2}{rgb}{0.85000,0.32500,0.09800}%
\definecolor{mycolor3}{rgb}{0.46600,0.67400,0.18800}%
\begin{tikzpicture}

\begin{axis}[%
width=2.25in,
height=0.9in,
scale only axis,
xmin=0,
xmax=55,
xlabel style={font=\color{white!15!black}},
xlabel={$y$\,[m]},
xtick=\empty,
xticklabel=\empty,
ymin=10,
ymax=55,
ylabel style={font=\color{white!15!black}},
ylabel={$z$\,[m]},
ytick=\empty,
yticklabel=\empty,
axis background/.style={fill=mycolor3},
axis x line*=bottom,
axis y line*=left,
legend style={legend cell align=left, align=left, draw=white!15!black, fill=mycolor3}
]
\addplot[fill=white, draw=none, forget plot] table[row sep=crcr]{%
0	45\\
0.320255082851593	44.8398724585742\\
0.640510165703186	44.6797449171484\\
0.960765248554779	44.5196173757226\\
1.28102033140637	44.3594898342968\\
1.60127541425797	44.199362292871\\
1.92153049710956	44.0392347514452\\
2.24178557996115	43.8791072100194\\
2.56204066281274	43.7189796685936\\
2.88229574566434	43.5588521271678\\
3.20255082851593	43.398724585742\\
3.52280591136752	43.2385970443162\\
3.84306099421912	43.0784695028904\\
4.16331607707071	42.9183419614646\\
4.4835711599223	42.7582144200388\\
4.8038262427739	42.5980868786131\\
5.12408132562549	42.4379593371873\\
5.44433640847708	42.2778317957615\\
5.76459149132867	42.1177042543357\\
6.08484657418027	41.9575767129099\\
6.40510165703186	41.7974491714841\\
6.72535673988345	41.6373216300583\\
7.04561182273505	41.4771940886325\\
7.36586690558664	41.3170665472067\\
7.68612198843823	41.1569390057809\\
8.00637707128982	40.9968114643551\\
8.32663215414142	40.8366839229293\\
8.64688723699301	40.6765563815035\\
8.9671423198446	40.5164288400777\\
9.2873974026962	40.3563012986519\\
9.60765248554779	40.1961737572261\\
9.92790756839938	40.0360462158003\\
10.248162651251	39.8759186743745\\
10.5684177341026	39.7157911329487\\
10.8886728169542	39.5556635915229\\
11.2089278998058	39.3955360500971\\
11.5291829826573	39.2354085086713\\
11.8494380655089	39.0752809672455\\
12.1696931483605	38.9151534258197\\
12.4899482312121	38.7550258843939\\
12.8102033140637	38.5948983429681\\
13.1304583969153	38.4347708015423\\
13.4507134797669	38.2746432601165\\
13.7709685626185	38.1145157186908\\
14.0912236454701	37.954388177265\\
14.4114787283217	37.7942606358392\\
14.7317338111733	37.6341330944134\\
15.0519888940249	37.4740055529876\\
15.3722439768765	37.3138780115618\\
15.6924990597281	37.153750470136\\
16.0127541425796	36.9936229287102\\
16.3330092254312	36.8334953872844\\
16.6532643082828	36.6733678458586\\
16.9735193911344	36.5132403044328\\
17.293774473986	36.353112763007\\
17.6140295568376	36.1929852215812\\
17.9342846396892	36.0328576801554\\
18.2545397225408	35.8727301387296\\
18.5747948053924	35.7126025973038\\
18.895049888244	35.552475055878\\
19.2153049710956	35.3923475144522\\
19.5355600539472	35.2322199730264\\
19.8558151367988	35.0720924316006\\
20.1760702196504	34.9119648901748\\
20.496325302502	34.751837348749\\
20.8165803853535	34.5917098073232\\
21.1368354682051	34.4315822658974\\
21.4570905510567	34.2714547244716\\
21.7773456339083	34.1113271830458\\
22.0976007167599	33.95119964162\\
22.4178557996115	33.7910721001942\\
22.7381108824631	33.6309445587684\\
23.0583659653147	33.4708170173427\\
23.3786210481663	33.3106894759169\\
23.6988761310179	33.1505619344911\\
24.0191312138695	32.9904343930653\\
24.3393862967211	32.8303068516395\\
24.6596413795727	32.6701793102137\\
24.9798964624243	32.5100517687879\\
25.3001515452758	32.3499242273621\\
25.6204066281274	32.1897966859363\\
25.940661710979	32.0296691445105\\
26.2609167938306	31.8695416030847\\
26.5811718766822	31.7094140616589\\
26.9014269595338	31.5492865202331\\
27.2216820423854	31.3891589788073\\
27.541937125237	31.2290314373815\\
27.8621922080886	31.0689038959557\\
28.1824472909402	30.9087763545299\\
28.5027023737918	30.7486488131041\\
28.8229574566434	30.5885212716783\\
29.143212539495	30.4283937302525\\
29.4634676223466	30.2682661888267\\
29.7837227051981	30.1081386474009\\
30.1039777880497	29.9480111059751\\
30.4242328709013	29.7878835645493\\
30.7444879537529	29.6277560231235\\
31.0647430366045	29.4676284816977\\
31.3849981194561	29.3075009402719\\
31.7052532023077	29.1473733988461\\
32.0255082851593	28.9872458574204\\
32.3457633680109	28.8271183159946\\
32.6660184508625	28.6669907745688\\
32.9862735337141	28.506863233143\\
33.3065286165657	28.3467356917172\\
33.6267836994173	28.1866081502914\\
33.9470387822689	28.0264806088656\\
34.2672938651205	27.8663530674398\\
34.587548947972	27.706225526014\\
34.9078040308236	27.5460979845882\\
35.2280591136752	27.3859704431624\\
35.5483141965268	27.2258429017366\\
35.8685692793784	27.0657153603108\\
36.18882436223	26.905587818885\\
36.5090794450816	26.7454602774592\\
36.8293345279332	26.5853327360334\\
37.1495896107848	26.4252051946076\\
37.4698446936364	26.2650776531818\\
37.790099776488	26.104950111756\\
38.1103548593396	25.9448225703302\\
38.4306099421912	25.7846950289044\\
38.7508650250427	25.6245674874786\\
39.0711201078943	25.4644399460528\\
39.3913751907459	25.304312404627\\
39.7116302735975	25.1441848632012\\
40.0318853564491	24.9840573217754\\
40.3521404393007	24.8239297803496\\
40.6723955221523	24.6638022389238\\
40.9926506050039	24.503674697498\\
41.3129056878555	24.3435471560723\\
41.6331607707071	24.1834196146465\\
41.9534158535587	24.0232920732207\\
42.2736709364103	23.8631645317949\\
42.5939260192619	23.7030369903691\\
42.9141811021135	23.5429094489433\\
43.2344361849651	23.3827819075175\\
43.5546912678166	23.2226543660917\\
43.8749463506682	23.0625268246659\\
44.1952014335198	22.9023992832401\\
44.5154565163714	22.7422717418143\\
44.835711599223	22.5821442003885\\
45.1559666820746	22.4220166589627\\
45.4762217649262	22.2618891175369\\
45.7964768477778	22.1017615761111\\
46.1167319306294	21.9416340346853\\
46.436987013481	21.7815064932595\\
46.7572420963326	21.6213789518337\\
47.0774971791842	21.4612514104079\\
47.3977522620358	21.3011238689821\\
47.7180073448874	21.1409963275563\\
48.0382624277389	20.9808687861305\\
48.3585175105905	20.8207412447047\\
48.6787725934421	20.6606137032789\\
48.9990276762937	20.5004861618531\\
49.3192827591453	20.3403586204273\\
49.6395378419969	20.1802310790015\\
49.9597929248485	20.0201035375757\\
50.2800480077001	19.8599759961499\\
50.6003030905517	19.6998484547242\\
50.9205581734033	19.5397209132984\\
51.2408132562549	19.3795933718726\\
51.5610683391065	19.2194658304468\\
51.8813234219581	19.059338289021\\
52.2015785048097	18.8992107475952\\
52.5218335876612	18.7390832061694\\
52.8420886705128	18.5789556647436\\
53.1623437533644	18.4188281233178\\
53.482598836216	18.258700581892\\
53.8028539190676	18.0985730404662\\
54.1231090019192	17.9384454990404\\
54.4433640847708	17.7783179576146\\
54.7636191676224	17.6181904161888\\
55.083874250474	17.458062874763\\
}
\closedcycle;
\addplot[fill=white, draw=none, forget plot] table[row sep=crcr]{%
0	99.15\\
0.320255082851593	96.2609996344702\\
0.640510165703186	93.4569270092921\\
0.960765248554779	90.7373008899489\\
1.28102033140637	88.1016015667726\\
1.60127541425797	85.5492708549445\\
1.92153049710956	83.0797120944943\\
2.24178557996115	80.692290150301\\
2.56204066281274	78.3863314120924\\
2.88229574566434	76.1611237944452\\
3.20255082851593	74.0159167367851\\
3.52280591136752	71.9499212033867\\
3.84306099421912	69.9623096833735\\
4.16331607707071	68.0522161907179\\
4.4835711599223	66.2187362642413\\
4.8038262427739	64.4609269676139\\
5.12408132562549	62.7778068893549\\
5.44433640847708	61.1683561428325\\
5.76459149132867	59.6315163662638\\
6.08484657418027	58.1661907227146\\
6.40510165703186	56.7712439001\\
6.72535673988345	55.4455021111836\\
7.04561182273505	54.1877530935783\\
7.36586690558664	52.9967461097457\\
7.68612198843823	51.8711919469965\\
8.00637707128982	50.8097629174901\\
8.32663215414142	49.8110928582349\\
8.64688723699301	48.8737771310884\\
8.9671423198446	47.9963726227569\\
9.2873974026962	47.1773977447955\\
9.60765248554779	46.4153324336083\\
9.92790756839938	45.7086181504486\\
10.248162651251	45.0556578814181\\
10.5684177341026	44.4548161374678\\
10.8886728169542	43.9044189543977\\
11.2089278998058	43.4027538928563\\
11.5291829826573	42.9480700383415\\
11.8494380655089	42.5385780011997\\
12.1696931483605	42.1724499166266\\
12.4899482312121	41.8478194446666\\
12.8102033140637	41.562781770213\\
13.1304583969153	41.3153936030082\\
13.4507134797669	41.1036731776433\\
13.7709685626185	40.9256002535586\\
14.0912236454701	40.7791161150431\\
14.4114787283217	40.6621235712348\\
14.7317338111733	40.5724869561205\\
15.0519888940249	40.5080321285363\\
15.3722439768765	40.4665464721667\\
15.6924990597281	40.4457788955456\\
16.0127541425796	40.4434398320554\\
16.3330092254312	40.4572012399279\\
16.6532643082828	40.4846966022433\\
16.9735193911344	40.5235209269312\\
17.293774473986	40.5712307467698\\
17.6140295568376	40.6253441193863\\
17.9342846396892	40.6833406272569\\
18.2545397225408	40.7426613777068\\
18.5747948053924	40.8007090029098\\
18.895049888244	40.8548476598889\\
19.2153049710956	40.902403030516\\
19.5355600539472	40.9406623215119\\
19.8558151367988	40.9668742644462\\
20.1760702196504	40.9782491157376\\
20.496325302502	40.9719586566536\\
20.8165803853535	40.9451361933108\\
21.1368354682051	40.8948765566743\\
21.4570905510567	40.8182361025588\\
21.7773456339083	40.7122327116273\\
22.0976007167599	40.573845789392\\
22.4178557996115	40.400016266214\\
22.7381108824631	40.1876465973033\\
23.0583659653147	39.933600762719\\
23.3786210481663	39.6347042673687\\
23.6988761310179	39.2877441410094\\
24.0191312138695	38.8894689382467\\
24.3393862967211	38.4365887385353\\
24.6596413795727	37.9257751461787\\
24.9798964624243	37.3536612903294\\
25.3001515452758	36.7168418249888\\
25.6204066281274	36.0118729290073\\
25.940661710979	35.2352723060841\\
26.2609167938306	34.3835191847674\\
26.5811718766822	33.4530543184542\\
26.9014269595338	32.4402799853906\\
27.2216820423854	31.3415599886717\\
27.541937125237	30.153219656241\\
27.8621922080886	28.8715458408917\\
28.1824472909402	27.4927869202653\\
28.5027023737918	26.0131527968525\\
28.8229574566434	24.4288148979929\\
29.143212539495	22.7359061758749\\
29.4634676223466	20.930521107536\\
29.7837227051981	19.0087156948626\\
30.1039777880497	16.9665074645897\\
30.4242328709013	14.7998754683018\\
30.7444879537529	12.5047602824318\\
31.0647430366045	10.0770640082619\\
}
\closedcycle;
\addplot [color=black, densely dashed, line width=2pt]
  table[row sep=crcr]{%
0	99.15\\
0.320255082851593	96.2609996344702\\
0.640510165703186	93.4569270092921\\
0.960765248554779	90.7373008899489\\
1.28102033140637	88.1016015667726\\
1.60127541425797	85.5492708549445\\
1.92153049710956	83.0797120944943\\
2.24178557996115	80.692290150301\\
2.56204066281274	78.3863314120924\\
2.88229574566434	76.1611237944452\\
3.20255082851593	74.0159167367851\\
3.52280591136752	71.9499212033867\\
3.84306099421912	69.9623096833735\\
4.16331607707071	68.0522161907179\\
4.4835711599223	66.2187362642413\\
4.8038262427739	64.4609269676139\\
5.12408132562549	62.7778068893549\\
5.44433640847708	61.1683561428325\\
5.76459149132867	59.6315163662638\\
6.08484657418027	58.1661907227146\\
6.40510165703186	56.7712439001\\
6.72535673988345	55.4455021111836\\
7.04561182273505	54.1877530935783\\
7.36586690558664	52.9967461097457\\
7.68612198843823	51.8711919469965\\
8.00637707128982	50.8097629174901\\
8.32663215414142	49.8110928582349\\
8.64688723699301	48.8737771310884\\
8.9671423198446	47.9963726227569\\
9.2873974026962	47.1773977447955\\
9.60765248554779	46.4153324336083\\
9.92790756839938	45.7086181504486\\
10.248162651251	45.0556578814181\\
10.5684177341026	44.4548161374678\\
10.8886728169542	43.9044189543977\\
11.2089278998058	43.4027538928563\\
11.5291829826573	42.9480700383415\\
11.8494380655089	42.5385780011997\\
12.1696931483605	42.1724499166266\\
12.4899482312121	41.8478194446666\\
12.8102033140637	41.562781770213\\
13.1304583969153	41.3153936030082\\
13.4507134797669	41.1036731776433\\
13.7709685626185	40.9256002535586\\
14.0912236454701	40.7791161150431\\
14.4114787283217	40.6621235712348\\
14.7317338111733	40.5724869561205\\
15.0519888940249	40.5080321285363\\
15.3722439768765	40.4665464721667\\
15.6924990597281	40.4457788955456\\
16.0127541425796	40.4434398320554\\
16.3330092254312	40.4572012399279\\
16.6532643082828	40.4846966022433\\
16.9735193911344	40.5235209269312\\
17.293774473986	40.5712307467698\\
17.6140295568376	40.6253441193863\\
17.9342846396892	40.6833406272569\\
18.2545397225408	40.7426613777068\\
18.5747948053924	40.8007090029098\\
18.895049888244	40.8548476598889\\
19.2153049710956	40.902403030516\\
19.5355600539472	40.9406623215119\\
19.8558151367988	40.9668742644462\\
20.1760702196504	40.9782491157376\\
20.496325302502	40.9719586566536\\
20.8165803853535	40.9451361933108\\
21.1368354682051	40.8948765566743\\
21.4570905510567	40.8182361025588\\
21.7773456339083	40.7122327116273\\
22.0976007167599	40.573845789392\\
22.4178557996115	40.400016266214\\
22.7381108824631	40.1876465973033\\
23.0583659653147	39.933600762719\\
23.3786210481663	39.6347042673687\\
23.6988761310179	39.2877441410094\\
24.0191312138695	38.8894689382467\\
24.3393862967211	38.4365887385353\\
24.6596413795727	37.9257751461787\\
24.9798964624243	37.3536612903294\\
25.3001515452758	36.7168418249888\\
25.6204066281274	36.0118729290073\\
25.940661710979	35.2352723060841\\
26.2609167938306	34.3835191847674\\
26.5811718766822	33.4530543184542\\
26.9014269595338	32.4402799853906\\
27.2216820423854	31.3415599886717\\
27.541937125237	30.153219656241\\
27.8621922080886	28.8715458408917\\
28.1824472909402	27.4927869202653\\
28.5027023737918	26.0131527968525\\
28.8229574566434	24.4288148979929\\
29.143212539495	22.7359061758749\\
29.4634676223466	20.930521107536\\
29.7837227051981	19.0087156948626\\
30.1039777880497	16.9665074645897\\
30.4242328709013	14.7998754683018\\
30.7444879537529	12.5047602824318\\
31.0647430366045	10.0770640082619\\
};
\addlegendentry{$h_{\text{jump}}$}

\addplot [color=black, loosely dotted, line width=2pt]
  table[row sep=crcr]{%
0	45\\
0.320255082851593	44.8398724585742\\
0.640510165703186	44.6797449171484\\
0.960765248554779	44.5196173757226\\
1.28102033140637	44.3594898342968\\
1.60127541425797	44.199362292871\\
1.92153049710956	44.0392347514452\\
2.24178557996115	43.8791072100194\\
2.56204066281274	43.7189796685936\\
2.88229574566434	43.5588521271678\\
3.20255082851593	43.398724585742\\
3.52280591136752	43.2385970443162\\
3.84306099421912	43.0784695028904\\
4.16331607707071	42.9183419614646\\
4.4835711599223	42.7582144200388\\
4.8038262427739	42.5980868786131\\
5.12408132562549	42.4379593371873\\
5.44433640847708	42.2778317957615\\
5.76459149132867	42.1177042543357\\
6.08484657418027	41.9575767129099\\
6.40510165703186	41.7974491714841\\
6.72535673988345	41.6373216300583\\
7.04561182273505	41.4771940886325\\
7.36586690558664	41.3170665472067\\
7.68612198843823	41.1569390057809\\
8.00637707128982	40.9968114643551\\
8.32663215414142	40.8366839229293\\
8.64688723699301	40.6765563815035\\
8.9671423198446	40.5164288400777\\
9.2873974026962	40.3563012986519\\
9.60765248554779	40.1961737572261\\
9.92790756839938	40.0360462158003\\
10.248162651251	39.8759186743745\\
10.5684177341026	39.7157911329487\\
10.8886728169542	39.5556635915229\\
11.2089278998058	39.3955360500971\\
11.5291829826573	39.2354085086713\\
11.8494380655089	39.0752809672455\\
12.1696931483605	38.9151534258197\\
12.4899482312121	38.7550258843939\\
12.8102033140637	38.5948983429681\\
13.1304583969153	38.4347708015423\\
13.4507134797669	38.2746432601165\\
13.7709685626185	38.1145157186908\\
14.0912236454701	37.954388177265\\
14.4114787283217	37.7942606358392\\
14.7317338111733	37.6341330944134\\
15.0519888940249	37.4740055529876\\
15.3722439768765	37.3138780115618\\
15.6924990597281	37.153750470136\\
16.0127541425796	36.9936229287102\\
16.3330092254312	36.8334953872844\\
16.6532643082828	36.6733678458586\\
16.9735193911344	36.5132403044328\\
17.293774473986	36.353112763007\\
17.6140295568376	36.1929852215812\\
17.9342846396892	36.0328576801554\\
18.2545397225408	35.8727301387296\\
18.5747948053924	35.7126025973038\\
18.895049888244	35.552475055878\\
19.2153049710956	35.3923475144522\\
19.5355600539472	35.2322199730264\\
19.8558151367988	35.0720924316006\\
20.1760702196504	34.9119648901748\\
20.496325302502	34.751837348749\\
20.8165803853535	34.5917098073232\\
21.1368354682051	34.4315822658974\\
21.4570905510567	34.2714547244716\\
21.7773456339083	34.1113271830458\\
22.0976007167599	33.95119964162\\
22.4178557996115	33.7910721001942\\
22.7381108824631	33.6309445587684\\
23.0583659653147	33.4708170173427\\
23.3786210481663	33.3106894759169\\
23.6988761310179	33.1505619344911\\
24.0191312138695	32.9904343930653\\
24.3393862967211	32.8303068516395\\
24.6596413795727	32.6701793102137\\
24.9798964624243	32.5100517687879\\
25.3001515452758	32.3499242273621\\
25.6204066281274	32.1897966859363\\
25.940661710979	32.0296691445105\\
26.2609167938306	31.8695416030847\\
26.5811718766822	31.7094140616589\\
26.9014269595338	31.5492865202331\\
27.2216820423854	31.3891589788073\\
27.541937125237	31.2290314373815\\
27.8621922080886	31.0689038959557\\
28.1824472909402	30.9087763545299\\
28.5027023737918	30.7486488131041\\
28.8229574566434	30.5885212716783\\
29.143212539495	30.4283937302525\\
29.4634676223466	30.2682661888267\\
29.7837227051981	30.1081386474009\\
30.1039777880497	29.9480111059751\\
30.4242328709013	29.7878835645493\\
30.7444879537529	29.6277560231235\\
31.0647430366045	29.4676284816977\\
31.3849981194561	29.3075009402719\\
31.7052532023077	29.1473733988461\\
32.0255082851593	28.9872458574204\\
32.3457633680109	28.8271183159946\\
32.6660184508625	28.6669907745688\\
32.9862735337141	28.506863233143\\
33.3065286165657	28.3467356917172\\
33.6267836994173	28.1866081502914\\
33.9470387822689	28.0264806088656\\
34.2672938651205	27.8663530674398\\
34.587548947972	27.706225526014\\
34.9078040308236	27.5460979845882\\
35.2280591136752	27.3859704431624\\
35.5483141965268	27.2258429017366\\
35.8685692793784	27.0657153603108\\
36.18882436223	26.905587818885\\
36.5090794450816	26.7454602774592\\
36.8293345279332	26.5853327360334\\
37.1495896107848	26.4252051946076\\
37.4698446936364	26.2650776531818\\
37.790099776488	26.104950111756\\
38.1103548593396	25.9448225703302\\
38.4306099421912	25.7846950289044\\
38.7508650250427	25.6245674874786\\
39.0711201078943	25.4644399460528\\
39.3913751907459	25.304312404627\\
39.7116302735975	25.1441848632012\\
40.0318853564491	24.9840573217754\\
40.3521404393007	24.8239297803496\\
40.6723955221523	24.6638022389238\\
40.9926506050039	24.503674697498\\
41.3129056878555	24.3435471560723\\
41.6331607707071	24.1834196146465\\
41.9534158535587	24.0232920732207\\
42.2736709364103	23.8631645317949\\
42.5939260192619	23.7030369903691\\
42.9141811021135	23.5429094489433\\
43.2344361849651	23.3827819075175\\
43.5546912678166	23.2226543660917\\
43.8749463506682	23.0625268246659\\
44.1952014335198	22.9023992832401\\
44.5154565163714	22.7422717418143\\
44.835711599223	22.5821442003885\\
45.1559666820746	22.4220166589627\\
45.4762217649262	22.2618891175369\\
45.7964768477778	22.1017615761111\\
46.1167319306294	21.9416340346853\\
46.436987013481	21.7815064932595\\
46.7572420963326	21.6213789518337\\
47.0774971791842	21.4612514104079\\
47.3977522620358	21.3011238689821\\
47.7180073448874	21.1409963275563\\
48.0382624277389	20.9808687861305\\
48.3585175105905	20.8207412447047\\
48.6787725934421	20.6606137032789\\
48.9990276762937	20.5004861618531\\
49.3192827591453	20.3403586204273\\
49.6395378419969	20.1802310790015\\
49.9597929248485	20.0201035375757\\
50.2800480077001	19.8599759961499\\
50.6003030905517	19.6998484547242\\
50.9205581734033	19.5397209132984\\
51.2408132562549	19.3795933718726\\
51.5610683391065	19.2194658304468\\
51.8813234219581	19.059338289021\\
52.2015785048097	18.8992107475952\\
52.5218335876612	18.7390832061694\\
52.8420886705128	18.5789556647436\\
53.1623437533644	18.4188281233178\\
53.482598836216	18.258700581892\\
53.8028539190676	18.0985730404662\\
54.1231090019192	17.9384454990404\\
54.4433640847708	17.7783179576146\\
54.7636191676224	17.6181904161888\\
55.083874250474	17.458062874763\\
};
\addlegendentry{$h_{\text{land}}$}

\addplot[fill=mycolor1, fill opacity=0.8, draw=none, forget plot] table[row sep=crcr]{%
0	45\\
0.320255082851593	44.8398724585742\\
0.640510165703186	44.6797449171484\\
0.960765248554779	44.5196173757226\\
1.28102033140637	44.3594898342968\\
1.60127541425797	44.199362292871\\
1.92153049710956	44.0392347514452\\
2.24178557996115	43.8791072100194\\
2.56204066281274	43.7189796685936\\
2.88229574566434	43.5588521271678\\
3.20255082851593	43.398724585742\\
3.52280591136752	43.2385970443162\\
3.84306099421912	43.0784695028904\\
4.16331607707071	42.9183419614646\\
4.4835711599223	42.7582144200388\\
4.8038262427739	42.5980868786131\\
5.12408132562549	42.4379593371873\\
5.44433640847708	42.2778317957615\\
5.76459149132867	42.1177042543357\\
6.08484657418027	41.9575767129099\\
6.40510165703186	41.7974491714841\\
6.72535673988345	41.6373216300583\\
7.04561182273505	41.4771940886325\\
7.36586690558664	41.3170665472067\\
7.68612198843823	41.1569390057809\\
8.00637707128982	40.9968114643551\\
8.32663215414142	40.8366839229293\\
8.64688723699301	40.6765563815035\\
8.9671423198446	40.5164288400777\\
9.2873974026962	40.3563012986519\\
9.60765248554779	40.1961737572261\\
9.92790756839938	40.0360462158003\\
10.248162651251	39.8759186743745\\
10.5684177341026	39.7157911329487\\
10.8886728169542	39.5556635915229\\
11.2089278998058	39.3955360500971\\
11.5291829826573	39.2354085086713\\
11.8494380655089	39.0752809672455\\
12.1696931483605	38.9151534258197\\
12.4899482312121	38.7550258843939\\
12.8102033140637	38.5948983429681\\
13.1304583969153	38.4347708015423\\
13.4507134797669	38.2746432601165\\
13.7709685626185	38.1145157186908\\
14.0912236454701	37.954388177265\\
14.4114787283217	37.7942606358392\\
14.7317338111733	37.6341330944134\\
15.0519888940249	37.4740055529876\\
15.3722439768765	37.3138780115618\\
15.6924990597281	37.153750470136\\
16.0127541425796	36.9936229287102\\
16.3330092254312	36.8334953872844\\
16.6532643082828	36.6733678458586\\
16.9735193911344	36.5132403044328\\
17.293774473986	36.353112763007\\
17.6140295568376	36.1929852215812\\
17.9342846396892	36.0328576801554\\
18.2545397225408	35.8727301387296\\
18.5747948053924	35.7126025973038\\
18.895049888244	35.552475055878\\
19.2153049710956	35.3923475144522\\
19.5355600539472	35.2322199730264\\
19.8558151367988	35.0720924316006\\
20.1760702196504	34.9119648901748\\
20.496325302502	34.751837348749\\
20.8165803853535	34.5917098073232\\
21.1368354682051	34.4315822658974\\
21.4570905510567	34.2714547244716\\
21.7773456339083	34.1113271830458\\
22.0976007167599	33.95119964162\\
22.4178557996115	33.7910721001942\\
22.7381108824631	33.6309445587684\\
23.0583659653147	33.4708170173427\\
23.3786210481663	33.3106894759169\\
23.6988761310179	33.1505619344911\\
24.0191312138695	32.9904343930653\\
24.3393862967211	32.8303068516395\\
24.6596413795727	32.6701793102137\\
24.9798964624243	32.5100517687879\\
25.3001515452758	32.3499242273621\\
25.6204066281274	32.1897966859363\\
25.940661710979	32.0296691445105\\
26.2609167938306	31.8695416030847\\
26.5811718766822	31.7094140616589\\
26.9014269595338	31.5492865202331\\
27.2216820423854	31.3891589788073\\
27.541937125237	31.2290314373815\\
27.8621922080886	31.0689038959557\\
28.1824472909402	30.9087763545299\\
28.5027023737918	30.7486488131041\\
28.8229574566434	30.5885212716783\\
29.143212539495	30.4283937302525\\
29.4634676223466	30.2682661888267\\
29.7837227051981	30.1081386474009\\
30.1039777880497	29.9480111059751\\
30.4242328709013	29.7878835645493\\
30.7444879537529	29.6277560231235\\
31.0647430366045	29.4676284816977\\
31.3849981194561	29.3075009402719\\
31.7052532023077	29.1473733988461\\
32.0255082851593	28.9872458574204\\
32.3457633680109	28.8271183159946\\
32.6660184508625	28.6669907745688\\
32.9862735337141	28.506863233143\\
33.3065286165657	28.3467356917172\\
33.6267836994173	28.1866081502914\\
33.9470387822689	28.0264806088656\\
34.2672938651205	27.8663530674398\\
34.587548947972	27.706225526014\\
34.9078040308236	27.5460979845882\\
35.2280591136752	27.3859704431624\\
35.5483141965268	27.2258429017366\\
35.8685692793784	27.0657153603108\\
36.18882436223	26.905587818885\\
36.5090794450816	26.7454602774592\\
36.8293345279332	26.5853327360334\\
37.1495896107848	26.4252051946076\\
37.4698446936364	26.2650776531818\\
37.790099776488	26.104950111756\\
38.1103548593396	25.9448225703302\\
38.4306099421912	25.7846950289044\\
38.7508650250427	25.6245674874786\\
39.0711201078943	25.4644399460528\\
39.3913751907459	25.304312404627\\
39.7116302735975	25.1441848632012\\
40.0318853564491	24.9840573217754\\
40.3521404393007	24.8239297803496\\
40.6723955221523	24.6638022389238\\
40.9926506050039	24.503674697498\\
41.3129056878555	24.3435471560723\\
41.6331607707071	24.1834196146465\\
41.9534158535587	24.0232920732207\\
42.2736709364103	23.8631645317949\\
42.5939260192619	23.7030369903691\\
42.9141811021135	23.5429094489433\\
43.2344361849651	23.3827819075175\\
43.5546912678166	23.2226543660917\\
43.8749463506682	23.0625268246659\\
44.1952014335198	22.9023992832401\\
44.5154565163714	22.7422717418143\\
44.835711599223	22.5821442003885\\
45.1559666820746	22.4220166589627\\
45.4762217649262	22.2618891175369\\
45.7964768477778	22.1017615761111\\
46.1167319306294	21.9416340346853\\
46.436987013481	21.7815064932595\\
46.7572420963326	21.6213789518337\\
47.0774971791842	21.4612514104079\\
47.3977522620358	21.3011238689821\\
47.7180073448874	21.1409963275563\\
48.0382624277389	20.9808687861305\\
48.3585175105905	20.8207412447047\\
48.6787725934421	20.6606137032789\\
48.9990276762937	20.5004861618531\\
49.3192827591453	20.3403586204273\\
49.6395378419969	20.1802310790015\\
49.9597929248485	20.0201035375757\\
50.2800480077001	19.8599759961499\\
50.6003030905517	19.6998484547242\\
50.9205581734033	19.5397209132984\\
51.2408132562549	19.3795933718726\\
51.5610683391065	19.2194658304468\\
51.8813234219581	19.059338289021\\
52.2015785048097	18.8992107475952\\
52.5218335876612	18.7390832061694\\
52.8420886705128	18.5789556647436\\
53.1623437533644	18.4188281233178\\
53.482598836216	18.258700581892\\
53.8028539190676	18.0985730404662\\
54.1231090019192	17.9384454990404\\
54.4433640847708	17.7783179576146\\
54.7636191676224	17.6181904161888\\
55.083874250474	17.458062874763\\
}
\closedcycle;
\addplot[fill=mycolor2, fill opacity=0.8, draw=none, forget plot] table[row sep=crcr]{%
0	99.15\\
0.320255082851593	96.2609996344702\\
0.640510165703186	93.4569270092921\\
0.960765248554779	90.7373008899489\\
1.28102033140637	88.1016015667726\\
1.60127541425797	85.5492708549445\\
1.92153049710956	83.0797120944943\\
2.24178557996115	80.692290150301\\
2.56204066281274	78.3863314120924\\
2.88229574566434	76.1611237944452\\
3.20255082851593	74.0159167367851\\
3.52280591136752	71.9499212033867\\
3.84306099421912	69.9623096833735\\
4.16331607707071	68.0522161907179\\
4.4835711599223	66.2187362642413\\
4.8038262427739	64.4609269676139\\
5.12408132562549	62.7778068893549\\
5.44433640847708	61.1683561428325\\
5.76459149132867	59.6315163662638\\
6.08484657418027	58.1661907227146\\
6.40510165703186	56.7712439001\\
6.72535673988345	55.4455021111836\\
7.04561182273505	54.1877530935783\\
7.36586690558664	52.9967461097457\\
7.68612198843823	51.8711919469965\\
8.00637707128982	50.8097629174901\\
8.32663215414142	49.8110928582349\\
8.64688723699301	48.8737771310884\\
8.9671423198446	47.9963726227569\\
9.2873974026962	47.1773977447955\\
9.60765248554779	46.4153324336083\\
9.92790756839938	45.7086181504486\\
10.248162651251	45.0556578814181\\
10.5684177341026	44.4548161374678\\
10.8886728169542	43.9044189543977\\
11.2089278998058	43.4027538928563\\
11.5291829826573	42.9480700383415\\
11.8494380655089	42.5385780011997\\
12.1696931483605	42.1724499166266\\
12.4899482312121	41.8478194446666\\
12.8102033140637	41.562781770213\\
13.1304583969153	41.3153936030082\\
13.4507134797669	41.1036731776433\\
13.7709685626185	40.9256002535586\\
14.0912236454701	40.7791161150431\\
14.4114787283217	40.6621235712348\\
14.7317338111733	40.5724869561205\\
15.0519888940249	40.5080321285363\\
15.3722439768765	40.4665464721667\\
15.6924990597281	40.4457788955456\\
16.0127541425796	40.4434398320554\\
16.3330092254312	40.4572012399279\\
16.6532643082828	40.4846966022433\\
16.9735193911344	40.5235209269312\\
17.293774473986	40.5712307467698\\
17.6140295568376	40.6253441193863\\
17.9342846396892	40.6833406272569\\
18.2545397225408	40.7426613777068\\
18.5747948053924	40.8007090029098\\
18.895049888244	40.8548476598889\\
19.2153049710956	40.902403030516\\
19.5355600539472	40.9406623215119\\
19.8558151367988	40.9668742644462\\
20.1760702196504	40.9782491157376\\
20.496325302502	40.9719586566536\\
20.8165803853535	40.9451361933108\\
21.1368354682051	40.8948765566743\\
21.4570905510567	40.8182361025588\\
21.7773456339083	40.7122327116273\\
22.0976007167599	40.573845789392\\
22.4178557996115	40.400016266214\\
22.7381108824631	40.1876465973033\\
23.0583659653147	39.933600762719\\
23.3786210481663	39.6347042673687\\
23.6988761310179	39.2877441410094\\
24.0191312138695	38.8894689382467\\
24.3393862967211	38.4365887385353\\
24.6596413795727	37.9257751461787\\
24.9798964624243	37.3536612903294\\
25.3001515452758	36.7168418249888\\
25.6204066281274	36.0118729290073\\
25.940661710979	35.2352723060841\\
26.2609167938306	34.3835191847674\\
26.5811718766822	33.4530543184542\\
26.9014269595338	32.4402799853906\\
27.2216820423854	31.3415599886717\\
27.541937125237	30.153219656241\\
27.8621922080886	28.8715458408917\\
28.1824472909402	27.4927869202653\\
28.5027023737918	26.0131527968525\\
28.8229574566434	24.4288148979929\\
29.143212539495	22.7359061758749\\
29.4634676223466	20.930521107536\\
29.7837227051981	19.0087156948626\\
30.1039777880497	16.9665074645897\\
30.4242328709013	14.7998754683018\\
30.7444879537529	12.5047602824318\\
31.0647430366045	10.0770640082619\\
}
\closedcycle;

\node (R1) [color=black, above, align=left]
at (axis cs:30, 40) {$R_1$};
\node (R2) [color=black, right, align=left]
at (axis cs:8, 27) {$R_2$};
\node (R3) [color=black, right, align=left]
at (axis cs:38, 17) {$R_3$};
\addplot[only marks, mark=*, mark options={}, mark size=2.5pt, color=red, fill=red, forget plot] table[row sep=crcr]{%
x	y\\
20.1760702196504	40.9782491157376\\
};
\node (q) [color=black, above, align=left]
at (axis cs:20.1760702196504,	40.9782491157376) {$q$};
\end{axis}
\end{tikzpicture}%

%% file: Graphics/tikz/pump_model.tex
%
%
\begin{tikzpicture}

\begin{axis}[%
width=2.25in,
height=1in,
scale only axis,
xmin=0,
xmax=4,
xlabel style={font=\color{white!15!black}},
xlabel={$y\,$[m]},
xtick=\empty,
xticklabel=\empty,
ymin=0,
ymax=1,
ytick=\empty,
yticklabel=\empty,
ylabel style={font=\color{white!15!black}},
ylabel={$z\,$[m]},
axis background/.style={fill=white},
legend style={legend cell align=left, align=left, draw=white!15!black}
]
\addplot [color=black, line width=1.5pt]
  table[row sep=crcr]{%
-5.77867381379646e-21	0\\
0.0403807678805997	0.000651888125948213\\
0.14896915667307	0.00881125431858437\\
0.188080606500601	0.0139836651540601\\
0.219960701648784	0.0190429724274414\\
0.304773939380283	0.0360186120781717\\
0.334981443405974	0.0432310575778511\\
0.335002590415998	0.0432363102707481\\
0.33506036512195	0.0432506623242105\\
0.335081512132075	0.0432559160636131\\
0.335081512146466	0.0432559160671885\\
0.335081512191968	0.0432559160784933\\
0.335081512210887	0.0432559160831935\\
0.335081541165613	0.0432559232768654\\
0.335081620271008	0.0432559429302485\\
0.33508164922545	0.0432559501238516\\
0.398325709719208	0.0601790153119863\\
0.561434253799376	0.113380257521995\\
0.617593934720888	0.134131251524562\\
0.635951416058013	0.14110741178846\\
0.685516474024654	0.160312515869508\\
0.703443133399206	0.167364580637138\\
0.721104198492855	0.174353228229787\\
0.768947133591517	0.193420775279281\\
0.786309527773309	0.200364545548487\\
0.803516406276199	0.207245711205301\\
0.850259222946039	0.225871720300797\\
0.867270461134348	0.232602769562717\\
0.88419001322756	0.239260123585705\\
0.930317007475938	0.257159337401859\\
0.947164758772977	0.263583684000824\\
0.963990358401288	0.269927543990819\\
1.00992832751983	0.2868238434986\\
1.02673165528281	0.292828626305973\\
1.04352754175246	0.298725989069557\\
1.08942744533873	0.314255017443283\\
1.10623262035674	0.319706647567837\\
1.13188842290484	0.327766612979281\\
1.20209568956132	0.348043670478757\\
1.2278351732663	0.354767039802275\\
1.25233755725364	0.360786555199534\\
1.31935887517584	0.375240127023616\\
1.34391950109778	0.379761610721282\\
1.3439195013771	0.379761610770257\\
1.3439195021402	0.379761610904057\\
1.34391950241951	0.379761610953031\\
1.35851755757464	0.382244215973228\\
1.39840018595013	0.388229136389234\\
1.41299824110526	0.39012230124917\\
1.43080535561526	0.392212085697019\\
1.47945529719279	0.396671997367925\\
1.49726241170279	0.397841000950732\\
1.51509624630267	0.39876028328246\\
1.56381918852331	0.399980528132585\\
1.58165302312319	0.399952854710283\\
1.59838443574899	0.399695635727088\\
1.64409550512506	0.397854738313292\\
1.66082691775086	0.396766547515032\\
1.6775740987954	0.395456709172965\\
1.72332824829264	0.390765555801601\\
1.74007542933718	0.388646901093894\\
1.75683583993185	0.386314704648317\\
1.8026261332322	0.378884364527983\\
1.81938654382687	0.375786168729894\\
1.83612770684379	0.372494366743647\\
1.8818654147838	0.362526848348802\\
1.89860657780072	0.358533958183855\\
1.91535421429355	0.354361711102923\\
1.96110960809865	0.342094549981789\\
1.97785724459148	0.33730143832635\\
1.99460744934674	0.332353527837004\\
2.04036985977529	0.318095320841145\\
2.05712006453054	0.312622737578413\\
2.07386123522565	0.307026824866896\\
2.11959896414288	0.29114580311634\\
2.13634013483799	0.285135180505717\\
2.15308757498745	0.279026820853144\\
2.19884243237251	0.2619070263783\\
2.21558987252197	0.255503517686161\\
2.2323402659424	0.249036599779944\\
2.27810319181379	0.231109682059621\\
2.29485358523422	0.224474859530176\\
2.29746437102326	0.223438065318771\\
2.3045971902801	0.220602308389816\\
2.30720799058826	0.219563267740549\\
2.30720794639063	0.21956328533482\\
2.30720840866808	0.219563101310579\\
2.30720879127629	0.2195629490012\\
2.35808555624472	0.199243575382436\\
2.50272144613595	0.142224544105973\\
2.55772550116584	0.121552635741589\\
2.6187495790192	0.0997383318013912\\
2.80101945190002	0.0446297284995498\\
2.87342618168455	0.0280823448436449\\
2.87342618168838	0.028082344842862\\
2.87342618169885	0.028082344840723\\
2.87342618170268	0.02808234483994\\
2.87342618259185	0.0280823446581972\\
2.87342618502112	0.0280823441616669\\
2.8734261859103	0.0280823439799243\\
2.89441566169949	0.0239449191310841\\
2.95292900039696	0.0140694651385368\\
2.97477426145481	0.0110284769455859\\
2.99614495461475	0.0084025099238525\\
3.05528272799449	0.00297236949997887\\
3.07720388959501	0.0016560746254877\\
3.11644222599968	0.000252964259429139\\
3.22304037919002	0.00264763046948602\\
3.26183730691369	0.00575569017959715\\
3.26183730691381	0.00575569017960897\\
3.26183730691415	0.00575569017964123\\
3.26183730691427	0.00575569017965299\\
3.2618373069144	0.00575569017966482\\
3.26183730691473	0.00575569017969707\\
3.26183730691486	0.00575569017970873\\
3.34170620751054	0.0158054939417487\\
3.54939568179581	0.0629145550485452\\
3.62156602857422	0.0852872785528315\\
3.62156602909721	0.0852872787241992\\
3.62156603044558	0.0852872791660117\\
3.62156603090966	0.0852872793180728\\
3.65515651363302	0.0965441758746963\\
3.74583676779589	0.129113313007161\\
3.77862883836301	0.141522219349656\\
3.81300906316795	0.154801162403156\\
3.90682233101098	0.191934793135847\\
3.94111818469121	0.205650195224765\\
3.94855252036893	0.20862200848442\\
3.96885623949837	0.216726608370417\\
3.97628525780601	0.219685849813515\\
3.98371855985915	0.222642447946365\\
4.00402752594086	0.230692911871686\\
};
\addlegendentry{$h_t$}

\addplot [color=black, dashed, line width=1.5pt, forget plot]
  table[row sep=crcr]{%
2	0.6\\
2	0.330728724172722\\
};
\addplot[fill=black, fill opacity=0.3, draw=black, forget plot] table[row sep=crcr]{%
-5.77867381379646e-21	0\\
0.0403807678805997	0.000651888125948213\\
0.14896915667307	0.00881125431858437\\
0.188080606500601	0.0139836651540601\\
0.219960701648784	0.0190429724274414\\
0.304773939380283	0.0360186120781717\\
0.334981443405974	0.0432310575778511\\
0.335002590415998	0.0432363102707481\\
0.33506036512195	0.0432506623242105\\
0.335081512132075	0.0432559160636131\\
0.335081512146466	0.0432559160671885\\
0.335081512191968	0.0432559160784933\\
0.335081512210887	0.0432559160831935\\
0.335081541165613	0.0432559232768654\\
0.335081620271008	0.0432559429302485\\
0.33508164922545	0.0432559501238516\\
0.398325709719208	0.0601790153119863\\
0.561434253799376	0.113380257521995\\
0.617593934720888	0.134131251524562\\
0.635951416058013	0.14110741178846\\
0.685516474024654	0.160312515869508\\
0.703443133399206	0.167364580637138\\
0.721104198492855	0.174353228229787\\
0.768947133591517	0.193420775279281\\
0.786309527773309	0.200364545548487\\
0.803516406276199	0.207245711205301\\
0.850259222946039	0.225871720300797\\
0.867270461134348	0.232602769562717\\
0.88419001322756	0.239260123585705\\
0.930317007475938	0.257159337401859\\
0.947164758772977	0.263583684000824\\
0.963990358401288	0.269927543990819\\
1.00992832751983	0.2868238434986\\
1.02673165528281	0.292828626305973\\
1.04352754175246	0.298725989069557\\
1.08942744533873	0.314255017443283\\
1.10623262035674	0.319706647567837\\
1.13188842290484	0.327766612979281\\
1.20209568956132	0.348043670478757\\
1.2278351732663	0.354767039802275\\
1.25233755725364	0.360786555199534\\
1.31935887517584	0.375240127023616\\
1.34391950109778	0.379761610721282\\
1.3439195013771	0.379761610770257\\
1.3439195021402	0.379761610904057\\
1.34391950241951	0.379761610953031\\
1.35851755757464	0.382244215973228\\
1.39840018595013	0.388229136389234\\
1.41299824110526	0.39012230124917\\
1.43080535561526	0.392212085697019\\
1.47945529719279	0.396671997367925\\
1.49726241170279	0.397841000950732\\
1.51509624630267	0.39876028328246\\
1.56381918852331	0.399980528132585\\
1.58165302312319	0.399952854710283\\
1.59838443574899	0.399695635727088\\
1.64409550512506	0.397854738313292\\
1.66082691775086	0.396766547515032\\
1.6775740987954	0.395456709172965\\
1.72332824829264	0.390765555801601\\
1.74007542933718	0.388646901093894\\
1.75683583993185	0.386314704648317\\
1.8026261332322	0.378884364527983\\
1.81938654382687	0.375786168729894\\
1.83612770684379	0.372494366743647\\
1.8818654147838	0.362526848348802\\
1.89860657780072	0.358533958183855\\
1.91535421429355	0.354361711102923\\
1.96110960809865	0.342094549981789\\
1.97785724459148	0.33730143832635\\
1.99460744934674	0.332353527837004\\
2.04036985977529	0.318095320841145\\
2.05712006453054	0.312622737578413\\
2.07386123522565	0.307026824866896\\
2.11959896414288	0.29114580311634\\
2.13634013483799	0.285135180505717\\
2.15308757498745	0.279026820853144\\
2.19884243237251	0.2619070263783\\
2.21558987252197	0.255503517686161\\
2.2323402659424	0.249036599779944\\
2.27810319181379	0.231109682059621\\
2.29485358523422	0.224474859530176\\
2.29746437102326	0.223438065318771\\
2.3045971902801	0.220602308389816\\
2.30720799058826	0.219563267740549\\
2.30720794639063	0.21956328533482\\
2.30720840866808	0.219563101310579\\
2.30720879127629	0.2195629490012\\
2.35808555624472	0.199243575382436\\
2.50272144613595	0.142224544105973\\
2.55772550116584	0.121552635741589\\
2.6187495790192	0.0997383318013912\\
2.80101945190002	0.0446297284995498\\
2.87342618168455	0.0280823448436449\\
2.87342618168838	0.028082344842862\\
2.87342618169885	0.028082344840723\\
2.87342618170268	0.02808234483994\\
2.87342618259185	0.0280823446581972\\
2.87342618502112	0.0280823441616669\\
2.8734261859103	0.0280823439799243\\
2.89441566169949	0.0239449191310841\\
2.95292900039696	0.0140694651385368\\
2.97477426145481	0.0110284769455859\\
2.99614495461475	0.0084025099238525\\
3.05528272799449	0.00297236949997887\\
3.07720388959501	0.0016560746254877\\
3.11644222599968	0.000252964259429139\\
3.22304037919002	0.00264763046948602\\
3.26183730691369	0.00575569017959715\\
3.26183730691381	0.00575569017960897\\
3.26183730691415	0.00575569017964123\\
3.26183730691427	0.00575569017965299\\
3.2618373069144	0.00575569017966482\\
3.26183730691473	0.00575569017969707\\
3.26183730691486	0.00575569017970873\\
3.34170620751054	0.0158054939417487\\
3.54939568179581	0.0629145550485452\\
3.62156602857422	0.0852872785528315\\
3.62156602909721	0.0852872787241992\\
3.62156603044558	0.0852872791660117\\
3.62156603090966	0.0852872793180728\\
3.65515651363302	0.0965441758746963\\
3.74583676779589	0.129113313007161\\
3.77862883836301	0.141522219349656\\
3.81300906316795	0.154801162403156\\
3.90682233101098	0.191934793135847\\
3.94111818469121	0.205650195224765\\
3.94855252036893	0.20862200848442\\
3.96885623949837	0.216726608370417\\
3.97628525780601	0.219685849813515\\
3.98371855985915	0.222642447946365\\
4.00402752594086	0.230692911871686\\
}
\closedcycle;
\addplot[only marks, mark=*, mark options={}, mark size=2.5pt, color=red, fill=red, forget plot] table[row sep=crcr]{%
x	y\\
2	0.6\\
};
\node (PB) [color=black, right, align=left]
at (axis cs:2, 0.6) {$P_b$};
\addplot[only marks, mark=*, mark options={}, mark size=2.5pt, color=red, fill=red, forget plot] table[row sep=crcr]{%
x	y\\
2	0.9\\
};
\node (PR) [color=black, right, align=left]
at (axis cs:2, 0.9) {$P_r$};
\addplot [color=red, line width=2pt, forget plot]
  table[row sep=crcr]{%
2	0.6\\
2	0.9\\
};
\addplot [color=black, dashed, line width=1.5pt, forget plot]
  table[row sep=crcr]{%
2	0.6\\
0	0.6\\
};
\node (q1) [color=black, above, align=left]
at (axis cs:1, 0.6) {$q_1$};
\node (q2) [color=black, left, align=left]
at (axis cs:2, 0.45) {$q_2$};
\node (q3) [color=black, left, align=left]
at (axis cs:2, 0.75) {$q_3$};

\node (R1) [color=black, above, align=left]
at (axis cs:3, 0.4) {$R_1$};
\node (R2) [color=black, right, align=left]
at (axis cs:1.25, 0.2) {$R_2$};
\end{axis}
\end{tikzpicture}%

%% file: Content/5_results.tex
\section{SIMULATION RESULTS}
\label{sec:results}
Implementations of the examples and MATLAB simulation code are publicly available at [\url{https://github.com/OptCon/PumpTrackOC}]. This repository includes the implementation of both examples using \textit{nosnoc}~\cite{nurkanovic2022nosnoc} and \textit{CasADi}~\cite{casad}. We employ the finite element with switch detection method~\cite{nurkanovic2022finite}, an event-driven integration scheme.

\subsubsection*{Ski jump}
\begin{figure}
    \centering
    \begin{subfigure}{\columnwidth}
        \centering
        \input{Graphics/tikz/ski_sim_45.tex}
        \caption{Trajectory for initial height of $45$m}
        \label{fig:ski_sim_45}
    \end{subfigure}
    \begin{subfigure}{\columnwidth}
        \centering
        \input{Graphics/tikz/ski_sim_50.tex}  
        \caption{Trajectory for initial height of $50$m}
        \label{fig:ski_sim_50}
    \end{subfigure}
    \caption{Ski jumping -- results for two different initial heights.}
    \label{fig:ski_sim_results}
\end{figure}
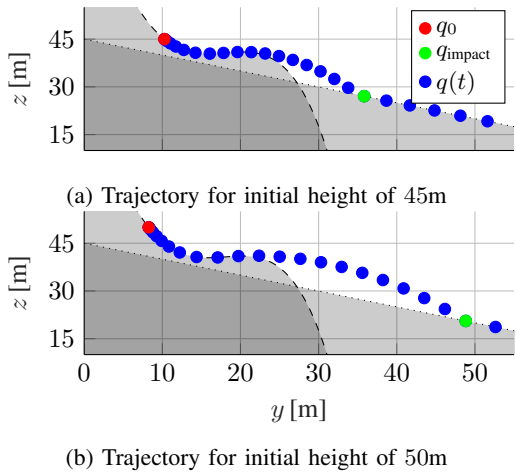

For the ski jump example, we use the automatic model reformulation provided by \textit{nosnoc}. Here, we provide the mass matrix, which is trivial in the case of a single point mass, the reduced dynamics~\eqref{eq:ski_free_dynamics}, and the impact surfaces~\eqref{eq:ski_impact_surfaces}. Simulation results are shown in Figure~\ref{fig:ski_sim_results}. 
The figure depicts the trajectories  with the starting and impact position highlighted. As anticipated, increasing the initial height shifts the point of impact further to the right. While these results are predictable, the simplicity of the problem is precisely what makes it well-suited for prototyping and testing new solution approaches. The complexity, in its current form, stems primarily from handling nonlinear impact surfaces.

\subsubsection*{Bicycle pumping}
As discussed in Section~\ref{subsec:time_freezing}, applying the time-freezing approach to the pump track problem involves the design of  auxiliary dynamics to simulate the impact behavior in its infeasible region while freezing physical time. For the extended state $\hat{x}(\tau)=(x(\tau), t(\tau)) \in \mathbb{R}^7$, we define the auxiliary dynamics as \begin{equation*}
    \label{eq:pump_aux_dyn}
    \hat{f}_{\text{aux}}(\hat{x})=k_n\begin{bmatrix}
        0,\:0,\:0,\:-h_t^\prime(q_1),\:1,\:0,\:0
    \end{bmatrix}^\top\in \mathbb{R}^7
\end{equation*} with $k_n>0$. By extending the reduced dynamics to the augmented state, we obtain $\hat{f}_{\text{free}}(\hat{x},u)=(f_{\text{free}}(x,u),\, 1)$, which incorporates the additional time dimension. Using this formulation, and the regions $R_1$ and $R_2$ from \eqref{eq:tf_regions}, we obtain the time-freezing system~\eqref{eq:tf_system_di}.

\begin{figure*}
    \centering
    \begin{subfigure}{0.33\textwidth}
        \input{Graphics/tikz/pump_sim_v1_10.tex}  
        \caption{$v_{1,0}=10\,$[km/h] without input}
        \label{fig:pump_sim_10}
    \end{subfigure}
    \begin{subfigure}{0.33\textwidth}
        \centering
        \input{Graphics/tikz/pump_sim_v1_15.tex}   
        \caption{$v_{1,0}=15\,$[km/h] without input}
        \label{fig:pump_sim_15}
    \end{subfigure}\begin{subfigure}{0.33\textwidth}
        \input{Graphics/tikz/pump_sim_v1_22.tex}   
        \caption{$v_{1,0}=22\,$[km/h] without input}
        \label{fig:pump_sim_22}
    \end{subfigure}
    \newline
    \begin{subfigure}{0.33\textwidth}
        \input{Graphics/tikz/pump_opt_v1_10.tex}  
        \caption{$v_{1,0}=10\,$[km/h] with optimal input $u^\star$}
        \label{fig:pump_opt_10}
    \end{subfigure}
    \begin{subfigure}{0.33\textwidth}
        \centering
        \input{Graphics/tikz/pump_opt_v1_15.tex}  
       \caption{$v_{1,0}=15\,$[km/h] with optimal input $u^\star$}
        \label{fig:pump_opt_15}
    \end{subfigure}\begin{subfigure}{0.33\textwidth}
        \input{Graphics/tikz/pump_opt_v1_22.tex}  
       \caption{$v_{1,0}=22\,$[km/h] with optimal input $u^\star$}
        \label{fig:pump_opt_22}
    \end{subfigure}
    \caption{Bicycle on pump track -- simulation results for three different scenarios. The first row shows the system without input, and the second row shows the system with the optimal input $u^\star$.}
    \label{fig:pump_sim_vs_opt_results}
\end{figure*}
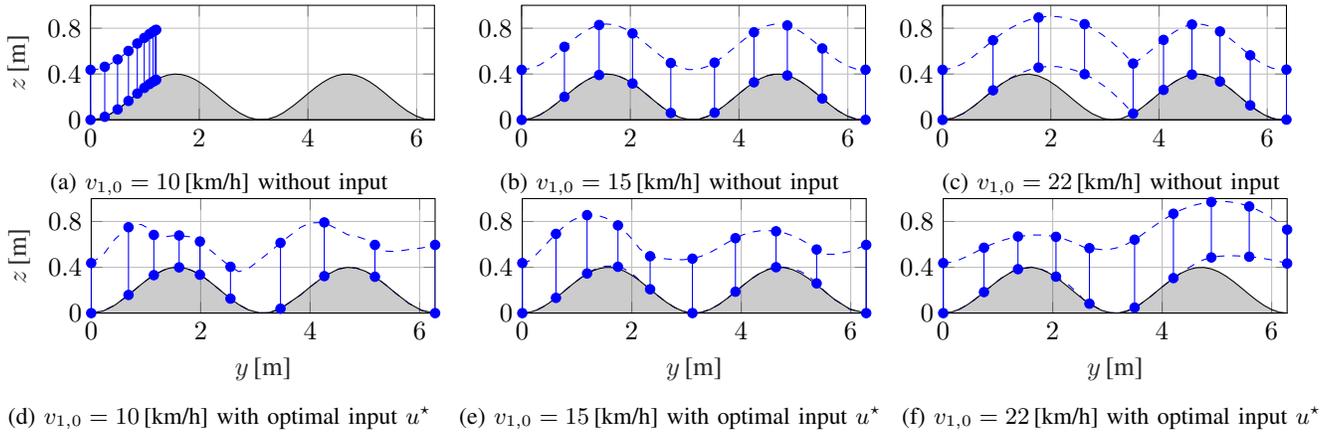

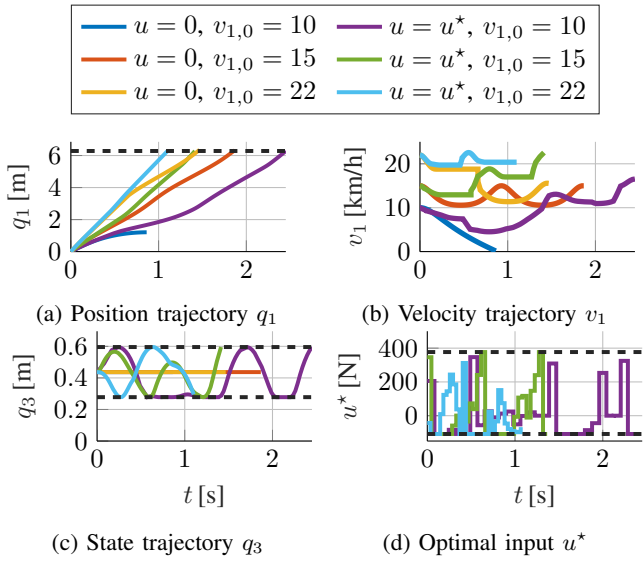
\begin{figure}
    \centering
    \begin{subfigure}{0.49\columnwidth}
        \centering
        \input{Graphics/tikz/pump_q1_over_t.tex}  
        \caption{Position trajectory $q_1$}
        \label{fig:pump_opt_q1}
    \end{subfigure}
    \begin{subfigure}{0.49\columnwidth}
        \centering
        \input{Graphics/tikz/pump_v1_over_t.tex}   
        \caption{Velocity trajectory $v_1$}
        \label{fig:pump_opt_v1}
    \end{subfigure}
   \newline
    \begin{subfigure}{0.49\columnwidth}
        \centering
        \input{Graphics/tikz/pump_s_over_t.tex}   
        \caption{State trajectory $q_3$}
        \label{fig:pump_opt_q3}
    \end{subfigure}
    \begin{subfigure}{0.49\columnwidth}
        \centering
        \input{Graphics/tikz/pump_u_over_t.tex}  
        \caption{Optimal input $u^\star$}
        \label{fig:pump_opt_u}
    \end{subfigure}
    \caption{Bicycle on pump track -- solution trajectories over time. The optimal input $u^\star$, the pumping motion, accelerates the system to reach the goal faster and to gain velocity.}
    \label{fig:pump_opt_time_results}
\end{figure}

Figure~\ref{fig:pump_sim_vs_opt_results} shows the simulation results for the pump track system. It compares the trajectories of the system for different initial velocities $v_{1,0}=\dot{q}_{1,0}$ without input and with the optimal input $u^\star$ to problem~\eqref{eq:pump_time_optimal_ocp}. The results without input in Figures~\ref{fig:pump_sim_10}--\ref{fig:pump_sim_22} show three different scenarios depending on the initial velocity. From left to right: (i) not reaching $q_{\text{goal}}$, (ii) reaching $q_{\text{goal}}$ while maintaining ground contact, and (iii) reaching $q_{\text{goal}}$ with a jump. From a rider's perspective, scenario (iii) represents a jump landing, which should be avoided because the impact with the incline significantly reduces speed. The optimal solutions in Figures~\ref{fig:pump_opt_10}-\ref{fig:pump_opt_22} show the trajectories for the same initial velocities with the optimal input $u^\star$. We can see, for scenario (i), the optimal input accelerates the system to successfully reach the goal. For scenario (iii), the optimal input avoids jumping and hitting the landing slope while still reaching the goal.
To analyze the velocities and times for each scenario, we show the solution trajectories in Figure~\ref{fig:pump_opt_time_results}. In scenario (ii), we observe the highest velocity gain through pumping, with $\Delta v_1=7.56\,$km/h, as shown in Figure~\ref{fig:pump_opt_v1}. The time required to reach the goal is decreased from $1.87\,$s to $1.42\,$s, see Figure~\ref{fig:pump_opt_q1}. Figures~\ref{fig:pump_opt_q3} and~\ref{fig:pump_opt_u} show the trajectories of $q_3(\cdot)$ and $u^\star(\cdot)$, satisfying the state and input constraints.

%% file: Graphics/tikz/ski_sim_45.tex
%
%
\begin{tikzpicture}

\begin{axis}[%
width=2.25in,
height=0.75in,
scale only axis,
xmin=0,
xmax=55,
xlabel style={font=\color{white!15!black}},
xticklabel=\empty,
ymin=10,
ymax=55,
ylabel style={font=\color{white!15!black}},
ylabel={$z$\,[m]},
ytick={15, 30, 45},
axis background/.style={fill=white},
xmajorgrids,
ymajorgrids,
axis x line*=bottom,
axis y line*=left,
legend style={legend cell align=left, align=left, font=\small}
]
\addplot[fill=black, fill opacity=0.2, draw=none, forget plot] table[row sep=crcr]{%
0	99.15\\
0.236077910910297	97.0121100382042\\
0.472155821820594	94.9204357652846\\
0.708233732730891	92.8747904897533\\
0.944311643641188	90.8749761590948\\
1.18038955455148	88.9207833597657\\
1.41646746546178	87.011991317195\\
1.65254537637208	85.1483678957835\\
1.88862328728238	83.3296695989047\\
2.12470119819267	81.555641568904\\
2.36077910910297	79.826017587099\\
2.59685702001327	78.1405200737796\\
2.83293493092356	76.498860088208\\
3.06901284183386	74.9007373286185\\
3.30509075274416	73.3458401322175\\
3.54116866365445	71.8338454751839\\
3.77724657456475	70.3644189726685\\
4.01332448547505	68.9372148787945\\
4.24940239638534	67.5518760866572\\
4.48548030729564	66.2080341283243\\
4.72155821820594	64.9053091748354\\
4.95763612911623	63.6433100362026\\
5.19371404002653	62.4216341614101\\
5.42979195093683	61.2398676384142\\
5.66586986184713	60.0975851941436\\
5.90194777275742	58.9943501944991\\
6.13802568366772	57.9297146443537\\
6.37410359457802	56.9032191875527\\
6.61018150548831	55.9143931069134\\
6.84625941639861	54.9627543242256\\
7.08233732730891	54.0478094002512\\
7.3184152382192	53.1690535347241\\
7.5544931491295	52.3259705663507\\
7.7905710600398	51.5180329728095\\
8.0266489709501	50.7447018707511\\
8.26272688186039	50.0054270157986\\
8.49880479277069	49.2996468025469\\
8.73488270368098	48.6267882645635\\
8.97096061459128	47.9862670743878\\
9.20703852550158	47.3774875435316\\
9.44311643641188	46.7998426224789\\
9.67919434732217	46.2527139006858\\
9.91527225823247	45.7354716065807\\
10.1513501691428	45.2474746075642\\
10.3874280800531	44.7880704100091\\
10.6235059909634	44.3565951592604\\
10.8595839018737	43.9523736396352\\
11.095661812784	43.5747192744231\\
11.3317397236943	43.2229341258856\\
11.5678176346045	42.8963088952566\\
11.8038955455148	42.594122922742\\
12.0399734564251	42.3156441875203\\
12.2760513673354	42.0601293077418\\
12.5121292782457	41.8268235405292\\
12.748207189156	41.6149607819774\\
12.9842851000663	41.4237635671535\\
13.2203630109766	41.2524430700968\\
13.4564409218869	41.1001991038187\\
13.6925188327972	40.9662201203031\\
13.9285967437075	40.8496832105058\\
14.1646746546178	40.749754104355\\
14.4007525655281	40.6655871707511\\
14.6368304764384	40.5963254175664\\
14.8729083873487	40.541100491646\\
15.108986298259	40.4990326788067\\
15.3450642091693	40.4692309038377\\
15.5811421200796	40.4507927305003\\
15.8172200309899	40.4428043615283\\
16.0532979419002	40.4443406386273\\
16.2893758528105	40.4544650424755\\
16.5254537637208	40.472229692723\\
16.7615316746311	40.4966753479923\\
16.9976095855414	40.5268314058781\\
17.2336874964517	40.5617159029471\\
17.469765407362	40.6003355147385\\
17.7058433182723	40.6416855557635\\
17.9419212291826	40.6847499795056\\
18.1779991400929	40.7285013784205\\
18.4140770510032	40.7719009839361\\
18.6501549619135	40.8138986664526\\
18.8862328728238	40.8534329353421\\
19.122310783734	40.8894309389493\\
19.3583886946443	40.9208084645909\\
19.5944666055546	40.9464699385559\\
19.8305445164649	40.9653084261053\\
20.0666224273752	40.9762056314726\\
20.3027003382855	40.9780318978633\\
20.5387782491958	40.9696462074552\\
20.7748561601061	40.9498961813983\\
21.0109340710164	40.9176180798149\\
21.2470119819267	40.8716368017992\\
21.483089892837	40.810765885418\\
21.7191678037473	40.7338075077101\\
21.9552457146576	40.6395524846864\\
22.1913236255679	40.5267802713303\\
22.4274015364782	40.3942589615972\\
22.6634794473885	40.2407452884147\\
22.8995573582988	40.0649846236828\\
23.1356352692091	39.8657109782735\\
23.3717131801194	39.6416470020311\\
23.6077910910297	39.3915039837721\\
23.84386900194	39.1139818512853\\
24.0799469128503	38.8077691713315\\
24.3160248237606	38.4715431496438\\
24.5521027346709	38.1039696309277\\
24.7881806455812	37.7037030988606\\
25.0242585564915	37.2693866760923\\
25.2603364674018	36.7996521242448\\
25.4964143783121	36.2931198439122\\
25.7324922892224	35.7483988746609\\
25.9685702001327	35.1640868950296\\
26.204648111043	34.5387702225289\\
26.4407260219533	33.871023813642\\
26.6768039328635	33.159411263824\\
26.9128818437738	32.4024848075024\\
27.1489597546841	31.5987853180768\\
27.3850376655944	30.7468423079191\\
27.6211155765047	29.8451739283732\\
27.857193487415	28.8922869697555\\
28.0932713983253	27.8866768613545\\
28.3293493092356	26.8268276714309\\
28.5654272201459	25.7112121072174\\
28.8015051310562	24.5382915149192\\
29.0375830419665	23.3065158797137\\
29.2736609528768	22.0143238257503\\
29.5097388637871	20.6601426161508\\
29.7458167746974	19.2423881530091\\
29.9818946856077	17.7594649773912\\
30.217972596518	16.2097662693359\\
30.4540505074283	14.5916738478533\\
30.6901284183386	12.9035581709264\\
30.9262063292489	11.1437783355101\\
31.1622842401592	9.31068207753171\\
}
\closedcycle;
\addplot [color=black, dashed, forget plot]
  table[row sep=crcr]{%
0	99.15\\
0.236077910910297	97.0121100382042\\
0.472155821820594	94.9204357652846\\
0.708233732730891	92.8747904897533\\
0.944311643641188	90.8749761590948\\
1.18038955455148	88.9207833597657\\
1.41646746546178	87.011991317195\\
1.65254537637208	85.1483678957835\\
1.88862328728238	83.3296695989047\\
2.12470119819267	81.555641568904\\
2.36077910910297	79.826017587099\\
2.59685702001327	78.1405200737796\\
2.83293493092356	76.498860088208\\
3.06901284183386	74.9007373286185\\
3.30509075274416	73.3458401322175\\
3.54116866365445	71.8338454751839\\
3.77724657456475	70.3644189726685\\
4.01332448547505	68.9372148787945\\
4.24940239638534	67.5518760866572\\
4.48548030729564	66.2080341283243\\
4.72155821820594	64.9053091748354\\
4.95763612911623	63.6433100362026\\
5.19371404002653	62.4216341614101\\
5.42979195093683	61.2398676384142\\
5.66586986184713	60.0975851941436\\
5.90194777275742	58.9943501944991\\
6.13802568366772	57.9297146443537\\
6.37410359457802	56.9032191875527\\
6.61018150548831	55.9143931069134\\
6.84625941639861	54.9627543242256\\
7.08233732730891	54.0478094002512\\
7.3184152382192	53.1690535347241\\
7.5544931491295	52.3259705663507\\
7.7905710600398	51.5180329728095\\
8.0266489709501	50.7447018707511\\
8.26272688186039	50.0054270157986\\
8.49880479277069	49.2996468025469\\
8.73488270368098	48.6267882645635\\
8.97096061459128	47.9862670743878\\
9.20703852550158	47.3774875435316\\
9.44311643641188	46.7998426224789\\
9.67919434732217	46.2527139006858\\
9.91527225823247	45.7354716065807\\
10.1513501691428	45.2474746075642\\
10.3874280800531	44.7880704100091\\
10.6235059909634	44.3565951592604\\
10.8595839018737	43.9523736396352\\
11.095661812784	43.5747192744231\\
11.3317397236943	43.2229341258856\\
11.5678176346045	42.8963088952566\\
11.8038955455148	42.594122922742\\
12.0399734564251	42.3156441875203\\
12.2760513673354	42.0601293077418\\
12.5121292782457	41.8268235405292\\
12.748207189156	41.6149607819774\\
12.9842851000663	41.4237635671535\\
13.2203630109766	41.2524430700968\\
13.4564409218869	41.1001991038187\\
13.6925188327972	40.9662201203031\\
13.9285967437075	40.8496832105058\\
14.1646746546178	40.749754104355\\
14.4007525655281	40.6655871707511\\
14.6368304764384	40.5963254175664\\
14.8729083873487	40.541100491646\\
15.108986298259	40.4990326788067\\
15.3450642091693	40.4692309038377\\
15.5811421200796	40.4507927305003\\
15.8172200309899	40.4428043615283\\
16.0532979419002	40.4443406386273\\
16.2893758528105	40.4544650424755\\
16.5254537637208	40.472229692723\\
16.7615316746311	40.4966753479923\\
16.9976095855414	40.5268314058781\\
17.2336874964517	40.5617159029471\\
17.469765407362	40.6003355147385\\
17.7058433182723	40.6416855557635\\
17.9419212291826	40.6847499795056\\
18.1779991400929	40.7285013784205\\
18.4140770510032	40.7719009839361\\
18.6501549619135	40.8138986664526\\
18.8862328728238	40.8534329353421\\
19.122310783734	40.8894309389493\\
19.3583886946443	40.9208084645909\\
19.5944666055546	40.9464699385559\\
19.8305445164649	40.9653084261053\\
20.0666224273752	40.9762056314726\\
20.3027003382855	40.9780318978633\\
20.5387782491958	40.9696462074552\\
20.7748561601061	40.9498961813983\\
21.0109340710164	40.9176180798149\\
21.2470119819267	40.8716368017992\\
21.483089892837	40.810765885418\\
21.7191678037473	40.7338075077101\\
21.9552457146576	40.6395524846864\\
22.1913236255679	40.5267802713303\\
22.4274015364782	40.3942589615972\\
22.6634794473885	40.2407452884147\\
22.8995573582988	40.0649846236828\\
23.1356352692091	39.8657109782735\\
23.3717131801194	39.6416470020311\\
23.6077910910297	39.3915039837721\\
23.84386900194	39.1139818512853\\
24.0799469128503	38.8077691713315\\
24.3160248237606	38.4715431496438\\
24.5521027346709	38.1039696309277\\
24.7881806455812	37.7037030988606\\
25.0242585564915	37.2693866760923\\
25.2603364674018	36.7996521242448\\
25.4964143783121	36.2931198439122\\
25.7324922892224	35.7483988746609\\
25.9685702001327	35.1640868950296\\
26.204648111043	34.5387702225289\\
26.4407260219533	33.871023813642\\
26.6768039328635	33.159411263824\\
26.9128818437738	32.4024848075024\\
27.1489597546841	31.5987853180768\\
27.3850376655944	30.7468423079191\\
27.6211155765047	29.8451739283732\\
27.857193487415	28.8922869697555\\
28.0932713983253	27.8866768613545\\
28.3293493092356	26.8268276714309\\
28.5654272201459	25.7112121072174\\
28.8015051310562	24.5382915149192\\
29.0375830419665	23.3065158797137\\
29.2736609528768	22.0143238257503\\
29.5097388637871	20.6601426161508\\
29.7458167746974	19.2423881530091\\
29.9818946856077	17.7594649773912\\
30.217972596518	16.2097662693359\\
30.4540505074283	14.5916738478533\\
30.6901284183386	12.9035581709264\\
30.9262063292489	11.1437783355101\\
31.1622842401592	9.31068207753171\\
};

\addplot[fill=black, fill opacity=0.2, draw=none, forget plot] table[row sep=crcr]{%
0	45\\
0.236077910910297	44.8819610445449\\
0.472155821820594	44.7639220890897\\
0.708233732730891	44.6458831336346\\
0.944311643641188	44.5278441781794\\
1.18038955455148	44.4098052227243\\
1.41646746546178	44.2917662672691\\
1.65254537637208	44.173727311814\\
1.88862328728238	44.0556883563588\\
2.12470119819267	43.9376494009037\\
2.36077910910297	43.8196104454485\\
2.59685702001327	43.7015714899934\\
2.83293493092356	43.5835325345382\\
3.06901284183386	43.4654935790831\\
3.30509075274416	43.3474546236279\\
3.54116866365445	43.2294156681728\\
3.77724657456475	43.1113767127176\\
4.01332448547505	42.9933377572625\\
4.24940239638534	42.8752988018073\\
4.48548030729564	42.7572598463522\\
4.72155821820594	42.639220890897\\
4.95763612911623	42.5211819354419\\
5.19371404002653	42.4031429799867\\
5.42979195093683	42.2851040245316\\
5.66586986184713	42.1670650690764\\
5.90194777275742	42.0490261136213\\
6.13802568366772	41.9309871581661\\
6.37410359457802	41.812948202711\\
6.61018150548831	41.6949092472558\\
6.84625941639861	41.5768702918007\\
7.08233732730891	41.4588313363455\\
7.3184152382192	41.3407923808904\\
7.5544931491295	41.2227534254353\\
7.7905710600398	41.1047144699801\\
8.0266489709501	40.986675514525\\
8.26272688186039	40.8686365590698\\
8.49880479277069	40.7505976036147\\
8.73488270368098	40.6325586481595\\
8.97096061459128	40.5145196927044\\
9.20703852550158	40.3964807372492\\
9.44311643641188	40.2784417817941\\
9.67919434732217	40.1604028263389\\
9.91527225823247	40.0423638708838\\
10.1513501691428	39.9243249154286\\
10.3874280800531	39.8062859599735\\
10.6235059909634	39.6882470045183\\
10.8595839018737	39.5702080490632\\
11.095661812784	39.452169093608\\
11.3317397236943	39.3341301381529\\
11.5678176346045	39.2160911826977\\
11.8038955455148	39.0980522272426\\
12.0399734564251	38.9800132717874\\
12.2760513673354	38.8619743163323\\
12.5121292782457	38.7439353608771\\
12.748207189156	38.625896405422\\
12.9842851000663	38.5078574499668\\
13.2203630109766	38.3898184945117\\
13.4564409218869	38.2717795390565\\
13.6925188327972	38.1537405836014\\
13.9285967437075	38.0357016281462\\
14.1646746546178	37.9176626726911\\
14.4007525655281	37.7996237172359\\
14.6368304764384	37.6815847617808\\
14.8729083873487	37.5635458063256\\
15.108986298259	37.4455068508705\\
15.3450642091693	37.3274678954154\\
15.5811421200796	37.2094289399602\\
15.8172200309899	37.0913899845051\\
16.0532979419002	36.9733510290499\\
16.2893758528105	36.8553120735948\\
16.5254537637208	36.7372731181396\\
16.7615316746311	36.6192341626845\\
16.9976095855414	36.5011952072293\\
17.2336874964517	36.3831562517742\\
17.469765407362	36.265117296319\\
17.7058433182723	36.1470783408639\\
17.9419212291826	36.0290393854087\\
18.1779991400929	35.9110004299536\\
18.4140770510032	35.7929614744984\\
18.6501549619135	35.6749225190433\\
18.8862328728238	35.5568835635881\\
19.122310783734	35.438844608133\\
19.3583886946443	35.3208056526778\\
19.5944666055546	35.2027666972227\\
19.8305445164649	35.0847277417675\\
20.0666224273752	34.9666887863124\\
20.3027003382855	34.8486498308572\\
20.5387782491958	34.7306108754021\\
20.7748561601061	34.6125719199469\\
21.0109340710164	34.4945329644918\\
21.2470119819267	34.3764940090366\\
21.483089892837	34.2584550535815\\
21.7191678037473	34.1404160981263\\
21.9552457146576	34.0223771426712\\
22.1913236255679	33.904338187216\\
22.4274015364782	33.7862992317609\\
22.6634794473885	33.6682602763057\\
22.8995573582988	33.5502213208506\\
23.1356352692091	33.4321823653954\\
23.3717131801194	33.3141434099403\\
23.6077910910297	33.1961044544852\\
23.84386900194	33.07806549903\\
24.0799469128503	32.9600265435749\\
24.3160248237606	32.8419875881197\\
24.5521027346709	32.7239486326646\\
24.7881806455812	32.6059096772094\\
25.0242585564915	32.4878707217543\\
25.2603364674018	32.3698317662991\\
25.4964143783121	32.251792810844\\
25.7324922892224	32.1337538553888\\
25.9685702001327	32.0157148999337\\
26.204648111043	31.8976759444785\\
26.4407260219533	31.7796369890234\\
26.6768039328635	31.6615980335682\\
26.9128818437738	31.5435590781131\\
27.1489597546841	31.4255201226579\\
27.3850376655944	31.3074811672028\\
27.6211155765047	31.1894422117476\\
27.857193487415	31.0714032562925\\
28.0932713983253	30.9533643008373\\
28.3293493092356	30.8353253453822\\
28.5654272201459	30.717286389927\\
28.8015051310562	30.5992474344719\\
29.0375830419665	30.4812084790167\\
29.2736609528768	30.3631695235616\\
29.5097388637871	30.2451305681064\\
29.7458167746974	30.1270916126513\\
29.9818946856077	30.0090526571961\\
30.217972596518	29.891013701741\\
30.4540505074283	29.7729747462858\\
30.6901284183386	29.6549357908307\\
30.9262063292489	29.5368968353756\\
31.1622842401592	29.4188578799204\\
31.3983621510695	29.3008189244653\\
31.6344400619798	29.1827799690101\\
31.8705179728901	29.064741013555\\
32.1065958838004	28.9467020580998\\
32.3426737947107	28.8286631026447\\
32.578751705621	28.7106241471895\\
32.8148296165313	28.5925851917344\\
33.0509075274416	28.4745462362792\\
33.2869854383519	28.3565072808241\\
33.5230633492622	28.2384683253689\\
33.7591412601725	28.1204293699138\\
33.9952191710828	28.0023904144586\\
34.231297081993	27.8843514590035\\
34.4673749929033	27.7663125035483\\
34.7034529038136	27.6482735480932\\
34.9395308147239	27.530234592638\\
35.1756087256342	27.4121956371829\\
35.4116866365445	27.2941566817277\\
35.6477645474548	27.1761177262726\\
35.8838424583651	27.0580787708174\\
36.1199203692754	26.9400398153623\\
36.3559982801857	26.8220008599071\\
36.592076191096	26.703961904452\\
36.8281541020063	26.5859229489968\\
37.0642320129166	26.4678839935417\\
37.3003099238269	26.3498450380865\\
37.5363878347372	26.2318060826314\\
37.7724657456475	26.1137671271762\\
38.0085436565578	25.9957281717211\\
38.2446215674681	25.8776892162659\\
38.4806994783784	25.7596502608108\\
38.7167773892887	25.6416113053557\\
38.952855300199	25.5235723499005\\
39.1889332111093	25.4055333944454\\
39.4250111220196	25.2874944389902\\
39.6610890329299	25.1694554835351\\
39.8971669438402	25.0514165280799\\
40.1332448547505	24.9333775726248\\
40.3693227656608	24.8153386171696\\
40.6054006765711	24.6972996617145\\
40.8414785874814	24.5792607062593\\
41.0775564983917	24.4612217508042\\
41.313634409302	24.343182795349\\
41.5497123202123	24.2251438398939\\
41.7857902311226	24.1071048844387\\
42.0218681420329	23.9890659289836\\
42.2579460529431	23.8710269735284\\
42.4940239638534	23.7529880180733\\
42.7301018747637	23.6349490626181\\
42.966179785674	23.516910107163\\
43.2022576965843	23.3988711517078\\
43.4383356074946	23.2808321962527\\
43.6744135184049	23.1627932407975\\
43.9104914293152	23.0447542853424\\
44.1465693402255	22.9267153298872\\
44.3826472511358	22.8086763744321\\
44.6187251620461	22.6906374189769\\
44.8548030729564	22.5725984635218\\
45.0908809838667	22.4545595080666\\
45.326958894777	22.3365205526115\\
45.5630368056873	22.2184815971563\\
45.7991147165976	22.1004426417012\\
46.0351926275079	21.9824036862461\\
46.2712705384182	21.8643647307909\\
46.5073484493285	21.7463257753358\\
46.7434263602388	21.6282868198806\\
46.9795042711491	21.5102478644255\\
47.2155821820594	21.3922089089703\\
47.4516600929697	21.2741699535152\\
47.68773800388	21.15613099806\\
47.9238159147903	21.0380920426049\\
48.1598938257006	20.9200530871497\\
48.3959717366109	20.8020141316946\\
48.6320496475212	20.6839751762394\\
48.8681275584315	20.5659362207843\\
49.1042054693418	20.4478972653291\\
49.3402833802521	20.329858309874\\
49.5763612911623	20.2118193544188\\
49.8124392020726	20.0937803989637\\
50.0485171129829	19.9757414435085\\
50.2845950238932	19.8577024880534\\
50.5206729348035	19.7396635325982\\
50.7567508457138	19.6216245771431\\
50.9928287566241	19.5035856216879\\
51.2289066675344	19.3855466662328\\
51.4649845784447	19.2675077107776\\
51.701062489355	19.1494687553225\\
51.9371404002653	19.0314297998673\\
52.1732183111756	18.9133908444122\\
52.4092962220859	18.795351888957\\
52.6453741329962	18.6773129335019\\
52.8814520439065	18.5592739780467\\
53.1175299548168	18.4412350225916\\
53.3536078657271	18.3231960671365\\
53.5896857766374	18.2051571116813\\
53.8257636875477	18.0871181562262\\
54.061841598458	17.969079200771\\
54.2979195093683	17.8510402453159\\
54.5339974202786	17.7330012898607\\
54.7700753311889	17.6149623344056\\
55.0061532420992	17.4969233789504\\
}
\closedcycle;
\addplot [color=black, dotted, forget plot]
  table[row sep=crcr]{%
0	45\\
0.236077910910297	44.8819610445449\\
0.472155821820594	44.7639220890897\\
0.708233732730891	44.6458831336346\\
0.944311643641188	44.5278441781794\\
1.18038955455148	44.4098052227243\\
1.41646746546178	44.2917662672691\\
1.65254537637208	44.173727311814\\
1.88862328728238	44.0556883563588\\
2.12470119819267	43.9376494009037\\
2.36077910910297	43.8196104454485\\
2.59685702001327	43.7015714899934\\
2.83293493092356	43.5835325345382\\
3.06901284183386	43.4654935790831\\
3.30509075274416	43.3474546236279\\
3.54116866365445	43.2294156681728\\
3.77724657456475	43.1113767127176\\
4.01332448547505	42.9933377572625\\
4.24940239638534	42.8752988018073\\
4.48548030729564	42.7572598463522\\
4.72155821820594	42.639220890897\\
4.95763612911623	42.5211819354419\\
5.19371404002653	42.4031429799867\\
5.42979195093683	42.2851040245316\\
5.66586986184713	42.1670650690764\\
5.90194777275742	42.0490261136213\\
6.13802568366772	41.9309871581661\\
6.37410359457802	41.812948202711\\
6.61018150548831	41.6949092472558\\
6.84625941639861	41.5768702918007\\
7.08233732730891	41.4588313363455\\
7.3184152382192	41.3407923808904\\
7.5544931491295	41.2227534254353\\
7.7905710600398	41.1047144699801\\
8.0266489709501	40.986675514525\\
8.26272688186039	40.8686365590698\\
8.49880479277069	40.7505976036147\\
8.73488270368098	40.6325586481595\\
8.97096061459128	40.5145196927044\\
9.20703852550158	40.3964807372492\\
9.44311643641188	40.2784417817941\\
9.67919434732217	40.1604028263389\\
9.91527225823247	40.0423638708838\\
10.1513501691428	39.9243249154286\\
10.3874280800531	39.8062859599735\\
10.6235059909634	39.6882470045183\\
10.8595839018737	39.5702080490632\\
11.095661812784	39.452169093608\\
11.3317397236943	39.3341301381529\\
11.5678176346045	39.2160911826977\\
11.8038955455148	39.0980522272426\\
12.0399734564251	38.9800132717874\\
12.2760513673354	38.8619743163323\\
12.5121292782457	38.7439353608771\\
12.748207189156	38.625896405422\\
12.9842851000663	38.5078574499668\\
13.2203630109766	38.3898184945117\\
13.4564409218869	38.2717795390565\\
13.6925188327972	38.1537405836014\\
13.9285967437075	38.0357016281462\\
14.1646746546178	37.9176626726911\\
14.4007525655281	37.7996237172359\\
14.6368304764384	37.6815847617808\\
14.8729083873487	37.5635458063256\\
15.108986298259	37.4455068508705\\
15.3450642091693	37.3274678954154\\
15.5811421200796	37.2094289399602\\
15.8172200309899	37.0913899845051\\
16.0532979419002	36.9733510290499\\
16.2893758528105	36.8553120735948\\
16.5254537637208	36.7372731181396\\
16.7615316746311	36.6192341626845\\
16.9976095855414	36.5011952072293\\
17.2336874964517	36.3831562517742\\
17.469765407362	36.265117296319\\
17.7058433182723	36.1470783408639\\
17.9419212291826	36.0290393854087\\
18.1779991400929	35.9110004299536\\
18.4140770510032	35.7929614744984\\
18.6501549619135	35.6749225190433\\
18.8862328728238	35.5568835635881\\
19.122310783734	35.438844608133\\
19.3583886946443	35.3208056526778\\
19.5944666055546	35.2027666972227\\
19.8305445164649	35.0847277417675\\
20.0666224273752	34.9666887863124\\
20.3027003382855	34.8486498308572\\
20.5387782491958	34.7306108754021\\
20.7748561601061	34.6125719199469\\
21.0109340710164	34.4945329644918\\
21.2470119819267	34.3764940090366\\
21.483089892837	34.2584550535815\\
21.7191678037473	34.1404160981263\\
21.9552457146576	34.0223771426712\\
22.1913236255679	33.904338187216\\
22.4274015364782	33.7862992317609\\
22.6634794473885	33.6682602763057\\
22.8995573582988	33.5502213208506\\
23.1356352692091	33.4321823653954\\
23.3717131801194	33.3141434099403\\
23.6077910910297	33.1961044544852\\
23.84386900194	33.07806549903\\
24.0799469128503	32.9600265435749\\
24.3160248237606	32.8419875881197\\
24.5521027346709	32.7239486326646\\
24.7881806455812	32.6059096772094\\
25.0242585564915	32.4878707217543\\
25.2603364674018	32.3698317662991\\
25.4964143783121	32.251792810844\\
25.7324922892224	32.1337538553888\\
25.9685702001327	32.0157148999337\\
26.204648111043	31.8976759444785\\
26.4407260219533	31.7796369890234\\
26.6768039328635	31.6615980335682\\
26.9128818437738	31.5435590781131\\
27.1489597546841	31.4255201226579\\
27.3850376655944	31.3074811672028\\
27.6211155765047	31.1894422117476\\
27.857193487415	31.0714032562925\\
28.0932713983253	30.9533643008373\\
28.3293493092356	30.8353253453822\\
28.5654272201459	30.717286389927\\
28.8015051310562	30.5992474344719\\
29.0375830419665	30.4812084790167\\
29.2736609528768	30.3631695235616\\
29.5097388637871	30.2451305681064\\
29.7458167746974	30.1270916126513\\
29.9818946856077	30.0090526571961\\
30.217972596518	29.891013701741\\
30.4540505074283	29.7729747462858\\
30.6901284183386	29.6549357908307\\
30.9262063292489	29.5368968353756\\
31.1622842401592	29.4188578799204\\
31.3983621510695	29.3008189244653\\
31.6344400619798	29.1827799690101\\
31.8705179728901	29.064741013555\\
32.1065958838004	28.9467020580998\\
32.3426737947107	28.8286631026447\\
32.578751705621	28.7106241471895\\
32.8148296165313	28.5925851917344\\
33.0509075274416	28.4745462362792\\
33.2869854383519	28.3565072808241\\
33.5230633492622	28.2384683253689\\
33.7591412601725	28.1204293699138\\
33.9952191710828	28.0023904144586\\
34.231297081993	27.8843514590035\\
34.4673749929033	27.7663125035483\\
34.7034529038136	27.6482735480932\\
34.9395308147239	27.530234592638\\
35.1756087256342	27.4121956371829\\
35.4116866365445	27.2941566817277\\
35.6477645474548	27.1761177262726\\
35.8838424583651	27.0580787708174\\
36.1199203692754	26.9400398153623\\
36.3559982801857	26.8220008599071\\
36.592076191096	26.703961904452\\
36.8281541020063	26.5859229489968\\
37.0642320129166	26.4678839935417\\
37.3003099238269	26.3498450380865\\
37.5363878347372	26.2318060826314\\
37.7724657456475	26.1137671271762\\
38.0085436565578	25.9957281717211\\
38.2446215674681	25.8776892162659\\
38.4806994783784	25.7596502608108\\
38.7167773892887	25.6416113053557\\
38.952855300199	25.5235723499005\\
39.1889332111093	25.4055333944454\\
39.4250111220196	25.2874944389902\\
39.6610890329299	25.1694554835351\\
39.8971669438402	25.0514165280799\\
40.1332448547505	24.9333775726248\\
40.3693227656608	24.8153386171696\\
40.6054006765711	24.6972996617145\\
40.8414785874814	24.5792607062593\\
41.0775564983917	24.4612217508042\\
41.313634409302	24.343182795349\\
41.5497123202123	24.2251438398939\\
41.7857902311226	24.1071048844387\\
42.0218681420329	23.9890659289836\\
42.2579460529431	23.8710269735284\\
42.4940239638534	23.7529880180733\\
42.7301018747637	23.6349490626181\\
42.966179785674	23.516910107163\\
43.2022576965843	23.3988711517078\\
43.4383356074946	23.2808321962527\\
43.6744135184049	23.1627932407975\\
43.9104914293152	23.0447542853424\\
44.1465693402255	22.9267153298872\\
44.3826472511358	22.8086763744321\\
44.6187251620461	22.6906374189769\\
44.8548030729564	22.5725984635218\\
45.0908809838667	22.4545595080666\\
45.326958894777	22.3365205526115\\
45.5630368056873	22.2184815971563\\
45.7991147165976	22.1004426417012\\
46.0351926275079	21.9824036862461\\
46.2712705384182	21.8643647307909\\
46.5073484493285	21.7463257753358\\
46.7434263602388	21.6282868198806\\
46.9795042711491	21.5102478644255\\
47.2155821820594	21.3922089089703\\
47.4516600929697	21.2741699535152\\
47.68773800388	21.15613099806\\
47.9238159147903	21.0380920426049\\
48.1598938257006	20.9200530871497\\
48.3959717366109	20.8020141316946\\
48.6320496475212	20.6839751762394\\
48.8681275584315	20.5659362207843\\
49.1042054693418	20.4478972653291\\
49.3402833802521	20.329858309874\\
49.5763612911623	20.2118193544188\\
49.8124392020726	20.0937803989637\\
50.0485171129829	19.9757414435085\\
50.2845950238932	19.8577024880534\\
50.5206729348035	19.7396635325982\\
50.7567508457138	19.6216245771431\\
50.9928287566241	19.5035856216879\\
51.2289066675344	19.3855466662328\\
51.4649845784447	19.2675077107776\\
51.701062489355	19.1494687553225\\
51.9371404002653	19.0314297998673\\
52.1732183111756	18.9133908444122\\
52.4092962220859	18.795351888957\\
52.6453741329962	18.6773129335019\\
52.8814520439065	18.5592739780467\\
53.1175299548168	18.4412350225916\\
53.3536078657271	18.3231960671365\\
53.5896857766374	18.2051571116813\\
53.8257636875477	18.0871181562262\\
54.061841598458	17.969079200771\\
54.2979195093683	17.8510402453159\\
54.5339974202786	17.7330012898607\\
54.7700753311889	17.6149623344056\\
55.0061532420992	17.4969233789504\\
};
\addplot[only marks, mark=*, mark options={}, mark size=2.1651pt, color=red, fill=red] table[row sep=crcr]{%
x	y\\
10.2759	45.0015810732957\\
};
\addlegendentry{$q_{0}$}

\addplot[only marks, mark=*, mark options={}, mark size=2.1651pt, color=green, fill=green] table[row sep=crcr]{%
x	y\\
35.8280435750106	27.0859781967575\\
};
\addlegendentry{$q_{\text{impact}}$}

\addplot[only marks, mark=*, mark options={}, mark size=2.1651pt, color=blue, fill=blue] table[row sep=crcr]{%
x	y\\
10.2759	45.0015810732957\\
};
\addlegendentry{$q(t)$}

\addplot[only marks, mark=*, mark options={}, mark size=2.1651pt, color=blue, fill=blue, forget plot] table[row sep=crcr]{%
x	y\\
10.3560760177448	44.8474558419\\
};
\addplot[only marks, mark=*, mark options={}, mark size=2.1651pt, color=blue, fill=blue, forget plot] table[row sep=crcr]{%
x	y\\
10.6028969581483	44.393165927445\\
};
\addplot[only marks, mark=*, mark options={}, mark size=2.1651pt, color=blue, fill=blue, forget plot] table[row sep=crcr]{%
x	y\\
11.0428664093539	43.6569058934547\\
};
\addplot[only marks, mark=*, mark options={}, mark size=2.1651pt, color=blue, fill=blue, forget plot] table[row sep=crcr]{%
x	y\\
11.7286127832909	42.6878743941855\\
};
\addplot[only marks, mark=*, mark options={}, mark size=2.1651pt, color=blue, fill=blue, forget plot] table[row sep=crcr]{%
x	y\\
12.752726607195	41.6111072523524\\
};
\addplot[only marks, mark=*, mark options={}, mark size=2.1651pt, color=blue, fill=blue, forget plot] table[row sep=crcr]{%
x	y\\
14.2374909230634	40.7221519137048\\
};
\addplot[only marks, mark=*, mark options={}, mark size=2.1651pt, color=blue, fill=blue, forget plot] table[row sep=crcr]{%
x	y\\
16.0766449950277	40.4448713686622\\
};
\addplot[only marks, mark=*, mark options={}, mark size=2.1651pt, color=blue, fill=blue, forget plot] table[row sep=crcr]{%
x	y\\
17.8335592769405	40.6648354019549\\
};
\addplot[only marks, mark=*, mark options={}, mark size=2.1651pt, color=blue, fill=blue, forget plot] table[row sep=crcr]{%
x	y\\
19.6136348303795	40.9482565398208\\
};
\addplot[only marks, mark=*, mark options={}, mark size=2.1651pt, color=blue, fill=blue, forget plot] table[row sep=crcr]{%
x	y\\
21.3845231909895	40.9161780373155\\
};
\addplot[only marks, mark=*, mark options={}, mark size=2.1651pt, color=blue, fill=blue, forget plot] table[row sep=crcr]{%
x	y\\
23.1554115403615	40.4916996599845\\
};
\addplot[only marks, mark=*, mark options={}, mark size=2.1651pt, color=blue, fill=blue, forget plot] table[row sep=crcr]{%
x	y\\
24.9262998897364	39.6748212826594\\
};
\addplot[only marks, mark=*, mark options={}, mark size=2.1651pt, color=blue, fill=blue, forget plot] table[row sep=crcr]{%
x	y\\
26.6971882391558	38.465542905423\\
};
\addplot[only marks, mark=*, mark options={}, mark size=2.1651pt, color=blue, fill=blue, forget plot] table[row sep=crcr]{%
x	y\\
28.4680765886355	36.8638645283073\\
};
\addplot[only marks, mark=*, mark options={}, mark size=2.1651pt, color=blue, fill=blue, forget plot] table[row sep=crcr]{%
x	y\\
30.2389649381828	34.8697861513268\\
};
\addplot[only marks, mark=*, mark options={}, mark size=2.1651pt, color=blue, fill=blue, forget plot] table[row sep=crcr]{%
x	y\\
32.0098532878136	32.4833077745135\\
};
\addplot[only marks, mark=*, mark options={}, mark size=2.1651pt, color=blue, fill=blue, forget plot] table[row sep=crcr]{%
x	y\\
33.7807416375682	29.7044293979478\\
};
\addplot[only marks, mark=*, mark options={}, mark size=2.1651pt, color=blue, fill=blue, forget plot] table[row sep=crcr]{%
x	y\\
35.8280435750106	27.0859781967575\\
};
\addplot[only marks, mark=*, mark options={}, mark size=2.1651pt, color=blue, fill=blue, forget plot] table[row sep=crcr]{%
x	y\\
38.6702255439322	25.6648870585257\\
};
\addplot[only marks, mark=*, mark options={}, mark size=2.1651pt, color=blue, fill=blue, forget plot] table[row sep=crcr]{%
x	y\\
41.6693675745392	24.1653160436646\\
};
\addplot[only marks, mark=*, mark options={}, mark size=2.1651pt, color=blue, fill=blue, forget plot] table[row sep=crcr]{%
x	y\\
44.8254696051461	22.5872650288034\\
};
\addplot[only marks, mark=*, mark options={}, mark size=2.1651pt, color=blue, fill=blue, forget plot] table[row sep=crcr]{%
x	y\\
48.1385316357532	20.9307340139422\\
};
\addplot[only marks, mark=*, mark options={}, mark size=2.1651pt, color=blue, fill=blue, forget plot] table[row sep=crcr]{%
x	y\\
51.6085536663603	19.1957229990809\\
};
\addplot[only marks, mark=*, mark options={}, mark size=2.1651pt, color=blue, fill=blue, forget plot] table[row sep=crcr]{%
x	y\\
55.2355356969673	17.3822319842196\\
};
\addplot[only marks, mark=*, mark options={}, mark size=2.1651pt, color=red, fill=red, forget plot] table[row sep=crcr]{%
x	y\\
10.2759	45.0015810732957\\
};
\addplot[only marks, mark=*, mark options={}, mark size=2.1651pt, color=green, fill=green, forget plot] table[row sep=crcr]{%
x	y\\
35.8280435750106	27.0859781967575\\
};
\end{axis}

\end{tikzpicture}%

%% file: Graphics/tikz/ski_sim_50.tex
%
%
\begin{tikzpicture}

\begin{axis}[%
width=2.25in,
height=0.75in,
scale only axis,
xmin=0,
xmax=55,
xlabel style={font=\color{white!15!black}},
xlabel={$y$\,[m]},
ymin=10,
ymax=55,
ylabel style={font=\color{white!15!black}},
ylabel={$z$\,[m]},
ytick={15, 30, 45},
axis background/.style={fill=white},
xmajorgrids,
ymajorgrids,
axis x line*=bottom,
axis y line*=left,
legend style={legend cell align=left, align=left, font=\small}
]

\addplot[fill=black, fill opacity=0.2, draw=none, forget plot] table[row sep=crcr]{%
0	99.15\\
0.259874151498117	96.7991814960542\\
0.519748302996233	94.5043427448215\\
0.77962245449435	92.2652324272557\\
1.03949660599247	90.0815825422939\\
1.29937075749058	87.9531084068568\\
1.5592449089887	85.8795086558485\\
1.81911906048682	83.8604652421569\\
2.07899321198493	81.895643436653\\
2.33886736348305	79.9846918281917\\
2.59874151498117	78.1272423236112\\
2.85861566647928	76.3229101477332\\
3.1184898179774	74.5712938433629\\
3.37836396947552	72.8719752712892\\
3.63823812097363	71.2245196102844\\
3.89811227247175	69.6284753571042\\
4.15798642396986	68.0833743264879\\
4.41786057546798	66.5887316511584\\
4.6777347269661	65.144045781822\\
4.93760887846421	63.7487984871686\\
5.19748302996233	62.4024548538714\\
5.45735718146045	61.1044632865875\\
5.71723133295856	59.854255507957\\
5.97710548445668	58.651246558604\\
6.2369796359548	57.4948347971359\\
6.49685378745291	56.3844019001435\\
6.75672793895103	55.3193128622012\\
7.01660209044915	54.2989159958671\\
7.27647624194726	53.3225429316826\\
7.53635039344538	52.3895086181726\\
7.7962245449435	51.4991113218456\\
8.05609869644161	50.6506326271935\\
8.31597284793973	49.843337436692\\
8.57584699943785	49.0764739707999\\
8.83572115093596	48.3492737679598\\
9.09559530243408	47.6609516845978\\
9.3554694539322	47.0107058951233\\
9.61534360543031	46.3977178919295\\
9.87521775692843	45.8211524853929\\
10.1350919084265	45.2801578038735\\
10.3949660599247	44.773865293715\\
10.6548402114228	44.3013897192445\\
10.9147143629209	43.8618291627726\\
11.174588514419	43.4542650245933\\
11.4344626659171	43.0777620229844\\
11.6943368174152	42.731368194207\\
11.9542109689134	42.4141148925057\\
12.2140851204115	42.1250167901088\\
12.4739592719096	41.8630718772278\\
12.7338334234077	41.627261462058\\
12.9937075749058	41.4165501707781\\
13.2535817264039	41.2298859475503\\
13.5134558779021	41.0662000545204\\
13.7733300294002	40.9244070718176\\
14.0332041808983	40.8034048975546\\
14.2930783323964	40.7020747478277\\
14.5529524838945	40.6192811567167\\
14.8128266353926	40.5538719762848\\
15.0727007868908	40.5046783765789\\
15.3325749383889	40.4705148456293\\
15.592449089887	40.4501791894498\\
15.8523232413851	40.4424525320377\\
16.1121973928832	40.4460993153738\\
16.3720715443813	40.4598672994227\\
16.6319456958795	40.482487562132\\
16.8918198473776	40.5126744994332\\
17.1516939988757	40.5491258252412\\
17.4115681503738	40.5905225714544\\
17.6714423018719	40.6355290879546\\
17.93131645337	40.6827930426074\\
18.1911906048682	40.7309454212617\\
18.4510647563663	40.7786005277498\\
18.7109389078644	40.8243559838879\\
18.9708130593625	40.8667927294753\\
19.2306872108606	40.904475022295\\
19.4905613623587	40.9359504381136\\
19.7504355138569	40.959749870681\\
20.010309665355	40.9743875317308\\
20.2701838168531	40.97836095098\\
20.5300579683512	40.9701509761291\\
20.7899321198493	40.9482217728622\\
21.0498062713474	40.9110208248469\\
21.3096804228456	40.8569789337343\\
21.5695545743437	40.7845102191588\\
21.8294287258418	40.6920121187387\\
22.0893028773399	40.5778653880756\\
22.349177028838	40.4404341007545\\
22.6090511803361	40.2780656483442\\
22.8689253318343	40.0890907403968\\
23.1287994833324	39.8718234044478\\
23.3886736348305	39.6245609860166\\
23.6485477863286	39.3455841486057\\
23.9084219378267	39.0331568737014\\
24.1682960893248	38.6855264607734\\
24.428170240823	38.300923527275\\
24.6880443923211	37.8775620086427\\
24.9479185438192	37.413639158297\\
25.2077926953173	36.9073355476415\\
25.4676668468154	36.3568150660634\\
25.7275409983135	35.7602249209337\\
25.9874151498117	35.1156956376065\\
26.2472893013098	34.4213410594197\\
26.5071634528079	33.6752583476946\\
26.767037604306	32.875527981736\\
27.0269117558041	32.0202137588322\\
27.2867859073022	31.1073627942552\\
27.5466600588004	30.1350055212602\\
27.8065342102985	29.101155691086\\
28.0664083617966	28.0038103729553\\
28.3262825132947	26.8409499540737\\
28.5861566647928	25.6105381396307\\
28.8460308162909	24.3105219527993\\
29.1059049677891	22.9388317347357\\
29.3657791192872	21.4933811445801\\
29.6256532707853	19.9720671594558\\
29.8855274222834	18.3727700744699\\
30.1454015737815	16.6933535027127\\
30.4052757252796	14.9316643752583\\
30.6651498767778	13.0855329411642\\
30.9250240282759	11.1527727674714\\
31.184898179774	9.13118073920435\\
}
\closedcycle;
\addplot [color=black, dashed]
  table[row sep=crcr]{%
0	99.15\\
0.259874151498117	96.7991814960542\\
0.519748302996233	94.5043427448215\\
0.77962245449435	92.2652324272557\\
1.03949660599247	90.0815825422939\\
1.29937075749058	87.9531084068568\\
1.5592449089887	85.8795086558485\\
1.81911906048682	83.8604652421569\\
2.07899321198493	81.895643436653\\
2.33886736348305	79.9846918281917\\
2.59874151498117	78.1272423236112\\
2.85861566647928	76.3229101477332\\
3.1184898179774	74.5712938433629\\
3.37836396947552	72.8719752712892\\
3.63823812097363	71.2245196102844\\
3.89811227247175	69.6284753571042\\
4.15798642396986	68.0833743264879\\
4.41786057546798	66.5887316511584\\
4.6777347269661	65.144045781822\\
4.93760887846421	63.7487984871686\\
5.19748302996233	62.4024548538714\\
5.45735718146045	61.1044632865875\\
5.71723133295856	59.854255507957\\
5.97710548445668	58.651246558604\\
6.2369796359548	57.4948347971359\\
6.49685378745291	56.3844019001435\\
6.75672793895103	55.3193128622012\\
7.01660209044915	54.2989159958671\\
7.27647624194726	53.3225429316826\\
7.53635039344538	52.3895086181726\\
7.7962245449435	51.4991113218456\\
8.05609869644161	50.6506326271935\\
8.31597284793973	49.843337436692\\
8.57584699943785	49.0764739707999\\
8.83572115093596	48.3492737679598\\
9.09559530243408	47.6609516845978\\
9.3554694539322	47.0107058951233\\
9.61534360543031	46.3977178919295\\
9.87521775692843	45.8211524853929\\
10.1350919084265	45.2801578038735\\
10.3949660599247	44.773865293715\\
10.6548402114228	44.3013897192445\\
10.9147143629209	43.8618291627726\\
11.174588514419	43.4542650245933\\
11.4344626659171	43.0777620229844\\
11.6943368174152	42.731368194207\\
11.9542109689134	42.4141148925057\\
12.2140851204115	42.1250167901088\\
12.4739592719096	41.8630718772278\\
12.7338334234077	41.627261462058\\
12.9937075749058	41.4165501707781\\
13.2535817264039	41.2298859475503\\
13.5134558779021	41.0662000545204\\
13.7733300294002	40.9244070718176\\
14.0332041808983	40.8034048975546\\
14.2930783323964	40.7020747478277\\
14.5529524838945	40.6192811567167\\
14.8128266353926	40.5538719762848\\
15.0727007868908	40.5046783765789\\
15.3325749383889	40.4705148456293\\
15.592449089887	40.4501791894498\\
15.8523232413851	40.4424525320377\\
16.1121973928832	40.4460993153738\\
16.3720715443813	40.4598672994227\\
16.6319456958795	40.482487562132\\
16.8918198473776	40.5126744994332\\
17.1516939988757	40.5491258252412\\
17.4115681503738	40.5905225714544\\
17.6714423018719	40.6355290879546\\
17.93131645337	40.6827930426074\\
18.1911906048682	40.7309454212617\\
18.4510647563663	40.7786005277498\\
18.7109389078644	40.8243559838879\\
18.9708130593625	40.8667927294753\\
19.2306872108606	40.904475022295\\
19.4905613623587	40.9359504381136\\
19.7504355138569	40.959749870681\\
20.010309665355	40.9743875317308\\
20.2701838168531	40.97836095098\\
20.5300579683512	40.9701509761291\\
20.7899321198493	40.9482217728622\\
21.0498062713474	40.9110208248469\\
21.3096804228456	40.8569789337343\\
21.5695545743437	40.7845102191588\\
21.8294287258418	40.6920121187387\\
22.0893028773399	40.5778653880756\\
22.349177028838	40.4404341007545\\
22.6090511803361	40.2780656483442\\
22.8689253318343	40.0890907403968\\
23.1287994833324	39.8718234044478\\
23.3886736348305	39.6245609860166\\
23.6485477863286	39.3455841486057\\
23.9084219378267	39.0331568737014\\
24.1682960893248	38.6855264607734\\
24.428170240823	38.300923527275\\
24.6880443923211	37.8775620086427\\
24.9479185438192	37.413639158297\\
25.2077926953173	36.9073355476415\\
25.4676668468154	36.3568150660634\\
25.7275409983135	35.7602249209337\\
25.9874151498117	35.1156956376065\\
26.2472893013098	34.4213410594197\\
26.5071634528079	33.6752583476946\\
26.767037604306	32.875527981736\\
27.0269117558041	32.0202137588322\\
27.2867859073022	31.1073627942552\\
27.5466600588004	30.1350055212602\\
27.8065342102985	29.101155691086\\
28.0664083617966	28.0038103729553\\
28.3262825132947	26.8409499540737\\
28.5861566647928	25.6105381396307\\
28.8460308162909	24.3105219527993\\
29.1059049677891	22.9388317347357\\
29.3657791192872	21.4933811445801\\
29.6256532707853	19.9720671594558\\
29.8855274222834	18.3727700744699\\
30.1454015737815	16.6933535027127\\
30.4052757252796	14.9316643752583\\
30.6651498767778	13.0855329411642\\
30.9250240282759	11.1527727674714\\
31.184898179774	9.13118073920435\\
};

\addplot[fill=black, fill opacity=0.2, draw=none, forget plot] table[row sep=crcr]{%
0	45\\
0.259874151498117	44.8700629242509\\
0.519748302996233	44.7401258485019\\
0.77962245449435	44.6101887727528\\
1.03949660599247	44.4802516970038\\
1.29937075749058	44.3503146212547\\
1.5592449089887	44.2203775455056\\
1.81911906048682	44.0904404697566\\
2.07899321198493	43.9605033940075\\
2.33886736348305	43.8305663182585\\
2.59874151498117	43.7006292425094\\
2.85861566647928	43.5706921667604\\
3.1184898179774	43.4407550910113\\
3.37836396947552	43.3108180152622\\
3.63823812097363	43.1808809395132\\
3.89811227247175	43.0509438637641\\
4.15798642396986	42.9210067880151\\
4.41786057546798	42.791069712266\\
4.6777347269661	42.661132636517\\
4.93760887846421	42.5311955607679\\
5.19748302996233	42.4012584850188\\
5.45735718146045	42.2713214092698\\
5.71723133295856	42.1413843335207\\
5.97710548445668	42.0114472577717\\
6.2369796359548	41.8815101820226\\
6.49685378745291	41.7515731062735\\
6.75672793895103	41.6216360305245\\
7.01660209044915	41.4916989547754\\
7.27647624194726	41.3617618790264\\
7.53635039344538	41.2318248032773\\
7.7962245449435	41.1018877275283\\
8.05609869644161	40.9719506517792\\
8.31597284793973	40.8420135760301\\
8.57584699943785	40.7120765002811\\
8.83572115093596	40.582139424532\\
9.09559530243408	40.452202348783\\
9.3554694539322	40.3222652730339\\
9.61534360543031	40.1923281972848\\
9.87521775692843	40.0623911215358\\
10.1350919084265	39.9324540457867\\
10.3949660599247	39.8025169700377\\
10.6548402114228	39.6725798942886\\
10.9147143629209	39.5426428185396\\
11.174588514419	39.4127057427905\\
11.4344626659171	39.2827686670414\\
11.6943368174152	39.1528315912924\\
11.9542109689134	39.0228945155433\\
12.2140851204115	38.8929574397943\\
12.4739592719096	38.7630203640452\\
12.7338334234077	38.6330832882961\\
12.9937075749058	38.5031462125471\\
13.2535817264039	38.373209136798\\
13.5134558779021	38.243272061049\\
13.7733300294002	38.1133349852999\\
14.0332041808983	37.9833979095509\\
14.2930783323964	37.8534608338018\\
14.5529524838945	37.7235237580527\\
14.8128266353926	37.5935866823037\\
15.0727007868908	37.4636496065546\\
15.3325749383889	37.3337125308056\\
15.592449089887	37.2037754550565\\
15.8523232413851	37.0738383793074\\
16.1121973928832	36.9439013035584\\
16.3720715443813	36.8139642278093\\
16.6319456958795	36.6840271520603\\
16.8918198473776	36.5540900763112\\
17.1516939988757	36.4241530005622\\
17.4115681503738	36.2942159248131\\
17.6714423018719	36.164278849064\\
17.93131645337	36.034341773315\\
18.1911906048682	35.9044046975659\\
18.4510647563663	35.7744676218169\\
18.7109389078644	35.6445305460678\\
18.9708130593625	35.5145934703187\\
19.2306872108606	35.3846563945697\\
19.4905613623587	35.2547193188206\\
19.7504355138569	35.1247822430716\\
20.010309665355	34.9948451673225\\
20.2701838168531	34.8649080915735\\
20.5300579683512	34.7349710158244\\
20.7899321198493	34.6050339400753\\
21.0498062713474	34.4750968643263\\
21.3096804228456	34.3451597885772\\
21.5695545743437	34.2152227128282\\
21.8294287258418	34.0852856370791\\
22.0893028773399	33.95534856133\\
22.349177028838	33.825411485581\\
22.6090511803361	33.6954744098319\\
22.8689253318343	33.5655373340829\\
23.1287994833324	33.4356002583338\\
23.3886736348305	33.3056631825848\\
23.6485477863286	33.1757261068357\\
23.9084219378267	33.0457890310866\\
24.1682960893248	32.9158519553376\\
24.428170240823	32.7859148795885\\
24.6880443923211	32.6559778038395\\
24.9479185438192	32.5260407280904\\
25.2077926953173	32.3961036523413\\
25.4676668468154	32.2661665765923\\
25.7275409983135	32.1362295008432\\
25.9874151498117	32.0062924250942\\
26.2472893013098	31.8763553493451\\
26.5071634528079	31.7464182735961\\
26.767037604306	31.616481197847\\
27.0269117558041	31.4865441220979\\
27.2867859073022	31.3566070463489\\
27.5466600588004	31.2266699705998\\
27.8065342102985	31.0967328948508\\
28.0664083617966	30.9667958191017\\
28.3262825132947	30.8368587433526\\
28.5861566647928	30.7069216676036\\
28.8460308162909	30.5769845918545\\
29.1059049677891	30.4470475161055\\
29.3657791192872	30.3171104403564\\
29.6256532707853	30.1871733646074\\
29.8855274222834	30.0572362888583\\
30.1454015737815	29.9272992131092\\
30.4052757252796	29.7973621373602\\
30.6651498767778	29.6674250616111\\
30.9250240282759	29.5374879858621\\
31.184898179774	29.407550910113\\
31.4447723312721	29.2776138343639\\
31.7046464827702	29.1476767586149\\
31.9645206342683	29.0177396828658\\
32.2243947857664	28.8878026071168\\
32.4842689372646	28.7578655313677\\
32.7441430887627	28.6279284556187\\
33.0040172402608	28.4979913798696\\
33.2638913917589	28.3680543041205\\
33.523765543257	28.2381172283715\\
33.7836396947552	28.1081801526224\\
34.0435138462533	27.9782430768734\\
34.3033879977514	27.8483060011243\\
34.5632621492495	27.7183689253753\\
34.8231363007476	27.5884318496262\\
35.0830104522457	27.4584947738771\\
35.3428846037438	27.3285576981281\\
35.602758755242	27.198620622379\\
35.8626329067401	27.06868354663\\
36.1225070582382	26.9387464708809\\
36.3823812097363	26.8088093951318\\
36.6422553612344	26.6788723193828\\
36.9021295127325	26.5489352436337\\
37.1620036642307	26.4189981678847\\
37.4218778157288	26.2890610921356\\
37.6817519672269	26.1591240163866\\
37.941626118725	26.0291869406375\\
38.2015002702231	25.8992498648884\\
38.4613744217212	25.7693127891394\\
38.7212485732194	25.6393757133903\\
38.9811227247175	25.5094386376413\\
39.2409968762156	25.3795015618922\\
39.5008710277137	25.2495644861431\\
39.7607451792118	25.1196274103941\\
40.0206193307099	24.989690334645\\
40.2804934822081	24.859753258896\\
40.5403676337062	24.7298161831469\\
40.8002417852043	24.5998791073979\\
41.0601159367024	24.4699420316488\\
41.3199900882005	24.3400049558997\\
41.5798642396986	24.2100678801507\\
41.8397383911968	24.0801308044016\\
42.0996125426949	23.9501937286526\\
42.359486694193	23.8202566529035\\
42.6193608456911	23.6903195771544\\
42.8792349971892	23.5603825014054\\
43.1391091486873	23.4304454256563\\
43.3989833001855	23.3005083499073\\
43.6588574516836	23.1705712741582\\
43.9187316031817	23.0406341984092\\
44.1786057546798	22.9106971226601\\
44.4384799061779	22.780760046911\\
44.698354057676	22.650822971162\\
44.9582282091742	22.5208858954129\\
45.2181023606723	22.3909488196639\\
45.4779765121704	22.2610117439148\\
45.7378506636685	22.1310746681657\\
45.9977248151666	22.0011375924167\\
46.2575989666647	21.8712005166676\\
46.5174731181629	21.7412634409186\\
46.777347269661	21.6113263651695\\
47.0372214211591	21.4813892894205\\
47.2970955726572	21.3514522136714\\
47.5569697241553	21.2215151379223\\
47.8168438756534	21.0915780621733\\
48.0767180271516	20.9616409864242\\
48.3365921786497	20.8317039106752\\
48.5964663301478	20.7017668349261\\
48.8563404816459	20.571829759177\\
49.116214633144	20.441892683428\\
49.3760887846421	20.3119556076789\\
49.6359629361403	20.1820185319299\\
49.8958370876384	20.0520814561808\\
50.1557112391365	19.9221443804318\\
50.4155853906346	19.7922073046827\\
50.6754595421327	19.6622702289336\\
50.9353336936308	19.5323331531846\\
51.195207845129	19.4023960774355\\
51.4550819966271	19.2724590016865\\
51.7149561481252	19.1425219259374\\
51.9748302996233	19.0125848501883\\
52.2347044511214	18.8826477744393\\
52.4945786026195	18.7527106986902\\
52.7544527541177	18.6227736229412\\
53.0143269056158	18.4928365471921\\
53.2742010571139	18.3628994714431\\
53.534075208612	18.232962395694\\
53.7939493601101	18.1030253199449\\
54.0538235116082	17.9730882441959\\
54.3136976631064	17.8431511684468\\
54.5735718146045	17.7132140926978\\
54.8334459661026	17.5832770169487\\
55.0933201176007	17.4533399411996\\
}
\closedcycle;
\addplot [color=black, dotted]
  table[row sep=crcr]{%
0	45\\
0.259874151498117	44.8700629242509\\
0.519748302996233	44.7401258485019\\
0.77962245449435	44.6101887727528\\
1.03949660599247	44.4802516970038\\
1.29937075749058	44.3503146212547\\
1.5592449089887	44.2203775455056\\
1.81911906048682	44.0904404697566\\
2.07899321198493	43.9605033940075\\
2.33886736348305	43.8305663182585\\
2.59874151498117	43.7006292425094\\
2.85861566647928	43.5706921667604\\
3.1184898179774	43.4407550910113\\
3.37836396947552	43.3108180152622\\
3.63823812097363	43.1808809395132\\
3.89811227247175	43.0509438637641\\
4.15798642396986	42.9210067880151\\
4.41786057546798	42.791069712266\\
4.6777347269661	42.661132636517\\
4.93760887846421	42.5311955607679\\
5.19748302996233	42.4012584850188\\
5.45735718146045	42.2713214092698\\
5.71723133295856	42.1413843335207\\
5.97710548445668	42.0114472577717\\
6.2369796359548	41.8815101820226\\
6.49685378745291	41.7515731062735\\
6.75672793895103	41.6216360305245\\
7.01660209044915	41.4916989547754\\
7.27647624194726	41.3617618790264\\
7.53635039344538	41.2318248032773\\
7.7962245449435	41.1018877275283\\
8.05609869644161	40.9719506517792\\
8.31597284793973	40.8420135760301\\
8.57584699943785	40.7120765002811\\
8.83572115093596	40.582139424532\\
9.09559530243408	40.452202348783\\
9.3554694539322	40.3222652730339\\
9.61534360543031	40.1923281972848\\
9.87521775692843	40.0623911215358\\
10.1350919084265	39.9324540457867\\
10.3949660599247	39.8025169700377\\
10.6548402114228	39.6725798942886\\
10.9147143629209	39.5426428185396\\
11.174588514419	39.4127057427905\\
11.4344626659171	39.2827686670414\\
11.6943368174152	39.1528315912924\\
11.9542109689134	39.0228945155433\\
12.2140851204115	38.8929574397943\\
12.4739592719096	38.7630203640452\\
12.7338334234077	38.6330832882961\\
12.9937075749058	38.5031462125471\\
13.2535817264039	38.373209136798\\
13.5134558779021	38.243272061049\\
13.7733300294002	38.1133349852999\\
14.0332041808983	37.9833979095509\\
14.2930783323964	37.8534608338018\\
14.5529524838945	37.7235237580527\\
14.8128266353926	37.5935866823037\\
15.0727007868908	37.4636496065546\\
15.3325749383889	37.3337125308056\\
15.592449089887	37.2037754550565\\
15.8523232413851	37.0738383793074\\
16.1121973928832	36.9439013035584\\
16.3720715443813	36.8139642278093\\
16.6319456958795	36.6840271520603\\
16.8918198473776	36.5540900763112\\
17.1516939988757	36.4241530005622\\
17.4115681503738	36.2942159248131\\
17.6714423018719	36.164278849064\\
17.93131645337	36.034341773315\\
18.1911906048682	35.9044046975659\\
18.4510647563663	35.7744676218169\\
18.7109389078644	35.6445305460678\\
18.9708130593625	35.5145934703187\\
19.2306872108606	35.3846563945697\\
19.4905613623587	35.2547193188206\\
19.7504355138569	35.1247822430716\\
20.010309665355	34.9948451673225\\
20.2701838168531	34.8649080915735\\
20.5300579683512	34.7349710158244\\
20.7899321198493	34.6050339400753\\
21.0498062713474	34.4750968643263\\
21.3096804228456	34.3451597885772\\
21.5695545743437	34.2152227128282\\
21.8294287258418	34.0852856370791\\
22.0893028773399	33.95534856133\\
22.349177028838	33.825411485581\\
22.6090511803361	33.6954744098319\\
22.8689253318343	33.5655373340829\\
23.1287994833324	33.4356002583338\\
23.3886736348305	33.3056631825848\\
23.6485477863286	33.1757261068357\\
23.9084219378267	33.0457890310866\\
24.1682960893248	32.9158519553376\\
24.428170240823	32.7859148795885\\
24.6880443923211	32.6559778038395\\
24.9479185438192	32.5260407280904\\
25.2077926953173	32.3961036523413\\
25.4676668468154	32.2661665765923\\
25.7275409983135	32.1362295008432\\
25.9874151498117	32.0062924250942\\
26.2472893013098	31.8763553493451\\
26.5071634528079	31.7464182735961\\
26.767037604306	31.616481197847\\
27.0269117558041	31.4865441220979\\
27.2867859073022	31.3566070463489\\
27.5466600588004	31.2266699705998\\
27.8065342102985	31.0967328948508\\
28.0664083617966	30.9667958191017\\
28.3262825132947	30.8368587433526\\
28.5861566647928	30.7069216676036\\
28.8460308162909	30.5769845918545\\
29.1059049677891	30.4470475161055\\
29.3657791192872	30.3171104403564\\
29.6256532707853	30.1871733646074\\
29.8855274222834	30.0572362888583\\
30.1454015737815	29.9272992131092\\
30.4052757252796	29.7973621373602\\
30.6651498767778	29.6674250616111\\
30.9250240282759	29.5374879858621\\
31.184898179774	29.407550910113\\
31.4447723312721	29.2776138343639\\
31.7046464827702	29.1476767586149\\
31.9645206342683	29.0177396828658\\
32.2243947857664	28.8878026071168\\
32.4842689372646	28.7578655313677\\
32.7441430887627	28.6279284556187\\
33.0040172402608	28.4979913798696\\
33.2638913917589	28.3680543041205\\
33.523765543257	28.2381172283715\\
33.7836396947552	28.1081801526224\\
34.0435138462533	27.9782430768734\\
34.3033879977514	27.8483060011243\\
34.5632621492495	27.7183689253753\\
34.8231363007476	27.5884318496262\\
35.0830104522457	27.4584947738771\\
35.3428846037438	27.3285576981281\\
35.602758755242	27.198620622379\\
35.8626329067401	27.06868354663\\
36.1225070582382	26.9387464708809\\
36.3823812097363	26.8088093951318\\
36.6422553612344	26.6788723193828\\
36.9021295127325	26.5489352436337\\
37.1620036642307	26.4189981678847\\
37.4218778157288	26.2890610921356\\
37.6817519672269	26.1591240163866\\
37.941626118725	26.0291869406375\\
38.2015002702231	25.8992498648884\\
38.4613744217212	25.7693127891394\\
38.7212485732194	25.6393757133903\\
38.9811227247175	25.5094386376413\\
39.2409968762156	25.3795015618922\\
39.5008710277137	25.2495644861431\\
39.7607451792118	25.1196274103941\\
40.0206193307099	24.989690334645\\
40.2804934822081	24.859753258896\\
40.5403676337062	24.7298161831469\\
40.8002417852043	24.5998791073979\\
41.0601159367024	24.4699420316488\\
41.3199900882005	24.3400049558997\\
41.5798642396986	24.2100678801507\\
41.8397383911968	24.0801308044016\\
42.0996125426949	23.9501937286526\\
42.359486694193	23.8202566529035\\
42.6193608456911	23.6903195771544\\
42.8792349971892	23.5603825014054\\
43.1391091486873	23.4304454256563\\
43.3989833001855	23.3005083499073\\
43.6588574516836	23.1705712741582\\
43.9187316031817	23.0406341984092\\
44.1786057546798	22.9106971226601\\
44.4384799061779	22.780760046911\\
44.698354057676	22.650822971162\\
44.9582282091742	22.5208858954129\\
45.2181023606723	22.3909488196639\\
45.4779765121704	22.2610117439148\\
45.7378506636685	22.1310746681657\\
45.9977248151666	22.0011375924167\\
46.2575989666647	21.8712005166676\\
46.5174731181629	21.7412634409186\\
46.777347269661	21.6113263651695\\
47.0372214211591	21.4813892894205\\
47.2970955726572	21.3514522136714\\
47.5569697241553	21.2215151379223\\
47.8168438756534	21.0915780621733\\
48.0767180271516	20.9616409864242\\
48.3365921786497	20.8317039106752\\
48.5964663301478	20.7017668349261\\
48.8563404816459	20.571829759177\\
49.116214633144	20.441892683428\\
49.3760887846421	20.3119556076789\\
49.6359629361403	20.1820185319299\\
49.8958370876384	20.0520814561808\\
50.1557112391365	19.9221443804318\\
50.4155853906346	19.7922073046827\\
50.6754595421327	19.6622702289336\\
50.9353336936308	19.5323331531846\\
51.195207845129	19.4023960774355\\
51.4550819966271	19.2724590016865\\
51.7149561481252	19.1425219259374\\
51.9748302996233	19.0125848501883\\
52.2347044511214	18.8826477744393\\
52.4945786026195	18.7527106986902\\
52.7544527541177	18.6227736229412\\
53.0143269056158	18.4928365471921\\
53.2742010571139	18.3628994714431\\
53.534075208612	18.232962395694\\
53.7939493601101	18.1030253199449\\
54.0538235116082	17.9730882441959\\
54.3136976631064	17.8431511684468\\
54.5735718146045	17.7132140926978\\
54.8334459661026	17.5832770169487\\
55.0933201176007	17.4533399411996\\
};

\addplot[only marks, mark=*, mark options={}, mark size=2.1651pt, color=red, fill=red] table[row sep=crcr]{%
x	y\\
8.264	50.0015315748611\\
};

\addplot[only marks, mark=*, mark options={}, mark size=2.1651pt, color=green, fill=green] table[row sep=crcr]{%
x	y\\
48.8307508835193	20.5846245388451\\
};

\addplot[only marks, mark=*, mark options={}, mark size=2.1651pt, color=blue, fill=blue] table[row sep=crcr]{%
x	y\\
8.264	50.0015315748611\\
};

\addplot[only marks, mark=*, mark options={}, mark size=2.1651pt, color=blue, fill=blue, forget plot] table[row sep=crcr]{%
x	y\\
8.32227524268865	49.8242674790752\\
};
\addplot[only marks, mark=*, mark options={}, mark size=2.1651pt, color=blue, fill=blue, forget plot] table[row sep=crcr]{%
x	y\\
8.50037481909624	49.2950638089152\\
};
\addplot[only marks, mark=*, mark options={}, mark size=2.1651pt, color=blue, fill=blue, forget plot] table[row sep=crcr]{%
x	y\\
8.81028066163199	48.4187336285228\\
};
\addplot[only marks, mark=*, mark options={}, mark size=2.1651pt, color=blue, fill=blue, forget plot] table[row sep=crcr]{%
x	y\\
9.26755843410105	47.2264089099472\\
};
\addplot[only marks, mark=*, mark options={}, mark size=2.1651pt, color=blue, fill=blue, forget plot] table[row sep=crcr]{%
x	y\\
9.91477581895102	45.7365283781963\\
};
\addplot[only marks, mark=*, mark options={}, mark size=2.1651pt, color=blue, fill=blue, forget plot] table[row sep=crcr]{%
x	y\\
10.8489222239293	43.9700483275146\\
};
\addplot[only marks, mark=*, mark options={}, mark size=2.1651pt, color=blue, fill=blue, forget plot] table[row sep=crcr]{%
x	y\\
12.2370821253948	42.1007572928369\\
};
\addplot[only marks, mark=*, mark options={}, mark size=2.1651pt, color=blue, fill=blue, forget plot] table[row sep=crcr]{%
x	y\\
14.3853609585761	40.6706064068703\\
};
\addplot[only marks, mark=*, mark options={}, mark size=2.1651pt, color=blue, fill=blue, forget plot] table[row sep=crcr]{%
x	y\\
17.0838843894301	40.5390589595126\\
};
\addplot[only marks, mark=*, mark options={}, mark size=2.1651pt, color=blue, fill=blue, forget plot] table[row sep=crcr]{%
x	y\\
19.7362050518722	40.9698184354298\\
};
\addplot[only marks, mark=*, mark options={}, mark size=2.1651pt, color=blue, fill=blue, forget plot] table[row sep=crcr]{%
x	y\\
22.3779982643229	41.0722879912515\\
};
\addplot[only marks, mark=*, mark options={}, mark size=2.1651pt, color=blue, fill=blue, forget plot] table[row sep=crcr]{%
x	y\\
25.0197914767736	40.7823575470732\\
};
\addplot[only marks, mark=*, mark options={}, mark size=2.1651pt, color=blue, fill=blue, forget plot] table[row sep=crcr]{%
x	y\\
27.6615846892244	40.1000271028947\\
};
\addplot[only marks, mark=*, mark options={}, mark size=2.1651pt, color=blue, fill=blue, forget plot] table[row sep=crcr]{%
x	y\\
30.3033779016751	39.0252966587163\\
};
\addplot[only marks, mark=*, mark options={}, mark size=2.1651pt, color=blue, fill=blue, forget plot] table[row sep=crcr]{%
x	y\\
32.9451711141429	37.5581662145721\\
};
\addplot[only marks, mark=*, mark options={}, mark size=2.1651pt, color=blue, fill=blue, forget plot] table[row sep=crcr]{%
x	y\\
35.5869643266559	35.698635770518\\
};
\addplot[only marks, mark=*, mark options={}, mark size=2.1651pt, color=blue, fill=blue, forget plot] table[row sep=crcr]{%
x	y\\
38.2287575392172	33.4467053265607\\
};
\addplot[only marks, mark=*, mark options={}, mark size=2.1651pt, color=blue, fill=blue, forget plot] table[row sep=crcr]{%
x	y\\
40.870550751833	30.8023748827123\\
};
\addplot[only marks, mark=*, mark options={}, mark size=2.1651pt, color=blue, fill=blue, forget plot] table[row sep=crcr]{%
x	y\\
43.5123439645154	27.7656444389971\\
};
\addplot[only marks, mark=*, mark options={}, mark size=2.1651pt, color=blue, fill=blue, forget plot] table[row sep=crcr]{%
x	y\\
46.1541371772918	24.3365139954701\\
};
\addplot[only marks, mark=*, mark options={}, mark size=2.1651pt, color=blue, fill=blue, forget plot] table[row sep=crcr]{%
x	y\\
48.8307508835193	20.5846245388451\\
};
\addplot[only marks, mark=*, mark options={}, mark size=2.1651pt, color=blue, fill=blue, forget plot] table[row sep=crcr]{%
x	y\\
52.6297576312733	18.6851211652731\\
};
\addplot[only marks, mark=*, mark options={}, mark size=2.1651pt, color=red, fill=red, forget plot] table[row sep=crcr]{%
x	y\\
8.264	50.0015315748611\\
};
\addplot[only marks, mark=*, mark options={}, mark size=2.1651pt, color=green, fill=green, forget plot] table[row sep=crcr]{%
x	y\\
48.8307508835193	20.5846245388451\\
};
\end{axis}
\end{tikzpicture}%

%% file: Graphics/tikz/pump_sim_v1_10.tex
%
%
\begin{tikzpicture}

\begin{axis}[%
width=1.8in,
height=0.6in,
scale only axis,
xmin=0,
xmax=6.32676701565019,
xlabel style={font=\color{white!15!black}},
ymin=0,
ymax=1,
ylabel style={font=\color{white!15!black}},
ylabel={$z$\,[m]},
ytick={0, 0.4, 0.8},
axis background/.style={fill=white},
xmajorgrids,
ymajorgrids
]
\addplot [color=blue, dashed, forget plot]
  table[row sep=crcr]{%
0	0\\
0.0532640231023162	0.00113374969234851\\
0.106269229176082	0.00450028053728098\\
0.158870545056015	0.010011285400809\\
0.210628675954945	0.0174848959947971\\
0.261550057835917	0.0267450748654208\\
0.311507758935552	0.0375754727137558\\
0.360062043707937	0.0496552197400115\\
0.407401617441647	0.0627976859463402\\
0.453433798388091	0.0767568445352903\\
0.497834587062522	0.0912117019193533\\
0.540836851966452	0.106029686654843\\
0.582385962513731	0.121007983692645\\
0.622289745517508	0.135907663585066\\
0.660740362785498	0.150651844775047\\
0.697713608013664	0.165105680851587\\
0.733117456118999	0.179125802949922\\
0.767081534003758	0.192674987649475\\
0.79960196535333	0.205680757535859\\
0.830649820532426	0.218075963968562\\
0.860302670439425	0.229849858733962\\
0.888569314736214	0.240976235601789\\
0.915455851290165	0.251438410865581\\
0.941006333421377	0.261243355708035\\
0.965237217077727	0.270394592749851\\
0.988171413618007	0.278904199762687\\
1.00983512908308	0.286790260237056\\
1.03024896488917	0.294072518001921\\
1.04944173151907	0.300776398726461\\
1.0674317292534	0.306925540463032\\
1.08424094310811	0.312546156407631\\
1.09989790971937	0.317667207115927\\
1.11441789653693	0.322313375690873\\
1.12782200972506	0.326511013007842\\
1.14013521882624	0.330286975508136\\
1.15137126128443	0.333663727535094\\
1.16154857841801	0.336664091523483\\
1.17068695518916	0.339309957560832\\
1.17879831752157	0.341619512824188\\
1.18589713029831	0.34361020265551\\
1.1919973316148	0.345297745376203\\
1.19710805752993	0.346694922999972\\
1.20123891707442	0.34781302985195\\
1.20439789560942	0.348661272256284\\
1.20659026217775	0.349246475836455\\
1.20782025557205	0.349573539505225\\
};
\addplot [color=blue, dashed, forget plot]
  table[row sep=crcr]{%
0	0.4368\\
0.0532640231023162	0.437933749692348\\
0.106269229176082	0.44130028053728\\
0.158870545056015	0.446811285400808\\
0.210628675954945	0.454284895994796\\
0.261550057835917	0.463545074865419\\
0.311507758935552	0.474375472713754\\
0.360062043707937	0.486455219740009\\
0.407401617441647	0.499597685946338\\
0.453433798388091	0.513556844535288\\
0.497834587062522	0.52801170191935\\
0.540836851966452	0.54282968665484\\
0.582385962513731	0.557807983692641\\
0.622289745517508	0.572707663585062\\
0.660740362785498	0.587451844775043\\
0.697713608013664	0.601905680851582\\
0.733117456118999	0.615925802949916\\
0.767081534003758	0.62947498764947\\
0.79960196535333	0.642480757535853\\
0.830649820532426	0.654875963968555\\
0.860302670439425	0.666649858733956\\
0.888569314736214	0.677776235601782\\
0.915455851290165	0.688238410865574\\
0.941006333421377	0.698043355708028\\
0.965237217077727	0.707194592749844\\
0.988171413618007	0.715704199762679\\
1.00983512908308	0.723590260237047\\
1.03024896488917	0.730872518001913\\
1.04944173151907	0.737576398726452\\
1.0674317292534	0.743725540463023\\
1.08424094310811	0.749346156407622\\
1.09989790971937	0.754467207115918\\
1.11441789653693	0.759113375690863\\
1.12782200972506	0.763311013007831\\
1.14013521882624	0.767086975508125\\
1.15137126128443	0.770463727535083\\
1.16154857841801	0.773464091523472\\
1.17068695518916	0.77610995756082\\
1.17879831752157	0.778419512824176\\
1.18589713029831	0.780410202655497\\
1.1919973316148	0.78209774537619\\
1.19710805752993	0.783494922999959\\
1.20123891707442	0.784613029851936\\
1.20439789560942	0.785461272256271\\
1.20659026217775	0.786046475836441\\
1.20782025557205	0.786373539505211\\
};
\addplot[fill=black, fill opacity=0.2, draw=black, forget plot] table[row sep=crcr]{%
0	0\\
0.170812099169403	0.0115576453834497\\
0.336782683861544	0.0436794709456657\\
0.497589275462136	0.0911293760741091\\
0.645476938193043	0.144759153751302\\
0.787462087675547	0.200825567366743\\
0.923608243883464	0.254582692273039\\
1.05248828809442	0.30182713249463\\
1.17897980506459	0.341670761453521\\
1.30349073904733	0.372093360585968\\
1.4258029605988	0.391649533866067\\
1.54801889538578	0.399792511333579\\
1.66987767775744	0.396085987645504\\
1.79248112171783	0.380662257661554\\
1.91605998939059	0.354182049138064\\
2.04070208168642	0.317988046270917\\
2.16990646629053	0.27280318985524\\
2.30301635863522	0.221231173120379\\
2.44048683040462	0.166442547465643\\
2.58892195577431	0.110234123537304\\
2.74441968589398	0.0598496941728275\\
2.90641898849603	0.0217178133317126\\
3.07777079746224	0.00162708076928292\\
3.2485078598767	0.00455494912007637\\
3.41889068677856	0.0299773557911501\\
3.5491051599619	0.0628299618344138\\
3.7320969136595	0.124001570596267\\
3.87836797833319	0.180581504224314\\
4.014230033924	0.234718902785784\\
4.14604663411222	0.284846040918872\\
4.27430016976514	0.328018507675508\\
4.39841560868949	0.361847093912904\\
4.52166062765036	0.385624665716526\\
4.64405658678381	0.398135178786292\\
4.76594323797774	0.398853872946082\\
4.88826206585726	0.387754504567346\\
5.01078739502764	0.36542800479171\\
5.13582426865394	0.332466196974521\\
5.26309035433946	0.290469107260682\\
5.39303183454903	0.241596248926718\\
5.53080007071276	0.186804421603012\\
5.67472847850648	0.130691502517037\\
5.8247730991157	0.0783312805146177\\
5.98772825867237	0.0339136471869849\\
6.15512262641353	0.00652423685480769\\
6.32676701565019	0.000759265235273393\\
}
\closedcycle;
\addplot[only marks, mark=*, mark options={}, mark size=1.7678pt, color=blue, fill=blue, forget plot] table[row sep=crcr]{%
x	y\\
0	0\\
};
\addplot[only marks, mark=*, mark options={}, mark size=1.7678pt, color=blue, fill=blue, forget plot] table[row sep=crcr]{%
x	y\\
0	0.4368\\
};
\addplot [color=blue, forget plot]
  table[row sep=crcr]{%
0	0\\
0	0.4368\\
};
\addplot[only marks, mark=*, mark options={}, mark size=1.7678pt, color=blue, fill=blue, forget plot] table[row sep=crcr]{%
x	y\\
0.261550057835917	0.0267450748654208\\
};
\addplot[only marks, mark=*, mark options={}, mark size=1.7678pt, color=blue, fill=blue, forget plot] table[row sep=crcr]{%
x	y\\
0.261550057835917	0.463545074865419\\
};
\addplot [color=blue, forget plot]
  table[row sep=crcr]{%
0.261550057835917	0.0267450748654208\\
0.261550057835917	0.463545074865419\\
};
\addplot[only marks, mark=*, mark options={}, mark size=1.7678pt, color=blue, fill=blue, forget plot] table[row sep=crcr]{%
x	y\\
0.497834587062522	0.0912117019193533\\
};
\addplot[only marks, mark=*, mark options={}, mark size=1.7678pt, color=blue, fill=blue, forget plot] table[row sep=crcr]{%
x	y\\
0.497834587062522	0.52801170191935\\
};
\addplot [color=blue, forget plot]
  table[row sep=crcr]{%
0.497834587062522	0.0912117019193533\\
0.497834587062522	0.52801170191935\\
};
\addplot[only marks, mark=*, mark options={}, mark size=1.7678pt, color=blue, fill=blue, forget plot] table[row sep=crcr]{%
x	y\\
0.697713608013664	0.165105680851587\\
};
\addplot[only marks, mark=*, mark options={}, mark size=1.7678pt, color=blue, fill=blue, forget plot] table[row sep=crcr]{%
x	y\\
0.697713608013664	0.601905680851582\\
};
\addplot [color=blue, forget plot]
  table[row sep=crcr]{%
0.697713608013664	0.165105680851587\\
0.697713608013664	0.601905680851582\\
};
\addplot[only marks, mark=*, mark options={}, mark size=1.7678pt, color=blue, fill=blue, forget plot] table[row sep=crcr]{%
x	y\\
0.860302670439425	0.229849858733962\\
};
\addplot[only marks, mark=*, mark options={}, mark size=1.7678pt, color=blue, fill=blue, forget plot] table[row sep=crcr]{%
x	y\\
0.860302670439425	0.666649858733956\\
};
\addplot [color=blue, forget plot]
  table[row sep=crcr]{%
0.860302670439425	0.229849858733962\\
0.860302670439425	0.666649858733956\\
};
\addplot[only marks, mark=*, mark options={}, mark size=1.7678pt, color=blue, fill=blue, forget plot] table[row sep=crcr]{%
x	y\\
0.988171413618007	0.278904199762687\\
};
\addplot[only marks, mark=*, mark options={}, mark size=1.7678pt, color=blue, fill=blue, forget plot] table[row sep=crcr]{%
x	y\\
0.988171413618007	0.715704199762679\\
};
\addplot [color=blue, forget plot]
  table[row sep=crcr]{%
0.988171413618007	0.278904199762687\\
0.988171413618007	0.715704199762679\\
};
\addplot[only marks, mark=*, mark options={}, mark size=1.7678pt, color=blue, fill=blue, forget plot] table[row sep=crcr]{%
x	y\\
1.08424094310811	0.312546156407631\\
};
\addplot[only marks, mark=*, mark options={}, mark size=1.7678pt, color=blue, fill=blue, forget plot] table[row sep=crcr]{%
x	y\\
1.08424094310811	0.749346156407622\\
};
\addplot [color=blue, forget plot]
  table[row sep=crcr]{%
1.08424094310811	0.312546156407631\\
1.08424094310811	0.749346156407622\\
};
\addplot[only marks, mark=*, mark options={}, mark size=1.7678pt, color=blue, fill=blue, forget plot] table[row sep=crcr]{%
x	y\\
1.15137126128443	0.333663727535094\\
};
\addplot[only marks, mark=*, mark options={}, mark size=1.7678pt, color=blue, fill=blue, forget plot] table[row sep=crcr]{%
x	y\\
1.15137126128443	0.770463727535083\\
};
\addplot [color=blue, forget plot]
  table[row sep=crcr]{%
1.15137126128443	0.333663727535094\\
1.15137126128443	0.770463727535083\\
};
\addplot[only marks, mark=*, mark options={}, mark size=1.7678pt, color=blue, fill=blue, forget plot] table[row sep=crcr]{%
x	y\\
1.1919973316148	0.345297745376203\\
};
\addplot[only marks, mark=*, mark options={}, mark size=1.7678pt, color=blue, fill=blue, forget plot] table[row sep=crcr]{%
x	y\\
1.1919973316148	0.78209774537619\\
};
\addplot [color=blue, forget plot]
  table[row sep=crcr]{%
1.1919973316148	0.345297745376203\\
1.1919973316148	0.78209774537619\\
};
\addplot[only marks, mark=*, mark options={}, mark size=1.7678pt, color=blue, fill=blue, forget plot] table[row sep=crcr]{%
x	y\\
1.20782025557205	0.349573539505225\\
};
\addplot[only marks, mark=*, mark options={}, mark size=1.7678pt, color=blue, fill=blue, forget plot] table[row sep=crcr]{%
x	y\\
1.20782025557205	0.786373539505211\\
};
\addplot [color=blue, forget plot]
  table[row sep=crcr]{%
1.20782025557205	0.349573539505225\\
1.20782025557205	0.786373539505211\\
};
\end{axis}

\end{tikzpicture}%

%% file: Graphics/tikz/pump_sim_v1_15.tex
%
%
\begin{tikzpicture}

\begin{axis}[%
width=1.8in,
height=0.6in,
scale only axis,
xmin=0,
xmax=6.32676701565019,
xlabel style={font=\color{white!15!black}},
ymin=0,
ymax=1,
ylabel style={font=\color{white!15!black}},
ytick={0, 0.4, 0.8},
axis background/.style={fill=white},
xmajorgrids,
ymajorgrids
]
\addplot [color=blue, dashed, forget plot]
  table[row sep=crcr]{%
0	0\\
0.170812099169403	0.0115576446309674\\
0.336782683861544	0.0436794680863015\\
0.497589275462136	0.0911293732918189\\
0.645476938193043	0.144759151105401\\
0.787462087675547	0.200825564928403\\
0.923608243883464	0.25458269012797\\
1.05248828809442	0.301827130734419\\
1.17897980506459	0.341670760210328\\
1.30349073904733	0.37209336000721\\
1.4258029605988	0.391649534099486\\
1.54801889538578	0.399792512507703\\
1.66987767775744	0.396085989754831\\
1.79248112171783	0.38066226059668\\
1.91605998939059	0.354182052725724\\
2.04070208168642	0.317988050315549\\
2.16990646629053	0.272803194204276\\
2.30301635863522	0.221231177644709\\
2.44048683040462	0.16644255195228\\
2.58892195577431	0.110234127919797\\
2.74441968589398	0.0598496971229264\\
2.90641898849603	0.0217178121258063\\
3.07777079746224	0.00162707871099803\\
3.2485078598767	0.00455494609554399\\
3.41889068677856	0.029977352389428\\
3.5491051599619	0.0628299578371897\\
3.7320969136595	0.124001566761919\\
3.87836797833319	0.18058150058317\\
4.014230033924	0.234718899405499\\
4.14604663411222	0.284846037897679\\
4.27430016976514	0.328018505131472\\
4.39841560868949	0.361847091971757\\
4.52166062765036	0.385624664546074\\
4.64405658678381	0.398135178520487\\
4.76594323797774	0.398853873645271\\
4.88826206585726	0.387754506178115\\
5.01078739502764	0.365428007140892\\
5.13582426865394	0.332466199884499\\
5.26309035433946	0.290469110568314\\
5.39303183454903	0.241596252478731\\
5.53080007071276	0.186804425275192\\
5.67472847850648	0.130691506166777\\
5.8247730991157	0.0783312836540657\\
5.98772825867237	0.0339136480753616\\
6.15512262641353	0.00652423511689626\\
6.32676701565019	0.000759262241091715\\
};
\addplot [color=blue, dashed, forget plot]
  table[row sep=crcr]{%
0	0.4368\\
0.170812099169403	0.448357644630967\\
0.336782683861544	0.480479468086301\\
0.497589275462136	0.527929373291818\\
0.645476938193043	0.581559151105399\\
0.787462087675547	0.637625564928402\\
0.923608243883464	0.691382690127968\\
1.05248828809442	0.738627130734417\\
1.17897980506459	0.778470760210326\\
1.30349073904733	0.808893360007208\\
1.4258029605988	0.828449534099483\\
1.54801889538578	0.836592512507701\\
1.66987767775744	0.832885989754828\\
1.79248112171783	0.817462260596677\\
1.91605998939059	0.79098205272572\\
2.04070208168642	0.754788050315546\\
2.16990646629053	0.709603194204272\\
2.30301635863522	0.658031177644705\\
2.44048683040462	0.603242551952275\\
2.58892195577431	0.547034127919792\\
2.74441968589398	0.496649697122921\\
2.90641898849603	0.458517812125801\\
3.07777079746224	0.438427078710992\\
3.2485078598767	0.441354946095538\\
3.41889068677856	0.466777352389422\\
3.5491051599619	0.499629957837183\\
3.7320969136595	0.560801566761912\\
3.87836797833319	0.617381500583163\\
4.014230033924	0.671518899405492\\
4.14604663411222	0.721646037897672\\
4.27430016976514	0.764818505131465\\
4.39841560868949	0.798647091971749\\
4.52166062765036	0.822424664546066\\
4.64405658678381	0.834935178520479\\
4.76594323797774	0.835653873645262\\
4.88826206585726	0.824554506178107\\
5.01078739502764	0.802228007140883\\
5.13582426865394	0.76926619988449\\
5.26309035433946	0.727269110568305\\
5.39303183454903	0.678396252478721\\
5.53080007071276	0.623604425275182\\
5.67472847850648	0.567491506166767\\
5.8247730991157	0.515131283654055\\
5.98772825867237	0.470713648075351\\
6.15512262641353	0.443324235116885\\
6.32676701565019	0.43755926224108\\
};
\addplot[fill=black, fill opacity=0.2, draw=black, forget plot] table[row sep=crcr]{%
0	0\\
0.170812099169403	0.0115576453834497\\
0.336782683861544	0.0436794709456657\\
0.497589275462136	0.0911293760741091\\
0.645476938193043	0.144759153751302\\
0.787462087675547	0.200825567366743\\
0.923608243883464	0.254582692273039\\
1.05248828809442	0.30182713249463\\
1.17897980506459	0.341670761453521\\
1.30349073904733	0.372093360585968\\
1.4258029605988	0.391649533866067\\
1.54801889538578	0.399792511333579\\
1.66987767775744	0.396085987645504\\
1.79248112171783	0.380662257661554\\
1.91605998939059	0.354182049138064\\
2.04070208168642	0.317988046270917\\
2.16990646629053	0.27280318985524\\
2.30301635863522	0.221231173120379\\
2.44048683040462	0.166442547465643\\
2.58892195577431	0.110234123537304\\
2.74441968589398	0.0598496941728275\\
2.90641898849603	0.0217178133317126\\
3.07777079746224	0.00162708076928292\\
3.2485078598767	0.00455494912007637\\
3.41889068677856	0.0299773557911501\\
3.5491051599619	0.0628299618344138\\
3.7320969136595	0.124001570596267\\
3.87836797833319	0.180581504224314\\
4.014230033924	0.234718902785784\\
4.14604663411222	0.284846040918872\\
4.27430016976514	0.328018507675508\\
4.39841560868949	0.361847093912904\\
4.52166062765036	0.385624665716526\\
4.64405658678381	0.398135178786292\\
4.76594323797774	0.398853872946082\\
4.88826206585726	0.387754504567346\\
5.01078739502764	0.36542800479171\\
5.13582426865394	0.332466196974521\\
5.26309035433946	0.290469107260682\\
5.39303183454903	0.241596248926718\\
5.53080007071276	0.186804421603012\\
5.67472847850648	0.130691502517037\\
5.8247730991157	0.0783312805146177\\
5.98772825867237	0.0339136471869849\\
6.15512262641353	0.00652423685480769\\
6.32676701565019	0.000759265235273393\\
}
\closedcycle;
\addplot[only marks, mark=*, mark options={}, mark size=1.7678pt, color=blue, fill=blue, forget plot] table[row sep=crcr]{%
x	y\\
0	0\\
};
\addplot[only marks, mark=*, mark options={}, mark size=1.7678pt, color=blue, fill=blue, forget plot] table[row sep=crcr]{%
x	y\\
0	0.4368\\
};
\addplot [color=blue, forget plot]
  table[row sep=crcr]{%
0	0\\
0	0.4368\\
};
\addplot[only marks, mark=*, mark options={}, mark size=1.7678pt, color=blue, fill=blue, forget plot] table[row sep=crcr]{%
x	y\\
0.787462087675547	0.200825564928403\\
};
\addplot[only marks, mark=*, mark options={}, mark size=1.7678pt, color=blue, fill=blue, forget plot] table[row sep=crcr]{%
x	y\\
0.787462087675547	0.637625564928402\\
};
\addplot [color=blue, forget plot]
  table[row sep=crcr]{%
0.787462087675547	0.200825564928403\\
0.787462087675547	0.637625564928402\\
};
\addplot[only marks, mark=*, mark options={}, mark size=1.7678pt, color=blue, fill=blue, forget plot] table[row sep=crcr]{%
x	y\\
1.4258029605988	0.391649534099486\\
};
\addplot[only marks, mark=*, mark options={}, mark size=1.7678pt, color=blue, fill=blue, forget plot] table[row sep=crcr]{%
x	y\\
1.4258029605988	0.828449534099483\\
};
\addplot [color=blue, forget plot]
  table[row sep=crcr]{%
1.4258029605988	0.391649534099486\\
1.4258029605988	0.828449534099483\\
};
\addplot[only marks, mark=*, mark options={}, mark size=1.7678pt, color=blue, fill=blue, forget plot] table[row sep=crcr]{%
x	y\\
2.04070208168642	0.317988050315549\\
};
\addplot[only marks, mark=*, mark options={}, mark size=1.7678pt, color=blue, fill=blue, forget plot] table[row sep=crcr]{%
x	y\\
2.04070208168642	0.754788050315546\\
};
\addplot [color=blue, forget plot]
  table[row sep=crcr]{%
2.04070208168642	0.317988050315549\\
2.04070208168642	0.754788050315546\\
};
\addplot[only marks, mark=*, mark options={}, mark size=1.7678pt, color=blue, fill=blue, forget plot] table[row sep=crcr]{%
x	y\\
2.74441968589398	0.0598496971229264\\
};
\addplot[only marks, mark=*, mark options={}, mark size=1.7678pt, color=blue, fill=blue, forget plot] table[row sep=crcr]{%
x	y\\
2.74441968589398	0.496649697122921\\
};
\addplot [color=blue, forget plot]
  table[row sep=crcr]{%
2.74441968589398	0.0598496971229264\\
2.74441968589398	0.496649697122921\\
};
\addplot[only marks, mark=*, mark options={}, mark size=1.7678pt, color=blue, fill=blue, forget plot] table[row sep=crcr]{%
x	y\\
3.5491051599619	0.0628299578371897\\
};
\addplot[only marks, mark=*, mark options={}, mark size=1.7678pt, color=blue, fill=blue, forget plot] table[row sep=crcr]{%
x	y\\
3.5491051599619	0.499629957837183\\
};
\addplot [color=blue, forget plot]
  table[row sep=crcr]{%
3.5491051599619	0.0628299578371897\\
3.5491051599619	0.499629957837183\\
};
\addplot[only marks, mark=*, mark options={}, mark size=1.7678pt, color=blue, fill=blue, forget plot] table[row sep=crcr]{%
x	y\\
4.27430016976514	0.328018505131472\\
};
\addplot[only marks, mark=*, mark options={}, mark size=1.7678pt, color=blue, fill=blue, forget plot] table[row sep=crcr]{%
x	y\\
4.27430016976514	0.764818505131465\\
};
\addplot [color=blue, forget plot]
  table[row sep=crcr]{%
4.27430016976514	0.328018505131472\\
4.27430016976514	0.764818505131465\\
};
\addplot[only marks, mark=*, mark options={}, mark size=1.7678pt, color=blue, fill=blue, forget plot] table[row sep=crcr]{%
x	y\\
4.88826206585726	0.387754506178115\\
};
\addplot[only marks, mark=*, mark options={}, mark size=1.7678pt, color=blue, fill=blue, forget plot] table[row sep=crcr]{%
x	y\\
4.88826206585726	0.824554506178107\\
};
\addplot [color=blue, forget plot]
  table[row sep=crcr]{%
4.88826206585726	0.387754506178115\\
4.88826206585726	0.824554506178107\\
};
\addplot[only marks, mark=*, mark options={}, mark size=1.7678pt, color=blue, fill=blue, forget plot] table[row sep=crcr]{%
x	y\\
5.53080007071276	0.186804425275192\\
};
\addplot[only marks, mark=*, mark options={}, mark size=1.7678pt, color=blue, fill=blue, forget plot] table[row sep=crcr]{%
x	y\\
5.53080007071276	0.623604425275182\\
};
\addplot [color=blue, forget plot]
  table[row sep=crcr]{%
5.53080007071276	0.186804425275192\\
5.53080007071276	0.623604425275182\\
};
\addplot[only marks, mark=*, mark options={}, mark size=1.7678pt, color=blue, fill=blue, forget plot] table[row sep=crcr]{%
x	y\\
6.32676701565019	0.000759262241091715\\
};
\addplot[only marks, mark=*, mark options={}, mark size=1.7678pt, color=blue, fill=blue, forget plot] table[row sep=crcr]{%
x	y\\
6.32676701565019	0.43755926224108\\
};
\addplot [color=blue, forget plot]
  table[row sep=crcr]{%
6.32676701565019	0.000759262241091715\\
6.32676701565019	0.43755926224108\\
};
\end{axis}

\end{tikzpicture}%

%% file: Graphics/tikz/pump_sim_v1_22.tex
%
%
\begin{tikzpicture}

\begin{axis}[%
width=1.8in,
height=0.6in,
scale only axis,
xmin=0,
xmax=6.35504238829741,
xlabel style={font=\color{white!15!black}},
ymin=0,
ymax=1,
ylabel style={font=\color{white!15!black}},
ytick={0, 0.4, 0.8},
axis background/.style={fill=white},
xmajorgrids,
ymajorgrids
]
\addplot [color=blue, dashed, forget plot]
  table[row sep=crcr]{%
0	0\\
0.195288343112986	0.0150620675538003\\
0.385298735521906	0.0565010723636668\\
0.574724603744389	0.118201972057842\\
0.745607681532994	0.184100598460833\\
0.932746829488645	0.258090045929528\\
1.09334354573301	0.315621606286307\\
1.26343884998757	0.366568251748995\\
1.43353415460852	0.407089232885684\\
1.60362945892219	0.437184551283246\\
1.77372476332688	0.45685420746088\\
1.94382006768292	0.466098200327331\\
2.11391537203901	0.464916527422136\\
2.28401067642996	0.453309185155567\\
2.4541059808072	0.431276169233908\\
2.62420128523312	0.398817475136088\\
2.76645729358833	0.363665456353507\\
2.964391894193	0.302623065077734\\
3.13448719860398	0.23888731876696\\
3.30458250321333	0.164725883617881\\
3.52152086937844	0.0550130172123048\\
3.61560328706064	0.0833416984420317\\
3.71250733671503	0.11681367582543\\
3.85814195660251	0.172547393918009\\
3.97370072286841	0.218656790318132\\
4.08348868074488	0.261582032048692\\
4.19145847207598	0.300922879947136\\
4.29787408713109	0.335118272482016\\
4.40184905284847	0.362650104126019\\
4.50565201160222	0.383146104773879\\
4.60912562867593	0.395749807673537\\
4.71211878808324	0.399999968269457\\
4.81534268279986	0.395775171422771\\
4.91836103278689	0.383268830099433\\
5.02239252564472	0.362774867899382\\
5.12698127090905	0.335095452907796\\
5.23215748790624	0.30132389818044\\
5.34072012053425	0.261798603787652\\
5.45156183422991	0.218463797575715\\
5.5649366103147	0.173220884182385\\
5.68537123336206	0.126714199016474\\
5.81008723166482	0.0830454047796916\\
5.93890203283705	0.0455684650734173\\
6.07575459634254	0.0169655607554871\\
6.2143799548546	0.00189068106810703\\
6.35504238829741	0.00206182012810204\\
};
\addplot [color=blue, dashed, forget plot]
  table[row sep=crcr]{%
0	0.4368\\
0.195288343112986	0.4518620675538\\
0.385298735521906	0.493301072363666\\
0.574724603744389	0.555001972057841\\
0.745607681532994	0.620900598460832\\
0.932746829488645	0.694890045929527\\
1.09334354573301	0.752421606286305\\
1.26343884998757	0.803368251748993\\
1.43353415460852	0.843889232885682\\
1.60362945892219	0.873984551283243\\
1.77372476332688	0.893654207460876\\
1.94382006768292	0.902898200327327\\
2.11391537203901	0.901716527422132\\
2.28401067642996	0.890109185155563\\
2.4541059808072	0.868076169233903\\
2.62420128523312	0.835617475136083\\
2.76645729358833	0.800465456353502\\
2.964391894193	0.739423065077729\\
3.13448719860398	0.675687318766954\\
3.30458250321333	0.601525883617875\\
3.52152086937844	0.491813017212299\\
3.61560328706064	0.520141698442025\\
3.71250733671503	0.553613675825424\\
3.85814195660251	0.609347393918002\\
3.97370072286841	0.655456790318126\\
4.08348868074488	0.698382032048686\\
4.19145847207598	0.737722879947129\\
4.29787408713109	0.77191827248201\\
4.40184905284847	0.799450104126012\\
4.50565201160222	0.819946104773872\\
4.60912562867593	0.83254980767353\\
4.71211878808324	0.83679996826945\\
4.81534268279986	0.832575171422763\\
4.91836103278689	0.820068830099426\\
5.02239252564472	0.799574867899374\\
5.12698127090905	0.771895452907787\\
5.23215748790624	0.738123898180432\\
5.34072012053425	0.698598603787643\\
5.45156183422991	0.655263797575706\\
5.5649366103147	0.610020884182376\\
5.68537123336206	0.563514199016464\\
5.81008723166482	0.519845404779682\\
5.93890203283705	0.482368465073407\\
6.07575459634254	0.453765560755477\\
6.2143799548546	0.438690681068097\\
6.35504238829741	0.438861820128092\\
};
\addplot[fill=black, fill opacity=0.2, draw=black, forget plot] table[row sep=crcr]{%
0	0\\
0.195288343112986	0.0150620686699846\\
0.385298735521906	0.0565010758297685\\
0.574724603744389	0.118201975512072\\
0.745607681532994	0.184100601820959\\
0.932746829488645	0.258090048718949\\
1.09334354573301	0.315537169244554\\
1.26343884998757	0.363387574731571\\
1.43353415460852	0.392510850315695\\
1.60362945892219	0.399568949100335\\
1.77372476332688	0.383752887145542\\
1.94382006768292	0.346875467223851\\
2.11391537203901	0.293163499339397\\
2.28401067642996	0.228773332912478\\
2.4541059808072	0.161085229386014\\
2.62420128523312	0.0978574525733746\\
2.76645729358833	0.0536991382367048\\
2.964391894193	0.0124291305629867\\
3.13448719860398	2.0194656359579e-05\\
3.30458250321333	0.010532510978961\\
3.52152086937844	0.0550130178728631\\
3.61560328706064	0.0833417036386168\\
3.71250733671503	0.11681368346461\\
3.85814195660251	0.172547401377376\\
3.97370072286841	0.218656797572174\\
4.08348868074488	0.261582039039943\\
4.19145847207598	0.30092288658887\\
4.29787408713109	0.335118278659382\\
4.40184905284847	0.362650109698523\\
4.50565201160222	0.383146109534216\\
4.60912562867593	0.395749811390529\\
4.71211878808324	0.399999970798449\\
4.81534268279986	0.395775172729868\\
4.91836103278689	0.383268830369334\\
5.02239252564472	0.362774867360123\\
5.12698127090905	0.335095451772836\\
5.23215748790624	0.301323896629412\\
5.34072012053425	0.2617986019425\\
5.45156183422991	0.218463795533101\\
5.5649366103147	0.173220882033998\\
5.68537123336206	0.126714196820896\\
5.81008723166482	0.0830454033052258\\
5.93890203283705	0.0455684665251707\\
6.07575459634254	0.0169655629813796\\
6.2143799548546	0.00189068416545993\\
6.35504238829741	0.00206182367398527\\
}
\closedcycle;
\addplot[only marks, mark=*, mark options={}, mark size=1.7678pt, color=blue, fill=blue, forget plot] table[row sep=crcr]{%
x	y\\
0	0\\
};
\addplot[only marks, mark=*, mark options={}, mark size=1.7678pt, color=blue, fill=blue, forget plot] table[row sep=crcr]{%
x	y\\
0	0.4368\\
};
\addplot [color=blue, forget plot]
  table[row sep=crcr]{%
0	0\\
0	0.4368\\
};
\addplot[only marks, mark=*, mark options={}, mark size=1.7678pt, color=blue, fill=blue, forget plot] table[row sep=crcr]{%
x	y\\
0.932746829488645	0.258090045929528\\
};
\addplot[only marks, mark=*, mark options={}, mark size=1.7678pt, color=blue, fill=blue, forget plot] table[row sep=crcr]{%
x	y\\
0.932746829488645	0.694890045929527\\
};
\addplot [color=blue, forget plot]
  table[row sep=crcr]{%
0.932746829488645	0.258090045929528\\
0.932746829488645	0.694890045929527\\
};
\addplot[only marks, mark=*, mark options={}, mark size=1.7678pt, color=blue, fill=blue, forget plot] table[row sep=crcr]{%
x	y\\
1.77372476332688	0.45685420746088\\
};
\addplot[only marks, mark=*, mark options={}, mark size=1.7678pt, color=blue, fill=blue, forget plot] table[row sep=crcr]{%
x	y\\
1.77372476332688	0.893654207460876\\
};
\addplot [color=blue, forget plot]
  table[row sep=crcr]{%
1.77372476332688	0.45685420746088\\
1.77372476332688	0.893654207460876\\
};
\addplot[only marks, mark=*, mark options={}, mark size=1.7678pt, color=blue, fill=blue, forget plot] table[row sep=crcr]{%
x	y\\
2.62420128523312	0.398817475136088\\
};
\addplot[only marks, mark=*, mark options={}, mark size=1.7678pt, color=blue, fill=blue, forget plot] table[row sep=crcr]{%
x	y\\
2.62420128523312	0.835617475136083\\
};
\addplot [color=blue, forget plot]
  table[row sep=crcr]{%
2.62420128523312	0.398817475136088\\
2.62420128523312	0.835617475136083\\
};
\addplot[only marks, mark=*, mark options={}, mark size=1.7678pt, color=blue, fill=blue, forget plot] table[row sep=crcr]{%
x	y\\
3.52152086937844	0.0550130172123048\\
};
\addplot[only marks, mark=*, mark options={}, mark size=1.7678pt, color=blue, fill=blue, forget plot] table[row sep=crcr]{%
x	y\\
3.52152086937844	0.491813017212299\\
};
\addplot [color=blue, forget plot]
  table[row sep=crcr]{%
3.52152086937844	0.0550130172123048\\
3.52152086937844	0.491813017212299\\
};
\addplot[only marks, mark=*, mark options={}, mark size=1.7678pt, color=blue, fill=blue, forget plot] table[row sep=crcr]{%
x	y\\
4.08348868074488	0.261582032048692\\
};
\addplot[only marks, mark=*, mark options={}, mark size=1.7678pt, color=blue, fill=blue, forget plot] table[row sep=crcr]{%
x	y\\
4.08348868074488	0.698382032048686\\
};
\addplot [color=blue, forget plot]
  table[row sep=crcr]{%
4.08348868074488	0.261582032048692\\
4.08348868074488	0.698382032048686\\
};
\addplot[only marks, mark=*, mark options={}, mark size=1.7678pt, color=blue, fill=blue, forget plot] table[row sep=crcr]{%
x	y\\
4.60912562867593	0.395749807673537\\
};
\addplot[only marks, mark=*, mark options={}, mark size=1.7678pt, color=blue, fill=blue, forget plot] table[row sep=crcr]{%
x	y\\
4.60912562867593	0.83254980767353\\
};
\addplot [color=blue, forget plot]
  table[row sep=crcr]{%
4.60912562867593	0.395749807673537\\
4.60912562867593	0.83254980767353\\
};
\addplot[only marks, mark=*, mark options={}, mark size=1.7678pt, color=blue, fill=blue, forget plot] table[row sep=crcr]{%
x	y\\
5.12698127090905	0.335095452907796\\
};
\addplot[only marks, mark=*, mark options={}, mark size=1.7678pt, color=blue, fill=blue, forget plot] table[row sep=crcr]{%
x	y\\
5.12698127090905	0.771895452907787\\
};
\addplot [color=blue, forget plot]
  table[row sep=crcr]{%
5.12698127090905	0.335095452907796\\
5.12698127090905	0.771895452907787\\
};
\addplot[only marks, mark=*, mark options={}, mark size=1.7678pt, color=blue, fill=blue, forget plot] table[row sep=crcr]{%
x	y\\
5.68537123336206	0.126714199016474\\
};
\addplot[only marks, mark=*, mark options={}, mark size=1.7678pt, color=blue, fill=blue, forget plot] table[row sep=crcr]{%
x	y\\
5.68537123336206	0.563514199016464\\
};
\addplot [color=blue, forget plot]
  table[row sep=crcr]{%
5.68537123336206	0.126714199016474\\
5.68537123336206	0.563514199016464\\
};
\addplot[only marks, mark=*, mark options={}, mark size=1.7678pt, color=blue, fill=blue, forget plot] table[row sep=crcr]{%
x	y\\
6.35504238829741	0.00206182012810204\\
};
\addplot[only marks, mark=*, mark options={}, mark size=1.7678pt, color=blue, fill=blue, forget plot] table[row sep=crcr]{%
x	y\\
6.35504238829741	0.438861820128092\\
};
\addplot [color=blue, forget plot]
  table[row sep=crcr]{%
6.35504238829741	0.00206182012810204\\
6.35504238829741	0.438861820128092\\
};
\end{axis}

\end{tikzpicture}%

%% file: Graphics/tikz/pump_opt_v1_10.tex
%
%
\begin{tikzpicture}

\begin{axis}[%
width=1.8in,
height=0.6in,
scale only axis,
xmin=0,
xmax=6.28318530717959,
xlabel style={font=\color{white!15!black}},
xlabel={$y$\,[m]},
ymin=0,
ymax=1,
ylabel style={font=\color{white!15!black}},
ylabel={$z$\,[m]},
ytick={0, 0.4, 0.8},
axis background/.style={fill=white},
xmajorgrids,
ymajorgrids
]
\addplot [color=blue, dashed, forget plot]
  table[row sep=crcr]{%
-9.62706427323694e-24	-6.99303746461171e-19\\
0.000127005798210759	2.60865811763621e-09\\
0.221052368176468	0.0192293622421004\\
0.425922693207235	0.0682808738692027\\
0.443976991631641	0.0737998761302241\\
0.681136057972706	0.158596739724414\\
0.801413459700505	0.206405025625742\\
0.91680425782384	0.251959446249862\\
1.03077186892922	0.294257054964382\\
1.14402282915273	0.3314628391626\\
1.14402282921332	0.331462839180865\\
1.32295497247782	0.375928833611403\\
1.39729246150958	0.388078903593729\\
1.4571638697038	0.394857251602008\\
1.53782824997031	0.399565392901568\\
1.60621181492927	0.399498499812259\\
1.67659710115573	0.395539153204627\\
1.7468222453302	0.387733426604861\\
1.8224334766836	0.375201609685627\\
1.90218715812968	0.357656734180174\\
1.98589888915621	0.334944872241789\\
2.07655182077291	0.306116109838106\\
2.17656630053017	0.27031563487173\\
2.28772805616167	0.227301066916627\\
2.40936893670183	0.178770292315131\\
2.54448519176729	0.126451284856773\\
2.69591688589616	0.0743277585438199\\
2.87187349721876	0.0284005352755322\\
3.06760869805847	0.00218546026358587\\
3.26296173614243	0.00586330929868684\\
3.45833445016167	0.0388059491448212\\
3.458334450343	0.0388059491877568\\
3.82246893234831	0.158495088397274\\
3.99948389144615	0.228895751875784\\
4.25538066298878	0.322113929241737\\
4.25802213828148	0.322949190532319\\
4.51355260545592	0.384392952026463\\
4.68095313457302	0.399604840594836\\
4.84713080049676	0.392781694105929\\
5.01316766852557	0.364891045022813\\
5.18364974274483	0.317549969028496\\
5.36120115714847	0.253957423499666\\
5.5663049060741	0.17267858780299\\
5.80154518081607	0.0858340458294689\\
6.03859483416988	0.0234563902860091\\
6.28318530717959	-3.18965773461219e-09\\
};
\addplot [color=blue, dashed, forget plot]
  table[row sep=crcr]{%
-9.62706427323694e-24	0.4368\\
0.000127005798210759	0.436800019805693\\
0.221052368176468	0.510887766338907\\
0.425922693207235	0.640780580765721\\
0.443976991631641	0.650779139953312\\
0.681136057972706	0.749844090467327\\
0.801413459700505	0.77393154503637\\
0.91680425782384	0.773313962571367\\
1.03077186892922	0.743602470149789\\
1.14402282915273	0.682803840158831\\
1.14402282921332	0.682803840117614\\
1.32295497247782	0.658690149919602\\
1.39729246150958	0.67025012561462\\
1.4571638697038	0.673117771358511\\
1.53782824997031	0.678548113034235\\
1.60621181492927	0.678022804020969\\
1.67659710115573	0.673855366610659\\
1.7468222453302	0.669148706346315\\
1.8224334766836	0.665273260816744\\
1.90218715812968	0.65325458839885\\
1.98589888915621	0.625388499182931\\
2.07655182077291	0.589550015270222\\
2.17656630053017	0.54894308943946\\
2.28772805616167	0.505331066924898\\
2.40936893670183	0.45680029235145\\
2.54448519176729	0.404481284895253\\
2.69591688589616	0.362161448639761\\
2.87187349721876	0.388873644393598\\
3.06760869805847	0.459976097234283\\
3.26296173614243	0.534409683960655\\
3.45833445016167	0.613549475340079\\
3.458334450343	0.613549475414323\\
3.82246893234831	0.74798982729155\\
3.99948389144615	0.786905655765131\\
4.25538066298878	0.792058309023211\\
4.25802213828148	0.791748120020262\\
4.51355260545592	0.730039076910705\\
4.68095313457302	0.685148020248985\\
4.84713080049676	0.6708162706954\\
5.01316766852557	0.642925210124594\\
5.18364974274483	0.595580077147399\\
5.36120115714847	0.542010556488013\\
5.5663049060741	0.53732186962708\\
5.80154518081607	0.555397413922245\\
6.03859483416988	0.568664064439404\\
6.28318530717959	0.595589996804763\\
};
\addplot[fill=black, fill opacity=0.2, draw=black, forget plot] table[row sep=crcr]{%
-9.62706427323694e-24	0\\
0.000127005798210759	6.45218906081269e-09\\
0.221052368176468	0.0192293660760419\\
0.425922693207235	0.0682808759502106\\
0.443976991631641	0.0737998780302949\\
0.681136057972706	0.158596738810149\\
0.801413459700505	0.206405023175068\\
0.91680425782384	0.251959440577199\\
1.03077186892922	0.294257044399745\\
1.14402282915273	0.331462842348175\\
1.14402282921332	0.33146284236644\\
1.32295497247782	0.375928839315761\\
1.39729246150958	0.388078909705281\\
1.4571638697038	0.394857258094248\\
1.53782824997031	0.399565399853205\\
1.60621181492927	0.399498506999689\\
1.67659710115573	0.395539160375352\\
1.7468222453302	0.387733433102562\\
1.8224334766836	0.375201615493347\\
1.90218715812968	0.357656738619786\\
1.98589888915621	0.334944875581285\\
2.07655182077291	0.306116112400372\\
2.17656630053017	0.270315636742406\\
2.28772805616167	0.227301067987322\\
2.40936893670183	0.178770292570312\\
2.54448519176729	0.126451284367331\\
2.69591688589616	0.0743277575857583\\
2.87187349721876	0.0284005340632723\\
3.06760869805847	0.00218545844097701\\
3.26296173614243	0.00586330693922071\\
3.45833445016167	0.0388059462096812\\
3.458334450343	0.0388059462526169\\
3.82246893234831	0.158495083704943\\
3.99948389144615	0.228895744793808\\
4.25538066298878	0.322113931788836\\
4.25802213828148	0.322949003637719\\
4.51355260545592	0.384392954560488\\
4.68095313457302	0.399604845230587\\
4.84713080049676	0.392781699452153\\
5.01316766852557	0.364891050715299\\
5.18364974274483	0.317549974902979\\
5.36120115714847	0.253957429046984\\
5.5663049060741	0.172678593038715\\
5.80154518081607	0.0858340500071622\\
6.03859483416988	0.023456393876427\\
6.28318530717959	0\\
}
\closedcycle;
\addplot[only marks, mark=*, mark options={}, mark size=1.7678pt, color=blue, fill=blue, forget plot] table[row sep=crcr]{%
x	y\\
-9.62706427323694e-24	-6.99303746461171e-19\\
};
\addplot[only marks, mark=*, mark options={}, mark size=1.7678pt, color=blue, fill=blue, forget plot] table[row sep=crcr]{%
x	y\\
-9.62706427323694e-24	0.4368\\
};
\addplot [color=blue, forget plot]
  table[row sep=crcr]{%
-9.62706427323694e-24	-6.99303746461171e-19\\
-9.62706427323694e-24	0.4368\\
};
\addplot[only marks, mark=*, mark options={}, mark size=1.7678pt, color=blue, fill=blue, forget plot] table[row sep=crcr]{%
x	y\\
0.681136057972706	0.158596739724414\\
};
\addplot[only marks, mark=*, mark options={}, mark size=1.7678pt, color=blue, fill=blue, forget plot] table[row sep=crcr]{%
x	y\\
0.681136057972706	0.749844090467327\\
};
\addplot [color=blue, forget plot]
  table[row sep=crcr]{%
0.681136057972706	0.158596739724414\\
0.681136057972706	0.749844090467327\\
};
\addplot[only marks, mark=*, mark options={}, mark size=1.7678pt, color=blue, fill=blue, forget plot] table[row sep=crcr]{%
x	y\\
1.14402282921332	0.331462839180865\\
};
\addplot[only marks, mark=*, mark options={}, mark size=1.7678pt, color=blue, fill=blue, forget plot] table[row sep=crcr]{%
x	y\\
1.14402282921332	0.682803840117614\\
};
\addplot [color=blue, forget plot]
  table[row sep=crcr]{%
1.14402282921332	0.331462839180865\\
1.14402282921332	0.682803840117614\\
};
\addplot[only marks, mark=*, mark options={}, mark size=1.7678pt, color=blue, fill=blue, forget plot] table[row sep=crcr]{%
x	y\\
1.60621181492927	0.399498499812259\\
};
\addplot[only marks, mark=*, mark options={}, mark size=1.7678pt, color=blue, fill=blue, forget plot] table[row sep=crcr]{%
x	y\\
1.60621181492927	0.678022804020969\\
};
\addplot [color=blue, forget plot]
  table[row sep=crcr]{%
1.60621181492927	0.399498499812259\\
1.60621181492927	0.678022804020969\\
};
\addplot[only marks, mark=*, mark options={}, mark size=1.7678pt, color=blue, fill=blue, forget plot] table[row sep=crcr]{%
x	y\\
1.98589888915621	0.334944872241789\\
};
\addplot[only marks, mark=*, mark options={}, mark size=1.7678pt, color=blue, fill=blue, forget plot] table[row sep=crcr]{%
x	y\\
1.98589888915621	0.625388499182931\\
};
\addplot [color=blue, forget plot]
  table[row sep=crcr]{%
1.98589888915621	0.334944872241789\\
1.98589888915621	0.625388499182931\\
};
\addplot[only marks, mark=*, mark options={}, mark size=1.7678pt, color=blue, fill=blue, forget plot] table[row sep=crcr]{%
x	y\\
2.54448519176729	0.126451284856773\\
};
\addplot[only marks, mark=*, mark options={}, mark size=1.7678pt, color=blue, fill=blue, forget plot] table[row sep=crcr]{%
x	y\\
2.54448519176729	0.404481284895253\\
};
\addplot [color=blue, forget plot]
  table[row sep=crcr]{%
2.54448519176729	0.126451284856773\\
2.54448519176729	0.404481284895253\\
};
\addplot[only marks, mark=*, mark options={}, mark size=1.7678pt, color=blue, fill=blue, forget plot] table[row sep=crcr]{%
x	y\\
3.45833445016167	0.0388059491448212\\
};
\addplot[only marks, mark=*, mark options={}, mark size=1.7678pt, color=blue, fill=blue, forget plot] table[row sep=crcr]{%
x	y\\
3.45833445016167	0.613549475340079\\
};
\addplot [color=blue, forget plot]
  table[row sep=crcr]{%
3.45833445016167	0.0388059491448212\\
3.45833445016167	0.613549475340079\\
};
\addplot[only marks, mark=*, mark options={}, mark size=1.7678pt, color=blue, fill=blue, forget plot] table[row sep=crcr]{%
x	y\\
4.25802213828148	0.322949190532319\\
};
\addplot[only marks, mark=*, mark options={}, mark size=1.7678pt, color=blue, fill=blue, forget plot] table[row sep=crcr]{%
x	y\\
4.25802213828148	0.791748120020262\\
};
\addplot [color=blue, forget plot]
  table[row sep=crcr]{%
4.25802213828148	0.322949190532319\\
4.25802213828148	0.791748120020262\\
};
\addplot[only marks, mark=*, mark options={}, mark size=1.7678pt, color=blue, fill=blue, forget plot] table[row sep=crcr]{%
x	y\\
5.18364974274483	0.317549969028496\\
};
\addplot[only marks, mark=*, mark options={}, mark size=1.7678pt, color=blue, fill=blue, forget plot] table[row sep=crcr]{%
x	y\\
5.18364974274483	0.595580077147399\\
};
\addplot [color=blue, forget plot]
  table[row sep=crcr]{%
5.18364974274483	0.317549969028496\\
5.18364974274483	0.595580077147399\\
};
\addplot[only marks, mark=*, mark options={}, mark size=1.7678pt, color=blue, fill=blue, forget plot] table[row sep=crcr]{%
x	y\\
6.28318530717959	-3.18965773461219e-09\\
};
\addplot[only marks, mark=*, mark options={}, mark size=1.7678pt, color=blue, fill=blue, forget plot] table[row sep=crcr]{%
x	y\\
6.28318530717959	0.595589996804763\\
};
\addplot [color=blue, forget plot]
  table[row sep=crcr]{%
6.28318530717959	-3.18965773461219e-09\\
6.28318530717959	0.595589996804763\\
};
\end{axis}

\end{tikzpicture}%

%% file: Graphics/tikz/pump_opt_v1_15.tex
%
%
\begin{tikzpicture}

\begin{axis}[%
width=1.8in,
height=0.6in,
scale only axis,
xmin=0,
xmax=6.28318530717959,
xlabel style={font=\color{white!15!black}},
xlabel={$y$\,[m]},
ymin=0,
ymax=1,
ylabel style={font=\color{white!15!black}},
ytick={0, 0.4, 0.8},
axis background/.style={fill=white},
xmajorgrids,
ymajorgrids
]
\addplot [color=blue, dashed, forget plot]
  table[row sep=crcr]{%
-4.13656700280411e-16	1.29651474913035e-10\\
0.1295360599254	0.00667438025926806\\
0.255194487578515	0.0254890907446892\\
0.379267204001909	0.0548310201389143\\
0.498020084791263	0.0912739741949195\\
0.615187796567442	0.133223261605183\\
0.731356102175702	0.178425244598899\\
0.843706911633274	0.223270677080602\\
0.915347153554697	0.251396406186148\\
1.07049485676855	0.308952597509686\\
1.18333692226776	0.346139598396879\\
1.29617911254225	0.377007131246742\\
1.40902153143894	0.398629312643673\\
1.52186355376788	0.410077287282354\\
1.63470580145705	0.411315084697873\\
1.7475482061074	0.404772998211153\\
1.86039029541048	0.387305564168503\\
1.97323260119216	0.355642888857079\\
2.08607488080358	0.311469542870126\\
2.14370825436581	0.282454859662213\\
2.33523391984724	0.208377491692831\\
2.47624426835671	0.152435853316317\\
2.63128158306504	0.0954285751368256\\
2.7864731389126	0.0483545094070272\\
2.94393537116162	0.0154206244139603\\
3.10808567606907	0.00044463763306166\\
3.27002642302896	0.00655761249943636\\
3.43347251307455	0.0331164531147514\\
3.58792712821032	0.0745285539579322\\
3.74086043664204	0.127251263311086\\
3.89427949828709	0.186920522667008\\
4.04226137895964	0.245700875363943\\
4.19103203204174	0.301617710172067\\
4.33936340969731	0.348946908201375\\
4.48755332733755	0.383255206409967\\
4.63602648734463	0.400790878144521\\
4.78435714585934	0.400981706339015\\
4.93268913714205	0.3880028756439\\
5.08181876814474	0.364611973191706\\
5.22935088202991	0.323487455743951\\
5.37768258941214	0.259764379401392\\
5.57072996976627	0.1709262582702\\
5.70862827074654	0.118140819851888\\
5.98987553309435	0.0334366686478824\\
5.98987553353437	0.0334366685174356\\
6.28318530717959	9.98085304063989e-10\\
};
\addplot [color=blue, dashed, forget plot]
  table[row sep=crcr]{%
-4.13656700280411e-16	0.436800000129651\\
0.1295360599254	0.457028983885476\\
0.255194487578515	0.512186402988117\\
0.379267204001909	0.574297634722636\\
0.498020084791263	0.634450556879507\\
0.615187796567442	0.691868868148358\\
0.731356102175702	0.744029398787887\\
0.843706911633274	0.787221160142054\\
0.915347153554697	0.809881627075707\\
1.07049485676855	0.843792755506033\\
1.18333692226776	0.855599756495217\\
1.29617911254225	0.857055535764526\\
1.40902153143894	0.848745222644133\\
1.52186355376788	0.830854636340089\\
1.63470580145705	0.80339091146031\\
1.7475482061074	0.765867946804639\\
1.86039029541048	0.718915022019296\\
1.97323260119216	0.663185893228704\\
2.08607488080358	0.598343750244998\\
2.14370825436581	0.562037166051766\\
2.33523391984724	0.496698870159565\\
2.47624426835671	0.479608803148764\\
2.63128158306504	0.473367614204237\\
2.7864731389126	0.467168097148157\\
2.94393537116162	0.466257733742117\\
3.10808567606907	0.47548970749349\\
3.27002642302896	0.49700434144152\\
3.43347251307455	0.530716186138171\\
3.58792712821032	0.570744547043696\\
3.74086043664204	0.613634079013147\\
3.89427949828709	0.65438134357811\\
4.04226137895964	0.686119193280209\\
4.19103203204174	0.707267019736415\\
4.33936340969731	0.718404542383216\\
4.48755332733755	0.720528039845521\\
4.63602648734463	0.714396955776181\\
4.78435714585934	0.700116727324748\\
4.93268913714205	0.676855093905368\\
5.08181876814474	0.643827710537851\\
5.22935088202991	0.603198420901959\\
5.37768258941214	0.555240150733738\\
5.57072996976627	0.514051602625003\\
5.70862827074654	0.514154071113241\\
5.98987553309435	0.53885098617238\\
5.98987553353437	0.538850986197228\\
6.28318530717959	0.595590000986308\\
};
\addplot[fill=black, fill opacity=0.2, draw=black, forget plot] table[row sep=crcr]{%
-4.13656700280411e-16	0\\
0.1295360599254	0.00667437959423139\\
0.255194487578515	0.0254890894389152\\
0.379267204001909	0.0548310180647608\\
0.498020084791263	0.0912739712945458\\
0.615187796567442	0.133223257753275\\
0.731356102175702	0.178425239530681\\
0.843706911633274	0.22327067002921\\
0.915347153554697	0.251396391950567\\
1.07049485676855	0.30795897022582\\
1.18333692226776	0.342895568462209\\
1.29617911254225	0.370584885326272\\
1.40902153143894	0.389622571580216\\
1.52186355376788	0.399042997676121\\
1.63470580145705	0.39836845474097\\
1.7475482061074	0.387633103627795\\
1.86039029541048	0.367381480780167\\
1.97323260119216	0.338640620025322\\
2.08607488080358	0.302868231392502\\
2.14370825436581	0.282459140439379\\
2.33523391984724	0.208381772638496\\
2.47624426835671	0.152440134381656\\
2.63128158306504	0.0954328562760204\\
2.7864731389126	0.0483587906155948\\
2.94393537116162	0.0154249056866513\\
3.10808567606907	0.000448918976229007\\
3.27002642302896	0.00656189392099157\\
3.43347251307455	0.0331207346235878\\
3.58792712821032	0.0745328355511492\\
3.74086043664204	0.127255544986373\\
3.89427949828709	0.186924804417162\\
4.04226137895964	0.245700872866484\\
4.19103203204174	0.30077558398342\\
4.33936340969731	0.346874970438616\\
4.48755332733755	0.38011800484567\\
4.63602648734463	0.397672038105468\\
4.78435714585934	0.39793180764191\\
4.93268913714205	0.380899160598492\\
5.08181876814474	0.347847410557917\\
5.22935088202991	0.302290203575389\\
5.37768258941214	0.247581147297748\\
5.57072996976627	0.170926253843866\\
5.70862827074654	0.118140815436844\\
5.98987553309435	0.0334366641215993\\
5.98987553353437	0.0334366640241694\\
6.28318530717959	0\\
}
\closedcycle;
\addplot[only marks, mark=*, mark options={}, mark size=1.7678pt, color=blue, fill=blue, forget plot] table[row sep=crcr]{%
x	y\\
-4.13656700280411e-16	1.29651474913035e-10\\
};
\addplot[only marks, mark=*, mark options={}, mark size=1.7678pt, color=blue, fill=blue, forget plot] table[row sep=crcr]{%
x	y\\
-4.13656700280411e-16	0.436800000129651\\
};
\addplot [color=blue, forget plot]
  table[row sep=crcr]{%
-4.13656700280411e-16	1.29651474913035e-10\\
-4.13656700280411e-16	0.436800000129651\\
};
\addplot[only marks, mark=*, mark options={}, mark size=1.7678pt, color=blue, fill=blue, forget plot] table[row sep=crcr]{%
x	y\\
0.615187796567442	0.133223261605183\\
};
\addplot[only marks, mark=*, mark options={}, mark size=1.7678pt, color=blue, fill=blue, forget plot] table[row sep=crcr]{%
x	y\\
0.615187796567442	0.691868868148358\\
};
\addplot [color=blue, forget plot]
  table[row sep=crcr]{%
0.615187796567442	0.133223261605183\\
0.615187796567442	0.691868868148358\\
};
\addplot[only marks, mark=*, mark options={}, mark size=1.7678pt, color=blue, fill=blue, forget plot] table[row sep=crcr]{%
x	y\\
1.18333692226776	0.346139598396879\\
};
\addplot[only marks, mark=*, mark options={}, mark size=1.7678pt, color=blue, fill=blue, forget plot] table[row sep=crcr]{%
x	y\\
1.18333692226776	0.855599756495217\\
};
\addplot [color=blue, forget plot]
  table[row sep=crcr]{%
1.18333692226776	0.346139598396879\\
1.18333692226776	0.855599756495217\\
};
\addplot[only marks, mark=*, mark options={}, mark size=1.7678pt, color=blue, fill=blue, forget plot] table[row sep=crcr]{%
x	y\\
1.7475482061074	0.404772998211153\\
};
\addplot[only marks, mark=*, mark options={}, mark size=1.7678pt, color=blue, fill=blue, forget plot] table[row sep=crcr]{%
x	y\\
1.7475482061074	0.765867946804639\\
};
\addplot [color=blue, forget plot]
  table[row sep=crcr]{%
1.7475482061074	0.404772998211153\\
1.7475482061074	0.765867946804639\\
};
\addplot[only marks, mark=*, mark options={}, mark size=1.7678pt, color=blue, fill=blue, forget plot] table[row sep=crcr]{%
x	y\\
2.33523391984724	0.208377491692831\\
};
\addplot[only marks, mark=*, mark options={}, mark size=1.7678pt, color=blue, fill=blue, forget plot] table[row sep=crcr]{%
x	y\\
2.33523391984724	0.496698870159565\\
};
\addplot [color=blue, forget plot]
  table[row sep=crcr]{%
2.33523391984724	0.208377491692831\\
2.33523391984724	0.496698870159565\\
};
\addplot[only marks, mark=*, mark options={}, mark size=1.7678pt, color=blue, fill=blue, forget plot] table[row sep=crcr]{%
x	y\\
3.10808567606907	0.00044463763306166\\
};
\addplot[only marks, mark=*, mark options={}, mark size=1.7678pt, color=blue, fill=blue, forget plot] table[row sep=crcr]{%
x	y\\
3.10808567606907	0.47548970749349\\
};
\addplot [color=blue, forget plot]
  table[row sep=crcr]{%
3.10808567606907	0.00044463763306166\\
3.10808567606907	0.47548970749349\\
};
\addplot[only marks, mark=*, mark options={}, mark size=1.7678pt, color=blue, fill=blue, forget plot] table[row sep=crcr]{%
x	y\\
3.89427949828709	0.186920522667008\\
};
\addplot[only marks, mark=*, mark options={}, mark size=1.7678pt, color=blue, fill=blue, forget plot] table[row sep=crcr]{%
x	y\\
3.89427949828709	0.65438134357811\\
};
\addplot [color=blue, forget plot]
  table[row sep=crcr]{%
3.89427949828709	0.186920522667008\\
3.89427949828709	0.65438134357811\\
};
\addplot[only marks, mark=*, mark options={}, mark size=1.7678pt, color=blue, fill=blue, forget plot] table[row sep=crcr]{%
x	y\\
4.63602648734463	0.400790878144521\\
};
\addplot[only marks, mark=*, mark options={}, mark size=1.7678pt, color=blue, fill=blue, forget plot] table[row sep=crcr]{%
x	y\\
4.63602648734463	0.714396955776181\\
};
\addplot [color=blue, forget plot]
  table[row sep=crcr]{%
4.63602648734463	0.400790878144521\\
4.63602648734463	0.714396955776181\\
};
\addplot[only marks, mark=*, mark options={}, mark size=1.7678pt, color=blue, fill=blue, forget plot] table[row sep=crcr]{%
x	y\\
5.37768258941214	0.259764379401392\\
};
\addplot[only marks, mark=*, mark options={}, mark size=1.7678pt, color=blue, fill=blue, forget plot] table[row sep=crcr]{%
x	y\\
5.37768258941214	0.555240150733738\\
};
\addplot [color=blue, forget plot]
  table[row sep=crcr]{%
5.37768258941214	0.259764379401392\\
5.37768258941214	0.555240150733738\\
};
\addplot[only marks, mark=*, mark options={}, mark size=1.7678pt, color=blue, fill=blue, forget plot] table[row sep=crcr]{%
x	y\\
6.28318530717959	9.98085304063989e-10\\
};
\addplot[only marks, mark=*, mark options={}, mark size=1.7678pt, color=blue, fill=blue, forget plot] table[row sep=crcr]{%
x	y\\
6.28318530717959	0.595590000986308\\
};
\addplot [color=blue, forget plot]
  table[row sep=crcr]{%
6.28318530717959	9.98085304063989e-10\\
6.28318530717959	0.595590000986308\\
};
\end{axis}

\end{tikzpicture}%

%% file: Graphics/tikz/pump_opt_v1_22.tex
%
%
\begin{tikzpicture}

\begin{axis}[%
width=1.8in,
height=0.6in,
scale only axis,
xmin=0,
xmax=6.28318530717959,
xlabel style={font=\color{white!15!black}},
xlabel={$y$\,[m]},
ymin=0,
ymax=1,
ylabel style={font=\color{white!15!black}},
ytick={0, 0.4, 0.8},
axis background/.style={fill=white},
xmajorgrids,
ymajorgrids
]
\addplot [color=blue, dashed, forget plot]
  table[row sep=crcr]{%
-3.40960555917643e-24	7.82223249237631e-18\\
0.14790502696887	0.00868673744950219\\
0.29322844358505	0.0334186584591152\\
0.439486864546021	0.072411641403486\\
0.575575352858523	0.118512631118087\\
0.74469681193413	0.183737434904287\\
0.852202969946719	0.226642537278387\\
0.852389380623489	0.226716439211999\\
1.05442952165095	0.30249473292875\\
1.25570984449089	0.361585171845931\\
1.3657650508701	0.38360069213365\\
1.60627627897235	0.401921701600646\\
1.65792776005541	0.39949824238041\\
1.79200204324915	0.383555101645089\\
1.92607463706476	0.355062249285597\\
2.06014567551715	0.318455299372845\\
2.19421950585331	0.277382610044\\
2.3013398353269	0.242698614370679\\
2.46236359099118	0.171668319641952\\
2.55659024358025	0.12197055573715\\
2.67108611615174	0.0822060824171294\\
2.89255021597901	0.0243001813399393\\
3.04574919960291	0.00366315364661961\\
3.19674140520834	0.0012153248763085\\
3.35011694659997	0.017142317527459\\
3.49644763022879	0.0482898226403388\\
3.59950287869394	0.0781719815390826\\
3.78786530362233	0.145065130874152\\
3.99771614407695	0.228195883359311\\
3.99771614903065	0.228195885321002\\
4.20573950740395	0.305812833498424\\
4.344235897232	0.351822924669413\\
4.48273228192262	0.395517146879996\\
4.62122866726907	0.435001507818688\\
4.75972505372952	0.46625041397845\\
4.89822144053597	0.485941316871944\\
5.03671782713811	0.496398495332977\\
5.17521421396178	0.500419581014063\\
5.31371060059481	0.50057698588906\\
5.45220698707121	0.497843671872267\\
5.59070337385189	0.492610308626597\\
5.72919976041735	0.485131054660661\\
5.86769614708861	0.475559804829019\\
6.00619253341664	0.463969666936754\\
6.14468890592679	0.450257540886381\\
6.28318530717959	0.43352006007573\\
};
\addplot [color=blue, dashed, forget plot]
  table[row sep=crcr]{%
-3.40960555917643e-24	0.4368\\
0.14790502696887	0.444471742658281\\
0.29322844358505	0.465806368577858\\
0.439486864546021	0.496400094394092\\
0.575575352858523	0.529459697829158\\
0.74469681193413	0.571051677159922\\
0.852202969946719	0.594615118292174\\
0.852389380623489	0.594652627371278\\
1.05442952165095	0.631295957880416\\
1.25570984449089	0.6583833374274\\
1.3657650508701	0.668859236791396\\
1.60627627897235	0.6811612885895\\
1.65792776005541	0.682112778713297\\
1.79200204324915	0.681625394387015\\
1.92607463706476	0.676591321891074\\
2.06014567551715	0.666123683292046\\
2.19421950585331	0.649492131764659\\
2.3013398353269	0.631506382616186\\
2.46236359099118	0.599773805207894\\
2.55659024358025	0.578110299843509\\
2.67108611615174	0.567041595844307\\
2.89255021597901	0.554185579833417\\
3.04574919960291	0.557280905183346\\
3.19674140520834	0.572901339222989\\
3.35011694659997	0.602087057814601\\
3.49644763022879	0.641086212227946\\
3.59950287869394	0.673545602798911\\
3.78786530362233	0.738348641799556\\
3.99771614407695	0.80812856586935\\
3.99771614903065	0.808128567376547\\
4.20573950740395	0.867864165117056\\
4.344235897232	0.900498191173505\\
4.48273228192262	0.926538752467007\\
4.62122866726907	0.946364649986389\\
4.75972505372952	0.960781002803606\\
4.89822144053597	0.97045232046591\\
5.03671782713811	0.97491374685388\\
5.17521421396178	0.973605755021519\\
5.31371060059481	0.966013861652572\\
5.45220698707121	0.951943473255797\\
5.59070337385189	0.931316454579596\\
5.72919976041735	0.904081972707283\\
5.86769614708861	0.870209247388298\\
6.00619253341664	0.829683656136678\\
6.14468890592679	0.78252582341186\\
6.28318530717959	0.728916405563935\\
};
\addplot[fill=black, fill opacity=0.2, draw=black, forget plot] table[row sep=crcr]{%
-3.40960555917643e-24	0\\
0.14790502696887	0.00868673730688177\\
0.29322844358505	0.0334186581618101\\
0.439486864546021	0.0724116409170344\\
0.575575352858523	0.118512630402466\\
0.74469681193413	0.183737433687544\\
0.852202969946719	0.226642489025273\\
0.852389380623489	0.226716386889054\\
1.05442952165095	0.302494680605814\\
1.25570984449089	0.361585119006752\\
1.3657650508701	0.383419177454334\\
1.60627627897235	0.399496680447992\\
1.65792776005541	0.396970922459432\\
1.79200204324915	0.380744381840844\\
1.92607463706476	0.351599786892708\\
2.06014567551715	0.311620560636816\\
2.19421950585331	0.263662709413082\\
2.3013398353269	0.221897872567492\\
2.46236359099118	0.157850767726924\\
2.55659024358025	0.121970550746352\\
2.67108611615174	0.0822060788614557\\
2.89255021597901	0.0243001776596114\\
3.04574919960291	0.00366314990630173\\
3.19674140520834	0.00121532108586064\\
3.35011694659997	0.0171423136836761\\
3.49644763022879	0.0482898187325907\\
3.59950287869394	0.0781719775741684\\
3.78786530362233	0.145065126704957\\
3.99771614407695	0.228195885657357\\
3.99771614903065	0.228195887619048\\
4.20573950740395	0.305812835963686\\
4.344235897232	0.348190844499241\\
4.48273228192262	0.379271420485424\\
4.62122866726907	0.396685116657645\\
4.75972505372952	0.399104387697975\\
4.89822144053597	0.386344798564642\\
5.03671782713811	0.35937908535212\\
5.17521421396178	0.320262997704654\\
5.31371060059481	0.271978577622963\\
5.45220698707121	0.218206821151877\\
5.59070337385189	0.163047055010936\\
5.72919976041735	0.110704422171458\\
5.86769614708861	0.0651692991864188\\
6.00619253341664	0.0299130876418716\\
6.14468890592679	0.00762357049010765\\
6.28318530717959	0\\
}
\closedcycle;
\addplot[only marks, mark=*, mark options={}, mark size=1.7678pt, color=blue, fill=blue, forget plot] table[row sep=crcr]{%
x	y\\
-3.40960555917643e-24	7.82223249237631e-18\\
};
\addplot[only marks, mark=*, mark options={}, mark size=1.7678pt, color=blue, fill=blue, forget plot] table[row sep=crcr]{%
x	y\\
-3.40960555917643e-24	0.4368\\
};
\addplot [color=blue, forget plot]
  table[row sep=crcr]{%
-3.40960555917643e-24	7.82223249237631e-18\\
-3.40960555917643e-24	0.4368\\
};
\addplot[only marks, mark=*, mark options={}, mark size=1.7678pt, color=blue, fill=blue, forget plot] table[row sep=crcr]{%
x	y\\
0.74469681193413	0.183737434904287\\
};
\addplot[only marks, mark=*, mark options={}, mark size=1.7678pt, color=blue, fill=blue, forget plot] table[row sep=crcr]{%
x	y\\
0.74469681193413	0.571051677159922\\
};
\addplot [color=blue, forget plot]
  table[row sep=crcr]{%
0.74469681193413	0.183737434904287\\
0.74469681193413	0.571051677159922\\
};
\addplot[only marks, mark=*, mark options={}, mark size=1.7678pt, color=blue, fill=blue, forget plot] table[row sep=crcr]{%
x	y\\
1.3657650508701	0.38360069213365\\
};
\addplot[only marks, mark=*, mark options={}, mark size=1.7678pt, color=blue, fill=blue, forget plot] table[row sep=crcr]{%
x	y\\
1.3657650508701	0.668859236791396\\
};
\addplot [color=blue, forget plot]
  table[row sep=crcr]{%
1.3657650508701	0.38360069213365\\
1.3657650508701	0.668859236791396\\
};
\addplot[only marks, mark=*, mark options={}, mark size=1.7678pt, color=blue, fill=blue, forget plot] table[row sep=crcr]{%
x	y\\
2.06014567551715	0.318455299372845\\
};
\addplot[only marks, mark=*, mark options={}, mark size=1.7678pt, color=blue, fill=blue, forget plot] table[row sep=crcr]{%
x	y\\
2.06014567551715	0.666123683292046\\
};
\addplot [color=blue, forget plot]
  table[row sep=crcr]{%
2.06014567551715	0.318455299372845\\
2.06014567551715	0.666123683292046\\
};
\addplot[only marks, mark=*, mark options={}, mark size=1.7678pt, color=blue, fill=blue, forget plot] table[row sep=crcr]{%
x	y\\
2.67108611615174	0.0822060824171294\\
};
\addplot[only marks, mark=*, mark options={}, mark size=1.7678pt, color=blue, fill=blue, forget plot] table[row sep=crcr]{%
x	y\\
2.67108611615174	0.567041595844307\\
};
\addplot [color=blue, forget plot]
  table[row sep=crcr]{%
2.67108611615174	0.0822060824171294\\
2.67108611615174	0.567041595844307\\
};
\addplot[only marks, mark=*, mark options={}, mark size=1.7678pt, color=blue, fill=blue, forget plot] table[row sep=crcr]{%
x	y\\
3.49644763022879	0.0482898226403388\\
};
\addplot[only marks, mark=*, mark options={}, mark size=1.7678pt, color=blue, fill=blue, forget plot] table[row sep=crcr]{%
x	y\\
3.49644763022879	0.641086212227946\\
};
\addplot [color=blue, forget plot]
  table[row sep=crcr]{%
3.49644763022879	0.0482898226403388\\
3.49644763022879	0.641086212227946\\
};
\addplot[only marks, mark=*, mark options={}, mark size=1.7678pt, color=blue, fill=blue, forget plot] table[row sep=crcr]{%
x	y\\
4.20573950740395	0.305812833498424\\
};
\addplot[only marks, mark=*, mark options={}, mark size=1.7678pt, color=blue, fill=blue, forget plot] table[row sep=crcr]{%
x	y\\
4.20573950740395	0.867864165117056\\
};
\addplot [color=blue, forget plot]
  table[row sep=crcr]{%
4.20573950740395	0.305812833498424\\
4.20573950740395	0.867864165117056\\
};
\addplot[only marks, mark=*, mark options={}, mark size=1.7678pt, color=blue, fill=blue, forget plot] table[row sep=crcr]{%
x	y\\
4.89822144053597	0.485941316871944\\
};
\addplot[only marks, mark=*, mark options={}, mark size=1.7678pt, color=blue, fill=blue, forget plot] table[row sep=crcr]{%
x	y\\
4.89822144053597	0.97045232046591\\
};
\addplot [color=blue, forget plot]
  table[row sep=crcr]{%
4.89822144053597	0.485941316871944\\
4.89822144053597	0.97045232046591\\
};
\addplot[only marks, mark=*, mark options={}, mark size=1.7678pt, color=blue, fill=blue, forget plot] table[row sep=crcr]{%
x	y\\
5.59070337385189	0.492610308626597\\
};
\addplot[only marks, mark=*, mark options={}, mark size=1.7678pt, color=blue, fill=blue, forget plot] table[row sep=crcr]{%
x	y\\
5.59070337385189	0.931316454579596\\
};
\addplot [color=blue, forget plot]
  table[row sep=crcr]{%
5.59070337385189	0.492610308626597\\
5.59070337385189	0.931316454579596\\
};
\addplot[only marks, mark=*, mark options={}, mark size=1.7678pt, color=blue, fill=blue, forget plot] table[row sep=crcr]{%
x	y\\
6.28318530717959	0.43352006007573\\
};
\addplot[only marks, mark=*, mark options={}, mark size=1.7678pt, color=blue, fill=blue, forget plot] table[row sep=crcr]{%
x	y\\
6.28318530717959	0.728916405563935\\
};
\addplot [color=blue, forget plot]
  table[row sep=crcr]{%
6.28318530717959	0.43352006007573\\
6.28318530717959	0.728916405563935\\
};
\end{axis}

\end{tikzpicture}%

%% file: Graphics/tikz/pump_q1_over_t.tex
%
%
\definecolor{mycolor1}{rgb}{0.00000,0.44700,0.74100}%
\definecolor{mycolor2}{rgb}{0.85000,0.32500,0.09800}%
\definecolor{mycolor3}{rgb}{0.92900,0.69400,0.12500}%
\definecolor{mycolor4}{rgb}{0.49400,0.18400,0.55600}%
\definecolor{mycolor5}{rgb}{0.46600,0.67400,0.18800}%
\definecolor{mycolor6}{rgb}{0.30100,0.74500,0.93300}%
\begin{tikzpicture}

\begin{axis}[%
width=1.125in,
height=0.575in,
scale only axis,
xmin=0,
xmax=2.44986470069434,
xlabel style={font=\color{white!15!black}},
ymin=0,
ymax=6.85504238829741,
ylabel style={font=\color{white!15!black}},
ylabel={$q_1\,$[m]},
axis background/.style={fill=white},
axis x line*=bottom,
axis y line*=left,
xmajorgrids,
ymajorgrids,
legend columns=2,
legend style={at={(0,1.25)}, anchor=south west, legend cell align=left, align=left, draw=white!15!black,/tikz/column 2/.style={
  column sep=5pt}}
]
\addplot [color=mycolor1, line width=1.5pt]
  table[row sep=crcr]{%
0	0\\
0.009592716877721	0.0266412132504369\\
0.0191900779512148	0.0532640231023162\\
0.0287966666666667	0.0798500193267117\\
0.0383764106625928	0.106269229176082\\
0.0479739637442059	0.132615470345708\\
0.0575933333333333	0.158870545056015\\
0.0671627228549565	0.184812285656659\\
0.0767606095380687	0.210628675954945\\
0.08639	0.236302092473168\\
0.0959526488130106	0.261550057835917\\
0.105550931985979	0.286624120523043\\
0.115186666666667	0.311507758935552\\
0.124746338438628	0.335893423045959\\
0.134345004160315	0.360062043707937\\
0.143983333333333	0.383998769698821\\
0.153543281578624	0.407401617441647\\
0.16314226277433	0.430550871928742\\
0.17278	0.453433798388091\\
0.182342628698134	0.475775705015803\\
0.191941835145447	0.497834587062522\\
0.201576666666667	0.519600050617471\\
0.211143479382871	0.540836851966452\\
0.220742821833048	0.561768174296668\\
0.230373333333333	0.582385962513731\\
0.239945059590388	0.602498650775133\\
0.249544463504279	0.622289745517508\\
0.25917	0.641753360742887\\
0.268746789804754	0.660740362785498\\
0.278346200007921	0.67939510457669\\
0.287966666666667	0.697713608013664\\
0.297548282388758	0.7155847432217\\
0.307147662129566	0.733117456118999\\
0.316763333333333	0.750309386548576\\
0.326349306882443	0.767081534003758\\
0.335948634449087	0.783512734255451\\
0.34556	0.79960196535333\\
0.355149748467138	0.815296179793919\\
0.364749013337943	0.830649820532426\\
0.374356666666667	0.845662952229338\\
0.383949571467124	0.860302670439425\\
0.393548770909703	0.874604480587347\\
0.403153333333333	0.888569314736214\\
0.412748791290704	0.902179305398066\\
0.422347927675713	0.915455851290165\\
0.43195	0.928400562504056\\
0.441547454245984	0.941006333421377\\
0.451146533041037	0.953284510318441\\
0.460746666666667	0.965237217077727\\
0.470345623455828	0.976864668403244\\
0.479944651772111	0.988171413618007\\
0.489543333333334	0.999159946882073\\
0.499143369151285	1.00983512908308\\
0.508742354715033	1.02019722229143\\
0.51834	1.03024896488917\\
0.527940762061876	1.03999785764467\\
0.537539712509576	1.04944173151907\\
0.547136666666667	1.05858345839456\\
0.556737869045283	1.0674317292534\\
0.566336791467303	1.07598325460383\\
0.575933333333334	1.08424094310811\\
0.585534750389224	1.09221367342601\\
0.595133651064085	1.09989790971937\\
0.60473	1.10729651378185\\
0.614331458382677	1.11441789653693\\
0.623930342654895	1.121258814009\\
0.633526666666667	1.12782200972506\\
0.643128036838697	1.13411503213973\\
0.652726909100316	1.14013521882624\\
0.662323333333334	1.14588513502359\\
0.671924521300065	1.15137126128443\\
0.681523385042213	1.15659162997341\\
0.69112	1.16154857841801\\
0.700720939698663	1.1662474465683\\
0.710319797605709	1.17068695518916\\
0.719916666666667	1.17486917340659\\
0.729517313280697	1.17879831752157\\
0.739116167346049	1.1824737135275\\
0.748713333333334	1.18589713029831\\
0.758313657653905	1.18907173560654\\
0.767912509302591	1.1919973316148\\
0.77751	1.19467536207509\\
0.78710998385631	1.19710805752993\\
0.796708834065	1.19929554253704\\
0.806306666666667	1.20123891707442\\
0.81590629938541	1.20293960739485\\
0.825505148794983	1.20439789560942\\
0.835103333333334	1.20561452466873\\
0.844702609159177	1.20659026217775\\
0.854301458178384	1.20732538000299\\
0.863900000000001	1.20782025557205\\
};
\addlegendentry{$u=0$, $v_{1,0}=10$}

\addplot [color=mycolor4, line width=1.5pt]
  table[row sep=crcr]{%
1.03765445735025e-23	-9.62706427323694e-24\\
1.29759256803881e-05	3.60442380010779e-05\\
4.5722087678193e-05	0.000127005798210759\\
0.0816621566898113	0.221052367878388\\
0.081662156805818	0.221052368176468\\
0.0816621572712544	0.221052369372387\\
0.163324313379623	0.425922693207235\\
0.167028304993184	0.434955113180177\\
0.170737717675291	0.443976991631641\\
0.244986470069434	0.619670242088233\\
0.271964567360142	0.681136057972706\\
0.299187492156961	0.741694388201402\\
0.326648626759245	0.801413459700505\\
0.353662051952616	0.859185642035196\\
0.380885176937209	0.91680425782384\\
0.408310783449057	0.97436751044137\\
0.435354770039054	1.03077186892922\\
0.462578037294868	1.08729872173704\\
0.489972940138868	1.14402282915273\\
0.489972940153131	1.14402282918224\\
0.489972940168157	1.14402282921332\\
0.571635096828679	1.28484836159283\\
0.598852936529752	1.32295497247782\\
0.626074321130299	1.36040598828928\\
0.653297253518491	1.39729246150958\\
0.676020978913014	1.4275339353887\\
0.698733818138956	1.4571638697038\\
0.734959410208302	1.50348377773117\\
0.762181488175541	1.53782824997031\\
0.789401305309124	1.57201691860837\\
0.816621566898114	1.60621181492927\\
0.845211526026102	1.64233116165396\\
0.872023431871864	1.67659710115573\\
0.898283723587925	1.71070444836046\\
0.925485653797257	1.7468222453302\\
0.952705907650651	1.78399688348865\\
0.979945880277736	1.8224334766836\\
1.00717673943004	1.86191212039163\\
1.03439903970783	1.90218715812968\\
1.06160803696755	1.94335070568513\\
1.08884025094209	1.98589888915621\\
1.11606511073861	2.03025792234562\\
1.14327019365736	2.07655182077291\\
1.17054907923392	2.12527892523614\\
1.19777925089046	2.17656630053017\\
1.22493235034717	2.23050319625533\\
1.25233369638613	2.28772805616167\\
1.27957011906067	2.34735323440664\\
1.30659450703698	2.40936893670183\\
1.33418150710684	2.4757546853175\\
1.36142249550661	2.54448519176729\\
1.38825666372679	2.61536082436431\\
1.4164170399381	2.69591688589616\\
1.4436144821058	2.78189198986281\\
1.4699188204166	2.87187349721876\\
1.49735967240834	2.96965704088524\\
1.52450695235364	3.06760869805847\\
1.55158097710642	3.16585040441984\\
1.57839278468934	3.26296173614243\\
1.60555495386431	3.36045572585944\\
1.63324313379623	3.45833445016167\\
1.63324313382221	3.45833445025266\\
1.633243133848	3.458334450343\\
1.71490529048604	3.73578252078651\\
1.7414102421353	3.82246893234831\\
1.76862220043701	3.91023339675532\\
1.79656744717585	3.99948389144615\\
1.7965674476926	3.99948389309106\\
1.87761945729279	4.25538066298878\\
1.87822960386566	4.2573029144454\\
1.87845789447624	4.25802213828148\\
1.9598917605231	4.51355260535443\\
1.95989176055547	4.51355260545592\\
1.98710039493629	4.59787884562856\\
2.01432528023036	4.68095313457302\\
2.04155391724528	4.76367890299767\\
2.06895709225749	4.84713080049676\\
2.09620646182811	4.93032647781795\\
2.1232160739351	5.01316766852557\\
2.1508658844963	5.09858983839076\\
2.17810727028869	5.18364974274483\\
2.20487823062491	5.26845259467902\\
2.2326573613876	5.36120115714847\\
2.25998017546977	5.46078100166038\\
2.28654038731472	5.5663049060741\\
2.31521362593548	5.68602294792418\\
2.34236322711022	5.80154518081607\\
2.36820254400453	5.91393860654749\\
2.39621722257968	6.03859483416988\\
2.42328732679484	6.16137239906726\\
2.44986470069434	6.28318530717959\\
};
\addlegendentry{$u=u^\star$, $v_{1,0}=10$}

\addplot [color=mycolor2, line width=1.5pt]
  table[row sep=crcr]{%
0	0\\
0.020519869148075	0.0853864169530715\\
0.0412109713237239	0.170812099169403\\
0.0622333333333333	0.256292498272927\\
0.0824485669088393	0.336782683861544\\
0.103171665983051	0.41724292511129\\
0.124466666666667	0.497589275462136\\
0.14468611581555	0.571656741065215\\
0.165433126732874	0.645476938193043\\
0.1867	0.719000431329459\\
0.207038796724843	0.787462087675547\\
0.227792105201102	0.855653955266385\\
0.248933333333333	0.923608243883464\\
0.269386628454529	0.988104191108491\\
0.290139773987879	1.05248828809442\\
0.311166666666667	1.11682163582994\\
0.331714962602325	1.17897980506459\\
0.352468384986028	1.24119171070097\\
0.3734	1.30349073904733\\
0.394041616121054	1.3645970238306\\
0.414796668348417	1.4258029605988\\
0.435633333333333	1.48709202140802\\
0.456376368975533	1.54801889538578\\
0.47713235606439	1.60896076721412\\
0.497866666666667	1.66987767775744\\
0.518711628400556	1.7312238527855\\
0.539466535418641	1.79248112171783\\
0.5601	1.85363679308301\\
0.581039365308133	1.91605998939059\\
0.601792696566584	1.97840268422449\\
0.622333333333334	2.04070208168642\\
0.643368879660057	2.10527094609194\\
0.664122069691324	2.16990646629053\\
0.684566666666667	2.23466979255132\\
0.705718609665232	2.30301635863522\\
0.726471788872435	2.37161309555437\\
0.7468	2.44048683040462\\
0.768075605672128	2.51455919931335\\
0.788821404239904	2.58892195577431\\
0.809033333333334	2.66351867644446\\
0.830319374085418	2.74441968589398\\
0.851039636418456	2.82541306057636\\
0.871266666666667	2.90641898849603\\
0.892250577839826	2.99212212216357\\
0.912939949964931	3.07777079746224\\
0.9335	3.16338858284777\\
0.953981391263067	3.2485078598767\\
0.974674482632224	3.33367308899314\\
0.995733333333334	3.41889068677856\\
1.00155554078243	3.44212757934863\\
1.02890356040202	3.5491051599619\\
1.05796666666667	3.65855196931981\\
1.07819436364834	3.7320969136595\\
1.0989424517636	3.8053829713909\\
1.1202	3.87836797833319\\
1.14054928394361	3.94642962195861\\
1.16130268066699	4.014230033924\\
1.18243333333333	4.08180637325484\\
1.20289517306704	4.14604663411222\\
1.22364828069633	4.21019102128797\\
1.24466666666667	4.27430016976514\\
1.26522261635357	4.33632625861933\\
1.28597614125336	4.39841560868949\\
1.3069	4.46059762014578\\
1.32754983742022	4.52166062765036\\
1.34830502945701	4.58282033755104\\
1.36913333333333	4.64405658678381\\
1.38988509797376	4.70499747034582\\
1.41064106866904	4.76594323797774\\
1.43136666666667	4.8268543263143\\
1.45221985954319	4.88826206585726\\
1.47297461761471	4.9495759834407\\
1.4936	5.01078739502764\\
1.51454712107175	5.0733422420377\\
1.53530037333913	5.13582426865394\\
1.55583333333333	5.19827430111386\\
1.57687769132888	5.26309035433946\\
1.59763093047664	5.32798927567787\\
1.61806666666667	5.39303183454903\\
1.63922944124474	5.46178508899453\\
1.65998244349012	5.53080007071276\\
1.6803	5.60009781133633\\
1.70158362961131	5.67472847850648\\
1.7223280741175	5.74964124745505\\
1.74253333333333	5.8247730991157\\
1.76380800337896	5.90621510624021\\
1.78452536294331	5.98772825867237\\
1.80476666666667	6.06923810324536\\
1.8257104081455	6.15512262641353\\
1.84639833322159	6.24095541733748\\
1.867	6.32676701565019\\
};
\addlegendentry{$u=0$, $v_{1,0}=15$}

\addplot [color=mycolor5, line width=1.5pt]
  table[row sep=crcr]{%
1.88778178659998e-15	-4.13656700280411e-16\\
0.0155939897237202	0.0648766815558863\\
0.0312773854505402	0.1295360599254\\
0.0471171013715421	0.193670280808162\\
0.0625657990961033	0.255194487578515\\
0.0782558381909422	0.317035492331779\\
0.0942342027430839	0.379267204001909\\
0.109607651057008	0.438397548311521\\
0.125304622478269	0.498020084791263\\
0.141351304114626	0.558222143675275\\
0.156707860775057	0.615187796567442\\
0.17240816446176	0.672857693600602\\
0.188468405486168	0.731356102175702\\
0.203854929086086	0.787040498849949\\
0.219581758198925	0.843706911633274\\
0.235585506857708	0.901231414970503\\
0.239514814788346	0.915347153554697\\
0.244240241596162	0.932322763225878\\
0.282702608229251	1.07049485676855\\
0.298408265869098	1.126915815397\\
0.314113964774542	1.18333692226776\\
0.329819709600794	1.23975819410419\\
0.345525356053103	1.29617911254225\\
0.361231055801983	1.35260022244267\\
0.376936810972336	1.40902153143894\\
0.392642454782218	1.46544244038411\\
0.408348155500798	1.52186355376788\\
0.424053912343878	1.5782848687731\\
0.43975956276184	1.63470580145705\\
0.455465264403518	1.69112691815697\\
0.47117101371542	1.7475482061074\\
0.486876675962217	1.80396918128563\\
0.502582376886865	1.86039029541048\\
0.518288115086962	1.91681154344428\\
0.533993800317605	1.97323260119216\\
0.54969950334503	2.02965372287621\\
0.565405216458504	2.08607488080358\\
0.581440581542824	2.14368027981351\\
0.581448252500978	2.14370825436581\\
0.612522317830046	2.26729416218209\\
0.628629124396359	2.33523391984724\\
0.644358288025614	2.40501111933911\\
0.659639419201587	2.47624426835671\\
0.67606844351598	2.55512296480263\\
0.691742768218663	2.63128158306504\\
0.706756520573129	2.70521311090348\\
0.723018706120004	2.7864731389126\\
0.738682136975106	2.86590447327915\\
0.753873621944671	2.94393537116162\\
0.769753968906879	3.02633981691129\\
0.78540501747099	3.10808567606907\\
0.800990723316213	3.18964847405079\\
0.816391958637602	3.27002642302896\\
0.832045889413516	3.35114559307\\
0.848107824687755	3.43347251307455\\
0.863199551172138	3.5098064629911\\
0.878867562068711	3.58792712821032\\
0.895224926059297	3.6682795838169\\
0.91019924941889	3.74086043664204\\
0.925873201131356	3.81600213642608\\
0.942342027430839	3.89427949828709\\
0.970975660938321	4.02958371666624\\
0.973660349062586	4.04226137895964\\
0.989459128802381	4.11686654161455\\
1.00516480124996	4.19103203204174\\
1.02087051198337	4.2651977032625\\
1.03657623017392	4.33936340969731\\
1.05226699075058	4.41345848313142\\
1.06795770278474	4.48755332733755\\
1.08369333154547	4.56186027777872\\
1.0993991562818	4.63602648734463\\
1.11510486200109	4.71019213488693\\
1.13081043291701	4.78435714585934\\
1.14651629045832	4.8585235103373\\
1.16222199178606	4.93268913714205\\
1.17792753428855	5.00685401394094\\
1.19380246301685	5.08181876814474\\
1.20976685026767	5.15720596431886\\
1.22504463566009	5.22935088202991\\
1.24075043971959	5.30351699396667\\
1.25645613439221	5.37768258941214\\
1.27216173703163	5.45184775098162\\
1.29474986135324	5.57072996976627\\
1.31927883837979	5.70862827061443\\
1.31927883840317	5.70862827074654\\
1.36639593972728	5.98987553295208\\
1.36639593975099	5.98987553309435\\
1.36639593977472	5.98987553323677\\
1.36639593982431	5.98987553353437\\
1.41351303578855	6.28318527361039\\
1.41351304114626	6.28318530717959\\
};
\addlegendentry{$u=u^\star$, $v_{1,0}=15$}

\addplot [color=mycolor3, line width=1.5pt]
  table[row sep=crcr]{%
0	0\\
0.015920945296901	0.0971653577177913\\
0.0321258936520201	0.195288343112986\\
0.0489	0.295329698230947\\
0.0642854062006987	0.385298735521906\\
0.0805204768474893	0.478176204197702\\
0.0978	0.574724603744389\\
0.11307141431855	0.658256488255371\\
0.12932394598711	0.745607681532994\\
0.1467	0.837656311769842\\
0.164830323576576	0.932746829488645\\
0.184390404414823	1.03485582865255\\
0.1956	1.09334354573301\\
0.211899999935731	1.17839119763994\\
0.228199999963172	1.26343884998757\\
0.2445	1.34848650237091\\
0.260800000009975	1.43353415460852\\
0.277100000010011	1.51858180679191\\
0.2934	1.60362945892219\\
0.309700000007707	1.6886771111452\\
0.326000000007322	1.77372476332688\\
0.3423	1.85877241547366\\
0.358600000004347	1.94382006768292\\
0.374900000003989	2.02886771986944\\
0.3912	2.11391537203901\\
0.40750000000107	2.19896302423714\\
0.42380000000068	2.28401067642996\\
0.4401	2.36905832862375\\
0.456399999996824	2.4541059808072\\
0.472699999996189	2.53915363300687\\
0.489	2.62420128523312\\
0.502632198246431	2.69532928862759\\
0.516264396791976	2.76645729358833\\
0.5379	2.87934424189091\\
0.554200000013211	2.964391894193\\
0.570500000012681	3.04943954642869\\
0.5868	3.13448719860398\\
0.603100000020809	3.21953485095419\\
0.619400000024854	3.30458250321333\\
0.6357	3.389630155312\\
0.660977812903598	3.52152086937844\\
0.660977813160093	3.52152087060639\\
0.6846	3.61560328706064\\
0.686923291342102	3.62467241185849\\
0.709828843886269	3.71250733671503\\
0.7335	3.80037398707842\\
0.749479393137354	3.85814195660251\\
0.765782568824999	3.91591247543931\\
0.7824	3.97370072286841\\
0.798443067976431	4.02855909858427\\
0.814747107071812	4.08348868074488\\
0.8313	4.13852689064684\\
0.847399197796279	4.19145847207598\\
0.863703510362119	4.24456335580207\\
0.8802	4.29787408713109\\
0.896352712667712	4.34974553925979\\
0.91265811731506	4.40184905284847\\
0.9291	4.45418781578298\\
0.945313158714986	4.50565201160222\\
0.961620279831891	4.55730965872998\\
0.978000000000001	4.60912562867593\\
0.994283493958938	4.66059352451119\\
1.01059168235684	4.71211878808324\\
1.0269	4.76364435408918\\
1.04325647835499	4.81534268279986\\
1.05956410332731	4.86693011188911\\
1.0758	4.91836103278689\\
1.09222212183495	4.97048751826854\\
1.10852808437842	5.02239252564472\\
1.1247	5.07406619339419\\
1.14117844444149	5.12698127090905\\
1.157483012281	5.17967006737841\\
1.1736	5.23215748790624\\
1.19013310408991	5.28651115887729\\
1.20643719460963	5.34072012053425\\
1.2225	5.394823314149\\
1.23909499690082	5.45156183422991\\
1.25539869146234	5.50825784427565\\
1.2714	5.5649366103147\\
1.28805956894809	5.62515346293227\\
1.30436034215165	5.68537123336206\\
1.3203	5.74558170132569\\
1.33698778186737	5.81008723166482\\
1.35327949562606	5.87453652222877\\
1.3692	5.93890203283705\\
1.38580134400808	6.00739296486197\\
1.40207845977338	6.07575459634254\\
1.4181	6.14399517505533\\
1.43446289304104	6.2143799548546\\
1.45073038203765	6.28471055366256\\
1.467	6.35504238829741\\
};
\addlegendentry{$u=0$, $v_{1,0}=22$}

\addplot [color=mycolor6, line width=1.5pt]
  table[row sep=crcr]{%
2.09574318576006e-23	-3.40960555917643e-24\\
0.0120721169075488	0.073721560669935\\
0.0242714268381766	0.14790502696887\\
0.0367252812185313	0.223012311983912\\
0.0484913529856822	0.29322844358505\\
0.060694710620305	0.365215021812853\\
0.0734505624370627	0.439486864546021\\
0.0850560484753475	0.506207792825166\\
0.0972619247134961	0.575575352858523\\
0.110175843655594	0.648202634098446\\
0.127494317485398	0.74469681193413\\
0.14651087139629	0.850042783658296\\
0.146901124874125	0.852202969946719\\
0.146916280563363	0.852286861862977\\
0.146934801328469	0.852389380623489\\
0.183626406092657	1.05442952165043\\
0.183626406092751	1.05442952165095\\
0.183626406092845	1.05442952165146\\
0.220351687311188	1.25570984449089\\
0.223820549736117	1.27470545970323\\
0.240449292339064	1.3657650508701\\
0.257076968529719	1.45681880232497\\
0.284369973883784	1.60627627897235\\
0.289086114691192	1.63210203525843\\
0.293802249748251	1.65792776005541\\
0.306044308258645	1.7249657117283\\
0.318286070906833	1.79200204324915\\
0.330527530966782	1.85903671778599\\
0.342769583561538	1.92607463706476\\
0.355011339391295	1.99311093124782\\
0.367252812185313	2.06014567551715\\
0.379494800418983	2.12718324235268\\
0.391736550645601	2.19421950585331\\
0.403978093403845	2.26125463325009\\
0.41129820651838	2.3013398353269\\
0.418618300441398	2.34142493231081\\
0.440703374622376	2.46236359099118\\
0.449306930359902	2.50947697093027\\
0.457910466503541	2.55659024358025\\
0.477428655840907	2.67108611610115\\
0.477428655849572	2.67108611615174\\
0.479886911136005	2.68564190889486\\
0.514153937059439	2.89255021597901\\
0.526662220569708	2.96979744822497\\
0.538857500875048	3.04574919960291\\
0.55087921827797	3.12098669624812\\
0.562970746756373	3.19674140520834\\
0.575164952351746	3.27292055207696\\
0.587604499496501	3.35011694659997\\
0.599352675612022	3.42232981010453\\
0.611551143264307	3.49644763022879\\
0.624329780715033	3.57306949282646\\
0.628779313063295	3.59950287869394\\
0.646099422501005	3.70125936050086\\
0.661055061933564	3.78786530362233\\
0.697780295556283	3.99771587788916\\
0.697780342292722	3.99771614407695\\
0.697780343152095	3.99771614897152\\
0.697780343162478	3.99771614903065\\
0.697780343172887	3.99771614908994\\
0.734505624370627	4.20573950740395\\
0.746747385504919	4.27498770483404\\
0.758989145749661	4.344235897232\\
0.771230905589158	4.4134840873374\\
0.783472666220637	4.48273228192262\\
0.795714426618693	4.55198047518724\\
0.807956186807689	4.62122866726907\\
0.820197947233686	4.6904768606914\\
0.832439707591771	4.75972505372952\\
0.844681468026221	4.82897324719989\\
0.856923228436784	4.89822144053597\\
0.869164988846335	4.96746963386766\\
0.881406749244752	5.03671782713811\\
0.893648509685202	5.10596602064838\\
0.905890270090455	5.17521421396178\\
0.918132030463284	5.24446240709408\\
0.930373790900825	5.31371060059481\\
0.942615551306088	5.38295879391536\\
0.954857311681815	5.45220698707121\\
0.967099072110391	5.52145518052841\\
0.97934083251491	5.59070337385189\\
0.991582592900346	5.65995156706973\\
1.00382435330832	5.72919976041735\\
1.01606611371088	5.79844795373654\\
1.02830787411888	5.86769614708861\\
1.04054963447237	5.93694434013445\\
1.05279139486729	6.00619253341664\\
1.06503315533741	6.07544072712614\\
1.07727491317159	6.14468890592679\\
1.089516674879	6.21393710663893\\
1.10175843655594	6.28318530717959\\
};
\addlegendentry{$u=u^\star$, $v_{1,0}=22$}

\addplot [color=white!15!black, dashed, line width=1.5pt, forget plot]
  table[row sep=crcr]{%
0	6.28318530717959\\
2.44986470069434	6.28318530717959\\
};
\end{axis}

\end{tikzpicture}%

%% file: Graphics/tikz/pump_v1_over_t.tex
%
%
\definecolor{mycolor1}{rgb}{0.00000,0.44700,0.74100}%
\definecolor{mycolor2}{rgb}{0.85000,0.32500,0.09800}%
\definecolor{mycolor3}{rgb}{0.92900,0.69400,0.12500}%
\definecolor{mycolor4}{rgb}{0.49400,0.18400,0.55600}%
\definecolor{mycolor5}{rgb}{0.46600,0.67400,0.18800}%
\definecolor{mycolor6}{rgb}{0.30100,0.74500,0.93300}%
\begin{tikzpicture}

\begin{axis}[%
width=1.125in,
height=0.575in,
scale only axis,
xmin=0,
xmax=2.44986470069434,
xlabel style={font=\color{white!15!black}},
ymin=0,
ymax=25,
ylabel style={font=\color{white!15!black}},
ylabel={$v_1\,$[km/h]},
axis background/.style={fill=white},
axis x line*=bottom,
axis y line*=left,
xmajorgrids,
ymajorgrids
]
\addplot [color=mycolor1, line width=2.0pt]
  table[row sep=crcr]{%
0	10\\
0.009592716877721	9.99412193670434\\
0.0191900779512148	9.97653515130569\\
0.0287966666666667	9.94738080344059\\
0.0383764106625928	9.90708449348877\\
0.0479739637442059	9.85586078783613\\
0.0575933333333333	9.79410029542677\\
0.0671627228549565	9.72286240957602\\
0.0767606095380687	9.64221185823041\\
0.08639	9.55270411532949\\
0.0959526488130106	9.45599175420817\\
0.105550931985979	9.35178767227464\\
0.115186666666667	9.24070892887056\\
0.124746338438628	9.12480455532421\\
0.134345004160315	9.00339586018139\\
0.143983333333333	8.87707232114936\\
0.153543281578624	8.748025629404\\
0.16314226277433	8.61527308643131\\
0.17278	8.47931978884539\\
0.182342628698134	8.34227685585715\\
0.191941835145447	8.20300193782954\\
0.201576666666667	8.06189287858745\\
0.211143479382871	7.92083086642319\\
0.220742821833048	7.77865454958815\\
0.230373333333333	7.6356568438379\\
0.239945059590388	7.49341636255159\\
0.249544463504279	7.35085781754075\\
0.25917	7.20818537170339\\
0.268746789804754	7.06666086333996\\
0.278346200007921	6.92535747550124\\
0.287966666666667	6.78441140247678\\
0.297548282388758	6.64478933684235\\
0.307147662129566	6.50573912944118\\
0.316763333333333	6.36734869135671\\
0.326349306882443	6.23032859369903\\
0.335948634449087	6.09410201288886\\
0.34556	5.95872451981298\\
0.355149748467138	5.82469483817972\\
0.364749013337943	5.69159654204047\\
0.374356666666667	5.55946443221616\\
0.383949571467124	5.42862810167357\\
0.393548770909703	5.29880856748703\\
0.403153333333333	5.17002749446281\\
0.412748791290704	5.04248330655443\\
0.422347927675713	4.91600861913154\\
0.43195	4.79061652093055\\
0.441547454245984	4.66640639142339\\
0.451146533041037	4.54329694501407\\
0.460746666666667	4.42129510038981\\
0.470345623455828	4.30042749962722\\
0.479944651772111	4.1806756937176\\
0.489543333333334	4.06204148358251\\
0.499143369151285	3.94449978819715\\
0.508742354715033	3.82807494757718\\
0.51834	3.71276395826583\\
0.527940762061876	3.59850644919024\\
0.537539712509576	3.48535308122311\\
0.547136666666667	3.37329604671418\\
0.556737869045283	3.26225227883363\\
0.566336791467303	3.15228582814197\\
0.575933333333334	3.04338403269234\\
0.585534750389224	2.93545058761442\\
0.595133651064085	2.82855323866861\\
0.60473	2.72267450726327\\
0.614331458382677	2.61771180334088\\
0.623930342654895	2.5137296642742\\
0.633526666666667	2.41070597201349\\
0.643128036838697	2.3085368495032\\
0.652726909100316	2.20727900723541\\
0.662323333333334	2.10690600339081\\
0.671924521300065	2.0073161844044\\
0.681523385042213	1.90855559677613\\
0.69112	1.81059390106354\\
0.700720939698663	1.71333407712475\\
0.710319797605709	1.61680999119993\\
0.719916666666667	1.52098790565605\\
0.729517313280697	1.42577705707366\\
0.739116167346049	1.33119854162066\\
0.748713333333334	1.23721576088491\\
0.758313657653905	1.14374529402754\\
0.767912509302591	1.05079548428801\\
0.77751	0.958327420111063\\
0.78710998385631	0.866265765389938\\
0.796708834065	0.774606488715799\\
0.806306666666667	0.683308905070873\\
0.81590629938541	0.592306308854214\\
0.825505148794983	0.501582854138621\\
0.835103333333334	0.411096605310447\\
0.844702609159177	0.320789945581175\\
0.854301458178384	0.23063583394465\\
0.863900000000001	0.14059158921106\\
};

\addplot [color=mycolor2, line width=2.0pt]
  table[row sep=crcr]{%
0	15\\
0.020519869148075	14.9407041253453\\
0.0412109713237239	14.7676233637162\\
0.0622333333333333	14.4945207945673\\
0.0824485669088393	14.1646748556316\\
0.103171665983051	13.7855927091689\\
0.124466666666667	13.3792688900397\\
0.14468611581555	12.9976440134746\\
0.165433126732874	12.6248728270458\\
0.1867	12.2724636375437\\
0.207038796724843	11.9689079334516\\
0.227792105201102	11.6952154637744\\
0.248933333333333	11.4540110419094\\
0.269386628454529	11.2555047425724\\
0.290139773987879	11.0869081991674\\
0.311166666666667	10.9469452075961\\
0.331714962602325	10.8370726407472\\
0.352468384986028	10.7498589925236\\
0.3734	10.6828308381721\\
0.394041616121054	10.6343256174262\\
0.414796668348417	10.6005363605631\\
0.435633333333333	10.5795982795949\\
0.456376368975533	10.5702116653338\\
0.47713235606439	10.5715562627633\\
0.497866666666667	10.5837254257234\\
0.518711628400556	10.6077753863288\\
0.539466535418641	10.6451354105489\\
0.5601	10.6977854137885\\
0.581039365308133	10.7698010379316\\
0.601792696566584	10.8627366633863\\
0.622333333333334	10.9790377367688\\
0.643368879660057	11.1265042390091\\
0.664122069691324	11.3032048801204\\
0.684566666666667	11.5102173681138\\
0.705718609665232	11.7608731527473\\
0.726471788872435	12.0438468964001\\
0.7468	12.3559890632717\\
0.768075605672128	12.7166286339955\\
0.788821404239904	13.0952580922735\\
0.809033333333334	13.4793859490372\\
0.830319374085418	13.8835881385495\\
0.851039636418456	14.2546660086685\\
0.871266666666667	14.5702443763476\\
0.892250577839826	14.8209004955381\\
0.912939949964931	14.9667697297116\\
0.9335	14.9961107488541\\
0.953981391263067	14.9074027941805\\
0.974674482632224	14.7082964787268\\
0.995733333333334	14.414397912081\\
1.00155554078243	14.3206133298061\\
1.02890356040202	13.8333409174518\\
1.05796666666667	13.2794753296287\\
1.07819436364834	12.9012533611455\\
1.0989424517636	12.5349530038321\\
1.1202	12.1911251680677\\
1.14054928394361	11.8963064685002\\
1.16130268066699	11.6319034661073\\
1.18243333333333	11.4000533588143\\
1.20289517306704	11.2099204789177\\
1.22364828069633	11.049248068635\\
1.24466666666667	10.9165791750027\\
1.26522261635357	10.812934350728\\
1.28597614125336	10.7312375829636\\
1.3069	10.668999961747\\
1.32754983742022	10.6244751100199\\
1.34830502945701	10.5941211540134\\
1.36913333333333	10.5761988229387\\
1.38988509797376	10.5695521516738\\
1.41064106866904	10.5735925804459\\
1.43136666666667	10.5885996004529\\
1.45221985954319	10.6158691411873\\
1.47297461761471	10.6569448391106\\
1.4936	10.7139108652103\\
1.51454712107175	10.7911157167511\\
1.53530037333913	10.8899804025303\\
1.55583333333333	11.0129098364999\\
1.57687769132888	11.1680862774118\\
1.59763093047664	11.3530639463514\\
1.61806666666667	11.5687031879875\\
1.63922944124474	11.8288879905667\\
1.65998244349012	12.1211052767203\\
1.6803	12.4416419869145\\
1.70158362961131	12.81009652643\\
1.7223280741175	13.193893640939\\
1.74253333333333	13.5795395988419\\
1.76380800337896	13.9801550330512\\
1.78452536294331	14.3415490663123\\
1.80476666666667	14.6411246175255\\
1.8257104081455	14.8677838551606\\
1.84639833322159	14.9854218996706\\
1.867	14.98447520914\\
};

\addplot [color=mycolor3, line width=2.0pt]
  table[row sep=crcr]{%
0	22\\
0.015920945296901	21.9126986297057\\
0.0321258936520201	21.6591304050112\\
0.0489	21.2639631332135\\
0.0642854062006987	20.8325421639003\\
0.0805204768474893	20.3577806666545\\
0.0978	19.8801946177508\\
0.11307141431855	19.5129418474717\\
0.12932394598711	19.1985355450199\\
0.1467	18.9602984004789\\
0.164830323576576	18.8205804487892\\
0.184390404414823	18.7835306541037\\
0.1956	18.7835305630744\\
0.211899999935731	18.7835305503756\\
0.228199999963172	18.7835305462081\\
0.2445	18.7835305444636\\
0.260800000009975	18.7835305436952\\
0.277100000010011	18.7835305434057\\
0.2934	18.7835305433831\\
0.309700000007707	18.7835305435234\\
0.326000000007322	18.7835305437707\\
0.3423	18.7835305440925\\
0.358600000004347	18.7835305444697\\
0.374900000003989	18.7835305448906\\
0.3912	18.7835305453487\\
0.40750000000107	18.7835305458412\\
0.42380000000068	18.7835305463687\\
0.4401	18.7835305469357\\
0.456399999996824	18.7835305475522\\
0.472699999996189	18.7835305482385\\
0.489	18.7835305490379\\
0.502632198246431	18.7835305501577\\
0.516264396791976	18.7835305519724\\
0.5379	18.7835305538224\\
0.554200000013211	18.7835305552201\\
0.570500000012681	18.7835305561488\\
0.5868	18.7835305565293\\
0.603100000020809	18.783530556216\\
0.619400000024854	18.7835305549123\\
0.6357	18.7835305518901\\
0.660977812903598	18.7835305288438\\
0.660977813160093	15.934013138484\\
0.6846	14.3238099081581\\
0.686923291342102	14.0297018592977\\
0.709828843886269	13.5827398943218\\
0.7335	13.1499790636047\\
0.749479393137354	12.8828799972325\\
0.765782568824999	12.6346630794535\\
0.7824	12.4083833680934\\
0.798443067976431	12.2158394235671\\
0.814747107071812	12.0457468675106\\
0.8313	11.8983016684913\\
0.847399197796279	11.7777564221415\\
0.863703510362119	11.6767842990095\\
0.8802	11.5941101039147\\
0.896352712667712	11.5299400730066\\
0.91265811731506	11.4797309213598\\
0.9291	11.4416427755159\\
0.945313158714986	11.4142915468651\\
0.961620279831891	11.3950890767135\\
0.978000000000001	11.3825627980168\\
0.994283493958938	11.3756001711043\\
1.01059168235684	11.3733462839276\\
1.0269	11.3755529635807\\
1.04325647835499	11.3825055670111\\
1.05956410332731	11.3949267345531\\
1.0758	11.4139511056965\\
1.09222212183495	11.4415788878609\\
1.10852808437842	11.4792812990207\\
1.1247	11.5288850995128\\
1.14117844444149	11.5942173552613\\
1.157483012281	11.6758479058181\\
1.1736	11.7753373709804\\
1.19013310408991	11.8989447983613\\
1.20643719460963	12.0441046086513\\
1.2225	12.2112343939754\\
1.23909499690082	12.4101828878627\\
1.25539869146234	12.6321646533\\
1.2714	12.8752907969052\\
1.28805956894809	13.1535512665392\\
1.30436034215165	13.4477250737623\\
1.3203	13.7517817095989\\
1.33698778186737	14.0803579380568\\
1.35327949562606	14.4016653684946\\
1.3692	14.7045397609162\\
1.38580134400808	14.9941961063957\\
1.40207845977338	15.2366143036295\\
1.4181	15.4200921138368\\
1.43446289304104	15.5380143302237\\
1.45073038203765	15.5768662457896\\
1.467	15.5345078299321\\
};

\addplot [color=mycolor4, line width=2.0pt]
  table[row sep=crcr]{%
1.03765445735025e-23	10\\
1.29759256803881e-05	9.99999999999997\\
4.5722087678193e-05	9.99999989570968\\
0.0816621566898113	9.2502194309963\\
0.081662156805818	9.2502194045667\\
0.0816621572712544	9.24992508143685\\
0.163324313379623	8.79034373486275\\
0.167028304993184	8.76731119837795\\
0.170737717675291	8.74422632450793\\
0.244986470069434	8.30330287978648\\
0.271964567360142	8.10315851889834\\
0.299187492156961	7.91616437076705\\
0.326648626759245	7.74441279127042\\
0.353662051952616	7.65666715707235\\
0.380885176937209	7.58514626847966\\
0.408310783449057	7.52949749395345\\
0.435354770039054	7.48955264063337\\
0.462578037294868	7.46274020129308\\
0.489972940138868	7.44741174771792\\
0.489972940153131	7.44741174654946\\
0.489972940168157	7.44741174528241\\
0.571635096828679	5.08799735335671\\
0.598852936529752	4.99448146805409\\
0.626074321130299	4.91330647014975\\
0.653297253518491	4.84463149272181\\
0.676020978913014	4.74051245703506\\
0.698733818138956	4.65538042673743\\
0.734959410208302	4.559004554197\\
0.762181488175541	4.52829649840988\\
0.789401305309124	4.51855398239829\\
0.816621566898114	4.52979594761366\\
0.845211526026102	4.57099380126593\\
0.872023431871864	4.63473572361166\\
0.898283723587925	4.72061235430967\\
0.925485653797257	4.84376584316861\\
0.952705907650651	4.99368615663814\\
0.979945880277736	5.17015908531357\\
1.00717673943004	5.27046131591581\\
1.03439903970783	5.38405241178145\\
1.06160803696755	5.51076405299751\\
1.08884025094209	5.74195470413991\\
1.11606511073861	5.99266223069419\\
1.14327019365736	6.26252378955465\\
1.17054907923392	6.60210717287474\\
1.19777925089046	6.96248255768222\\
1.22493235034717	7.34314194209798\\
1.25233369638613	7.69675605938423\\
1.27957011906067	8.06863472356859\\
1.30659450703698	8.45710153339774\\
1.33418150710684	8.87192727193882\\
1.36142249550661	9.29602927133839\\
1.38825666372679	9.72150641950519\\
1.4164170399381	10.8634693670954\\
1.4436144821058	11.8781710256085\\
1.4699188204166	12.7241805653809\\
1.49735967240834	12.9220000231922\\
1.52450695235364	13.0421251022851\\
1.55158097710642	13.0673891359363\\
1.57839278468934	12.9950966633413\\
1.60555495386431	12.8353307049573\\
1.63324313379623	12.6088446126597\\
1.63324313382221	12.6088446124163\\
1.633243133848	12.6088446121821\\
1.71490529048604	11.8704144624271\\
1.7414102421353	11.6851814806576\\
1.76862220043701	11.5454192939646\\
1.79656744717585	11.4594930092647\\
1.7965674476926	11.4594929992729\\
1.87761945729279	11.341709601446\\
1.87822960386566	11.3417096014077\\
1.87845789447624	11.3417096011711\\
1.9598917605231	11.2845454529493\\
1.95989176055547	11.2845454526747\\
1.98710039493629	11.0505409438658\\
2.01432528023036	10.940420176962\\
2.04155391724528	10.9554444155269\\
2.06895709225749	10.9739504022782\\
2.09620646182811	11.0122674016669\\
2.1232160739351	11.0756555028593\\
2.1508658844963	11.1747111835691\\
2.17810727028869	11.3146960452283\\
2.20487823062491	11.5017178822608\\
2.2326573613876	12.5561973973333\\
2.25998017546977	13.7027768617514\\
2.28654038731472	14.9189256070762\\
2.31521362593548	15.1625014015259\\
2.34236322711022	15.4854922973286\\
2.36820254400453	15.8347929745232\\
2.39621722257968	16.1926113582337\\
2.42328732679484	16.4410411182214\\
2.44986470069434	16.5298225728655\\
};

\addplot [color=mycolor5, line width=2.0pt]
  table[row sep=crcr]{%
1.88778178659998e-15	15\\
0.0155939897237202	14.9320529827185\\
0.0312773854505402	14.730483290839\\
0.0471171013715421	14.4023238816116\\
0.0625657990961033	14.2672419915246\\
0.0782558381909422	14.1080380521807\\
0.0942342027430839	13.9327463175159\\
0.109607651057008	13.7604552810045\\
0.125304622478269	13.5890087407108\\
0.141351304114626	13.4256527209383\\
0.156707860775057	13.2860998845877\\
0.17240816446176	13.1645725014183\\
0.188468405486168	13.0653968566381\\
0.203854929086086	12.9959168547005\\
0.219581758198925	12.95135353632\\
0.235585506857708	12.9330796436416\\
0.239514814788346	12.9326302410218\\
0.244240241596162	12.9326294842761\\
0.282702608229251	12.9326294842501\\
0.298408265869098	12.9326294842251\\
0.314113964774542	12.9326294842008\\
0.329819709600794	12.9326294841774\\
0.345525356053103	12.9326294841554\\
0.361231055801983	12.9326294841353\\
0.376936810972336	12.9326294841177\\
0.392642454782218	12.9326294841036\\
0.408348155500798	12.9326294840941\\
0.424053912343878	12.9326294840912\\
0.43975956276184	12.9326294840976\\
0.455465264403518	12.9326294841171\\
0.47117101371542	12.9326294841564\\
0.486876675962217	12.9326294842267\\
0.502582376886865	12.9326294843509\\
0.518288115086962	12.9326294845764\\
0.533993800317605	12.9326294850215\\
0.54969950334503	12.9326294861037\\
0.565405216458504	12.9326294909097\\
0.581440581542824	12.9326297577921\\
0.581448252500978	13.8603197449719\\
0.612522317830046	14.8049142047983\\
0.628629124396359	15.5739442456967\\
0.644358288025614	16.3743913821688\\
0.659639419201587	17.1950122253647\\
0.67606844351598	17.3826373954935\\
0.691742768218663	17.6071749978161\\
0.706756520573129	17.8508190438281\\
0.723018706120004	18.1268257738132\\
0.738682136975106	18.3813614105344\\
0.753873621944671	18.5937491379872\\
0.769753968906879	18.7555645984049\\
0.78540501747099	18.8359669661401\\
0.800990723316213	18.8280024965255\\
0.816391958637602	18.734772549249\\
0.832045889413516	18.5646229282364\\
0.848107824687755	18.3318806851004\\
0.863199551172138	18.0827553147734\\
0.878867562068711	17.8171546173969\\
0.895224926059297	17.5560467975299\\
0.91019924941889	17.3487007033247\\
0.925873201131356	17.1770053742013\\
0.942342027430839	17.0556629351142\\
0.970975660938321	16.9999575905159\\
0.973660349062586	16.9999575904778\\
0.989459128802381	16.999957590439\\
1.00516480124996	16.9999575904003\\
1.02087051198337	16.9999575903623\\
1.03657623017392	16.9999575903259\\
1.05226699075058	16.9999575902918\\
1.06795770278474	16.9999575902613\\
1.08369333154547	16.9999575902361\\
1.0993991562818	16.9999575902184\\
1.11510486200109	16.9999575902112\\
1.13081043291701	16.9999575902194\\
1.14651629045832	16.99995759025\\
1.16222199178606	16.9999575903137\\
1.17792753428855	16.9999575904278\\
1.19380246301685	16.9999575906253\\
1.20976685026767	16.9999575909784\\
1.22504463566009	16.9999575916427\\
1.24075043971959	16.9999575945383\\
1.25645613439221	16.9999576365593\\
1.27216173703163	16.999957848483\\
1.29474986135324	18.8772461298079\\
1.31927883837979	20.3400469751465\\
1.31927883840317	20.3400469754446\\
1.36639593972728	21.6034147472221\\
1.36639593975099	21.60341474739\\
1.36639593977472	21.6034147475585\\
1.36639593982431	21.6034147478818\\
1.41351303578855	22.5561244703922\\
1.41351304114626	22.5561244888434\\
};

\addplot [color=mycolor6, line width=2.0pt]
  table[row sep=crcr]{%
2.09574318576006e-23	22\\
0.0120721169075488	21.9531763502146\\
0.0242714268381766	21.8154042681056\\
0.0367252812185313	21.5947608120657\\
0.0484913529856822	21.36656910086\\
0.060694710620305	21.1027833714498\\
0.0734505624370627	20.8204826962563\\
0.0850560484753475	20.5762755136601\\
0.0972619247134961	20.3483012653632\\
0.110175843655594	20.1529856757116\\
0.127494317485398	19.9821482710119\\
0.14651087139629	19.9272295572681\\
0.146901124874125	19.9272295571468\\
0.146916280563363	19.9272295570815\\
0.146934801328469	19.9272295570177\\
0.183626406092657	19.7976868954376\\
0.183626406092751	19.7976868954375\\
0.183626406092845	19.7976868954374\\
0.220351687311188	19.7137296840291\\
0.223820549736117	19.7137291753306\\
0.240449292339064	19.7137291753244\\
0.257076968529719	19.71372917532\\
0.284369973883784	19.7137291753188\\
0.289086114691192	19.7137291753206\\
0.293802249748251	19.7137291753235\\
0.306044308258645	19.7137291753285\\
0.318286070906833	19.7137291753374\\
0.330527530966782	19.7137291753514\\
0.342769583561538	19.7137291753721\\
0.355011339391295	19.713729175402\\
0.367252812185313	19.7137291754452\\
0.379494800418983	19.7137291755082\\
0.391736550645601	19.7137291756021\\
0.403978093403845	19.713729175748\\
0.41129820651838	19.7137291759581\\
0.418618300441398	19.7137291762798\\
0.440703374622376	19.713729177171\\
0.449306930359902	19.7137291798564\\
0.457910466503541	19.7137291832499\\
0.477428655840907	21.0171789158615\\
0.477428655849572	21.0171789160052\\
0.479886911136005	21.3435642768063\\
0.514153937059439	22.1159861981419\\
0.526662220569708	22.3391180041396\\
0.538857500875048	22.4895582234401\\
0.55087921827797	22.5571130248471\\
0.562970746756373	22.5367615449976\\
0.575164952351746	22.4288607486016\\
0.587604499496501	22.2406751306978\\
0.599352675612022	22.0087167835192\\
0.611551143264307	21.7347802319488\\
0.624329780715033	21.4374308923007\\
0.628779313063295	21.3360648904999\\
0.646099422501005	20.975261487142\\
0.661055061933564	20.7314561662068\\
0.697780295556283	20.5038311329398\\
0.697780342292722	20.5038311314754\\
0.697780343152095	20.5038311308093\\
0.697780343162478	20.5038311307967\\
0.697780343172887	20.5038311307839\\
0.734505624370627	20.3641868203365\\
0.746747385504919	20.36418682028\\
0.758989145749661	20.3641868202222\\
0.771230905589158	20.3641868201635\\
0.783472666220637	20.3641868201047\\
0.795714426618693	20.3641868200469\\
0.807956186807689	20.3641868199941\\
0.820197947233686	20.3641868199568\\
0.832439707591771	20.3641868199838\\
0.844681468026221	20.3641868201653\\
0.856923228436784	20.3641868205019\\
0.869164988846335	20.3641868209618\\
0.881406749244752	20.3641868215236\\
0.893648509685202	20.364186822147\\
0.905890270090455	20.3641868228035\\
0.918132030463284	20.3641868234842\\
0.930373790900825	20.3641868241775\\
0.942615551306088	20.3641868248744\\
0.954857311681815	20.3641868255714\\
0.967099072110391	20.3641868262637\\
0.97934083251491	20.3641868269471\\
0.991582592900346	20.364186827619\\
1.00382435330832	20.3641868282763\\
1.01606611371088	20.3641868289157\\
1.02830787411888	20.3641868295353\\
1.04054963447237	20.3641868301331\\
1.05279139486729	20.3641868307087\\
1.06503315533741	20.3641868312665\\
1.07727491317159	20.364186831832\\
1.089516674879	20.3641868322698\\
1.10175843655594	20.3641868323926\\
};

\end{axis}

\end{tikzpicture}%

%% file: Graphics/tikz/pump_s_over_t.tex
%
%
\definecolor{mycolor1}{rgb}{0.00000,0.44700,0.74100}%
\definecolor{mycolor2}{rgb}{0.85000,0.32500,0.09800}%
\definecolor{mycolor3}{rgb}{0.92900,0.69400,0.12500}%
\definecolor{mycolor4}{rgb}{0.49400,0.18400,0.55600}%
\definecolor{mycolor5}{rgb}{0.46600,0.67400,0.18800}%
\definecolor{mycolor6}{rgb}{0.30100,0.74500,0.93300}%
\begin{tikzpicture}

\begin{axis}[%
width=1.125in,
height=0.575in,
scale only axis,
xmin=0,
xmax=2.44986470069434,
xlabel style={font=\color{white!15!black}},
xlabel={$t\,$[s]},
ymin=0,
ymax=0.695589999999881,
ylabel style={font=\color{white!15!black}},
ylabel={$q_3\,$[m]},
axis background/.style={fill=white},
axis x line*=bottom,
axis y line*=left,
xmajorgrids,
ymajorgrids
]
\addplot [color=mycolor1, line width=1.5pt, forget plot]
  table[row sep=crcr]{%
0	0.4368\\
0.009592716877721	0.4368\\
0.0191900779512148	0.4368\\
0.0287966666666667	0.436799999999999\\
0.0383764106625928	0.436799999999999\\
0.0479739637442059	0.436799999999999\\
0.0575933333333333	0.436799999999999\\
0.0671627228549565	0.436799999999999\\
0.0767606095380687	0.436799999999999\\
0.08639	0.436799999999999\\
0.0959526488130106	0.436799999999998\\
0.105550931985979	0.436799999999998\\
0.115186666666667	0.436799999999998\\
0.124746338438628	0.436799999999998\\
0.134345004160315	0.436799999999998\\
0.143983333333333	0.436799999999998\\
0.153543281578624	0.436799999999998\\
0.16314226277433	0.436799999999997\\
0.17278	0.436799999999997\\
0.182342628698134	0.436799999999997\\
0.191941835145447	0.436799999999997\\
0.201576666666667	0.436799999999996\\
0.211143479382871	0.436799999999996\\
0.220742821833048	0.436799999999996\\
0.230373333333333	0.436799999999996\\
0.239945059590388	0.436799999999996\\
0.249544463504279	0.436799999999996\\
0.25917	0.436799999999996\\
0.268746789804754	0.436799999999995\\
0.278346200007921	0.436799999999995\\
0.287966666666667	0.436799999999995\\
0.297548282388758	0.436799999999995\\
0.307147662129566	0.436799999999995\\
0.316763333333333	0.436799999999995\\
0.326349306882443	0.436799999999995\\
0.335948634449087	0.436799999999994\\
0.34556	0.436799999999994\\
0.355149748467138	0.436799999999994\\
0.364749013337943	0.436799999999994\\
0.374356666666667	0.436799999999993\\
0.383949571467124	0.436799999999993\\
0.393548770909703	0.436799999999993\\
0.403153333333333	0.436799999999993\\
0.412748791290704	0.436799999999993\\
0.422347927675713	0.436799999999993\\
0.43195	0.436799999999993\\
0.441547454245984	0.436799999999993\\
0.451146533041037	0.436799999999993\\
0.460746666666667	0.436799999999992\\
0.470345623455828	0.436799999999992\\
0.479944651772111	0.436799999999992\\
0.489543333333334	0.436799999999992\\
0.499143369151285	0.436799999999992\\
0.508742354715033	0.436799999999992\\
0.51834	0.436799999999992\\
0.527940762061876	0.436799999999992\\
0.537539712509576	0.436799999999991\\
0.547136666666667	0.436799999999991\\
0.556737869045283	0.436799999999991\\
0.566336791467303	0.436799999999991\\
0.575933333333334	0.43679999999999\\
0.585534750389224	0.43679999999999\\
0.595133651064085	0.43679999999999\\
0.60473	0.43679999999999\\
0.614331458382677	0.43679999999999\\
0.623930342654895	0.43679999999999\\
0.633526666666667	0.43679999999999\\
0.643128036838697	0.436799999999989\\
0.652726909100316	0.436799999999989\\
0.662323333333334	0.436799999999989\\
0.671924521300065	0.436799999999989\\
0.681523385042213	0.436799999999988\\
0.69112	0.436799999999988\\
0.700720939698663	0.436799999999988\\
0.710319797605709	0.436799999999988\\
0.719916666666667	0.436799999999988\\
0.729517313280697	0.436799999999988\\
0.739116167346049	0.436799999999988\\
0.748713333333334	0.436799999999988\\
0.758313657653905	0.436799999999987\\
0.767912509302591	0.436799999999987\\
0.77751	0.436799999999987\\
0.78710998385631	0.436799999999987\\
0.796708834065	0.436799999999987\\
0.806306666666667	0.436799999999987\\
0.81590629938541	0.436799999999986\\
0.825505148794983	0.436799999999986\\
0.835103333333334	0.436799999999986\\
0.844702609159177	0.436799999999986\\
0.854301458178384	0.436799999999986\\
0.863900000000001	0.436799999999986\\
};
\addplot [color=mycolor2, line width=1.5pt, forget plot]
  table[row sep=crcr]{%
0	0.4368\\
0.020519869148075	0.4368\\
0.0412109713237239	0.4368\\
0.0622333333333333	0.436799999999999\\
0.0824485669088393	0.436799999999999\\
0.103171665983051	0.436799999999999\\
0.124466666666667	0.436799999999999\\
0.14468611581555	0.436799999999999\\
0.165433126732874	0.436799999999999\\
0.1867	0.436799999999999\\
0.207038796724843	0.436799999999998\\
0.227792105201102	0.436799999999998\\
0.248933333333333	0.436799999999999\\
0.269386628454529	0.436799999999998\\
0.290139773987879	0.436799999999998\\
0.311166666666667	0.436799999999998\\
0.331714962602325	0.436799999999998\\
0.352468384986028	0.436799999999998\\
0.3734	0.436799999999998\\
0.394041616121054	0.436799999999998\\
0.414796668348417	0.436799999999998\\
0.435633333333333	0.436799999999998\\
0.456376368975533	0.436799999999997\\
0.47713235606439	0.436799999999997\\
0.497866666666667	0.436799999999997\\
0.518711628400556	0.436799999999997\\
0.539466535418641	0.436799999999997\\
0.5601	0.436799999999997\\
0.581039365308133	0.436799999999997\\
0.601792696566584	0.436799999999996\\
0.622333333333334	0.436799999999996\\
0.643368879660057	0.436799999999996\\
0.664122069691324	0.436799999999996\\
0.684566666666667	0.436799999999996\\
0.705718609665232	0.436799999999996\\
0.726471788872435	0.436799999999995\\
0.7468	0.436799999999995\\
0.768075605672128	0.436799999999995\\
0.788821404239904	0.436799999999995\\
0.809033333333334	0.436799999999995\\
0.830319374085418	0.436799999999994\\
0.851039636418456	0.436799999999994\\
0.871266666666667	0.436799999999994\\
0.892250577839826	0.436799999999994\\
0.912939949964931	0.436799999999994\\
0.9335	0.436799999999994\\
0.953981391263067	0.436799999999994\\
0.974674482632224	0.436799999999994\\
0.995733333333334	0.436799999999994\\
1.00155554078243	0.436799999999994\\
1.02890356040202	0.436799999999994\\
1.05796666666667	0.436799999999993\\
1.07819436364834	0.436799999999993\\
1.0989424517636	0.436799999999993\\
1.1202	0.436799999999993\\
1.14054928394361	0.436799999999993\\
1.16130268066699	0.436799999999993\\
1.18243333333333	0.436799999999993\\
1.20289517306704	0.436799999999993\\
1.22364828069633	0.436799999999993\\
1.24466666666667	0.436799999999993\\
1.26522261635357	0.436799999999993\\
1.28597614125336	0.436799999999993\\
1.3069	0.436799999999992\\
1.32754983742022	0.436799999999992\\
1.34830502945701	0.436799999999992\\
1.36913333333333	0.436799999999992\\
1.38988509797376	0.436799999999992\\
1.41064106866904	0.436799999999992\\
1.43136666666667	0.436799999999992\\
1.45221985954319	0.436799999999992\\
1.47297461761471	0.436799999999992\\
1.4936	0.436799999999991\\
1.51454712107175	0.436799999999991\\
1.53530037333913	0.436799999999991\\
1.55583333333333	0.436799999999991\\
1.57687769132888	0.436799999999991\\
1.59763093047664	0.436799999999991\\
1.61806666666667	0.43679999999999\\
1.63922944124474	0.43679999999999\\
1.65998244349012	0.43679999999999\\
1.6803	0.43679999999999\\
1.70158362961131	0.43679999999999\\
1.7223280741175	0.43679999999999\\
1.74253333333333	0.43679999999999\\
1.76380800337896	0.436799999999989\\
1.78452536294331	0.436799999999989\\
1.80476666666667	0.436799999999989\\
1.8257104081455	0.436799999999989\\
1.84639833322159	0.436799999999988\\
1.867	0.436799999999988\\
};
\addplot [color=mycolor3, line width=1.5pt, forget plot]
  table[row sep=crcr]{%
0	0.4368\\
0.015920945296901	0.4368\\
0.0321258936520201	0.4368\\
0.0489	0.436799999999999\\
0.0642854062006987	0.436799999999999\\
0.0805204768474893	0.436799999999999\\
0.0978	0.436799999999999\\
0.11307141431855	0.436799999999999\\
0.12932394598711	0.436799999999999\\
0.1467	0.436799999999999\\
0.164830323576576	0.436799999999998\\
0.184390404414823	0.436799999999998\\
0.1956	0.436799999999998\\
0.211899999935731	0.436799999999998\\
0.228199999963172	0.436799999999998\\
0.2445	0.436799999999998\\
0.260800000009975	0.436799999999998\\
0.277100000010011	0.436799999999997\\
0.2934	0.436799999999997\\
0.309700000007707	0.436799999999997\\
0.326000000007322	0.436799999999997\\
0.3423	0.436799999999996\\
0.358600000004347	0.436799999999996\\
0.374900000003989	0.436799999999996\\
0.3912	0.436799999999996\\
0.40750000000107	0.436799999999996\\
0.42380000000068	0.436799999999996\\
0.4401	0.436799999999996\\
0.456399999996824	0.436799999999995\\
0.472699999996189	0.436799999999995\\
0.489	0.436799999999995\\
0.502632198246431	0.436799999999995\\
0.516264396791976	0.436799999999995\\
0.5379	0.436799999999995\\
0.554200000013211	0.436799999999995\\
0.570500000012681	0.436799999999995\\
0.5868	0.436799999999995\\
0.603100000020809	0.436799999999995\\
0.619400000024854	0.436799999999995\\
0.6357	0.436799999999994\\
0.660977812903598	0.436799999999994\\
0.660977813160093	0.436799999999994\\
0.6846	0.436799999999994\\
0.686923291342102	0.436799999999994\\
0.709828843886269	0.436799999999993\\
0.7335	0.436799999999993\\
0.749479393137354	0.436799999999993\\
0.765782568824999	0.436799999999993\\
0.7824	0.436799999999993\\
0.798443067976431	0.436799999999993\\
0.814747107071812	0.436799999999993\\
0.8313	0.436799999999993\\
0.847399197796279	0.436799999999993\\
0.863703510362119	0.436799999999993\\
0.8802	0.436799999999993\\
0.896352712667712	0.436799999999993\\
0.91265811731506	0.436799999999993\\
0.9291	0.436799999999993\\
0.945313158714986	0.436799999999993\\
0.961620279831891	0.436799999999993\\
0.978000000000001	0.436799999999993\\
0.994283493958938	0.436799999999993\\
1.01059168235684	0.436799999999993\\
1.0269	0.436799999999993\\
1.04325647835499	0.436799999999993\\
1.05956410332731	0.436799999999993\\
1.0758	0.436799999999992\\
1.09222212183495	0.436799999999992\\
1.10852808437842	0.436799999999992\\
1.1247	0.436799999999992\\
1.14117844444149	0.436799999999992\\
1.157483012281	0.436799999999992\\
1.1736	0.436799999999992\\
1.19013310408991	0.436799999999992\\
1.20643719460963	0.436799999999991\\
1.2225	0.436799999999991\\
1.23909499690082	0.436799999999991\\
1.25539869146234	0.436799999999991\\
1.2714	0.436799999999991\\
1.28805956894809	0.436799999999991\\
1.30436034215165	0.436799999999991\\
1.3203	0.43679999999999\\
1.33698778186737	0.43679999999999\\
1.35327949562606	0.43679999999999\\
1.3692	0.43679999999999\\
1.38580134400808	0.43679999999999\\
1.40207845977338	0.43679999999999\\
1.4181	0.43679999999999\\
1.43446289304104	0.43679999999999\\
1.45073038203765	0.43679999999999\\
1.467	0.43679999999999\\
};
\addplot [color=mycolor4, line width=1.5pt, forget plot]
  table[row sep=crcr]{%
1.03765445735025e-23	0.4368\\
1.29759256803881e-05	0.43680000138509\\
4.5722087678193e-05	0.436800017197035\\
0.0816621566898113	0.491658403940946\\
0.081662156805818	0.491658404096806\\
0.0816621572712544	0.491658404722141\\
0.163324313379623	0.572499706896518\\
0.167028304993184	0.574797340129675\\
0.170737717675291	0.576979263823088\\
0.244986470069434	0.595589999999881\\
0.271964567360142	0.591247350742913\\
0.299187492156961	0.581936028541763\\
0.326648626759245	0.567526519410628\\
0.353662051952616	0.54771401546119\\
0.380885176937209	0.521354516321505\\
0.408310783449057	0.488309265128745\\
0.435354770039054	0.449345415185407\\
0.462578037294868	0.403726417759465\\
0.489972940138868	0.351341000996231\\
0.489972940153131	0.351341000967264\\
0.489972940168157	0.351341000936749\\
0.571635096828679	0.27803\\
0.598852936529752	0.282761316308199\\
0.626074321130299	0.284142070904578\\
0.653297253518491	0.282171222020891\\
0.676020978913014	0.279671426857151\\
0.698733818138956	0.278260519756503\\
0.734959410208302	0.278260519753057\\
0.762181488175541	0.278982720132667\\
0.789401305309124	0.27907062421127\\
0.816621566898114	0.27852430420871\\
0.845211526026102	0.278030000051142\\
0.872023431871864	0.278316213406032\\
0.898283723587925	0.279300009149551\\
0.925485653797257	0.281415279741453\\
0.952705907650651	0.285004736231712\\
0.979945880277736	0.290071651131117\\
1.00717673943004	0.294354514697556\\
1.03439903970783	0.295597854218675\\
1.06160803696755	0.293805112109331\\
1.08884025094209	0.290443626941142\\
1.11606511073861	0.286985983789598\\
1.14327019365736	0.283433905432116\\
1.17054907923392	0.280426606349267\\
1.19777925089046	0.27862745456773\\
1.22493235034717	0.278030000014464\\
1.25233369638613	0.278030000008271\\
1.27957011906067	0.278030000015654\\
1.30659450703698	0.278030000036319\\
1.33418150710684	0.278030000050717\\
1.36142249550661	0.27803000003848\\
1.38825666372679	0.278030000000724\\
1.4164170399381	0.287833690095941\\
1.4436144821058	0.315915288902549\\
1.4699188204166	0.360473109118065\\
1.49735967240834	0.412619166913239\\
1.52450695235364	0.457790636970697\\
1.55158097710642	0.496483822581779\\
1.57839278468934	0.528546374661968\\
1.60555495386431	0.554679904358585\\
1.63324313379623	0.574743526195257\\
1.63324313382221	0.574743526210968\\
1.633243133848	0.574743526226566\\
1.71490529048604	0.595252219964017\\
1.7414102421353	0.589494738894276\\
1.76862220043701	0.577254335608548\\
1.79656744717585	0.558009903889347\\
1.7965674476926	0.558009903470959\\
1.87761945729279	0.469944379781474\\
1.87822960386566	0.469111227266541\\
1.87845789447624	0.468798929487943\\
1.9598917605231	0.345646124937865\\
1.95989176055547	0.345646124884242\\
1.98710039493629	0.308090860783054\\
2.01432528023036	0.285543179654149\\
2.04155391724528	0.278030000002033\\
2.06895709225749	0.278034576589471\\
2.09620646182811	0.278035943488783\\
2.1232160739351	0.278034165101782\\
2.1508658844963	0.278031439811913\\
2.17810727028869	0.278030108118903\\
2.20487823062491	0.278030108098896\\
2.2326573613876	0.288053132988348\\
2.25998017546977	0.317465022197779\\
2.28654038731472	0.36464328182409\\
2.31521362593548	0.421905939775889\\
2.34236322711022	0.469563368092776\\
2.36820254400453	0.508992145820307\\
2.39621722257968	0.545207674153394\\
2.42328732679484	0.573745436059027\\
2.44986470069434	0.595589999994421\\
};
\addplot [color=mycolor5, line width=1.5pt, forget plot]
  table[row sep=crcr]{%
1.88778178659998e-15	0.4368\\
0.0155939897237202	0.440169305765188\\
0.0312773854505402	0.450354603626208\\
0.0471171013715421	0.467559768853174\\
0.0625657990961033	0.486697312243428\\
0.0782558381909422	0.50401832515275\\
0.0942342027430839	0.519466614583722\\
0.109607651057008	0.532243058497845\\
0.125304622478269	0.543176582684588\\
0.141351304114626	0.552148076362427\\
0.156707860775057	0.558645606543175\\
0.17240816446176	0.563177262657971\\
0.188468405486168	0.565604154188988\\
0.203854929086086	0.565834140935086\\
0.219581758198925	0.563950483061452\\
0.235585506857708	0.559834848063334\\
0.239514814788346	0.558485220889559\\
0.244240241596162	0.556685059760685\\
0.282702608229251	0.534840157996347\\
0.298408265869098	0.522727285834771\\
0.314113964774542	0.509460158098338\\
0.329819709600794	0.495038761243529\\
0.345525356053103	0.480048404517784\\
0.361231055801983	0.4650741213896\\
0.376936810972336	0.45011591000046\\
0.392642454782218	0.435306261257646\\
0.408348155500798	0.420777349057735\\
0.424053912343878	0.406529177157425\\
0.43975956276184	0.392075826762437\\
0.455465264403518	0.376931084942195\\
0.47117101371542	0.361094948593487\\
0.486876675962217	0.345632723372797\\
0.502582376886865	0.331609457850793\\
0.518288115086962	0.319025163715516\\
0.533993800317605	0.307543004371625\\
0.54969950334503	0.296826016491676\\
0.565405216458504	0.286874207374871\\
0.581440581542824	0.2795846064038\\
0.581448252500978	0.279582306389553\\
0.612522317830046	0.279582306335433\\
0.628629124396359	0.288321378466733\\
0.644358288025614	0.304404407072033\\
0.659639419201587	0.327172949832447\\
0.67606844351598	0.354267745218182\\
0.691742768218663	0.377939039067412\\
0.706756520573129	0.39861772155376\\
0.723018706120004	0.41881358774113\\
0.738682136975106	0.436100585620792\\
0.753873621944671	0.450837109328157\\
0.769753968906879	0.464105306413926\\
0.78540501747099	0.475045069860428\\
0.800990723316213	0.483831113151628\\
0.816391958637602	0.490446728942084\\
0.832045889413516	0.495065923943842\\
0.848107824687755	0.49759973302342\\
0.863199551172138	0.497944685515465\\
0.878867562068711	0.496215993085764\\
0.895224926059297	0.492142968943274\\
0.91019924941889	0.486382815702061\\
0.925873201131356	0.478273493747267\\
0.942342027430839	0.467460820911102\\
0.970975660938321	0.443069329114459\\
0.973660349062586	0.440418317916266\\
0.989459128802381	0.423553288927368\\
1.00516480124996	0.405649309564348\\
1.02087051198337	0.387617413037142\\
1.03657623017392	0.369457634181841\\
1.05226699075058	0.352308243143446\\
1.06795770278474	0.337272833435554\\
1.08369333154547	0.324317393639622\\
1.0993991562818	0.31360607763166\\
1.11510486200109	0.3052119693248\\
1.13081043291701	0.299135020985733\\
1.14651629045832	0.294105057410245\\
1.16222199178606	0.288852218261468\\
1.17792753428855	0.283376511120309\\
1.19380246301685	0.279215737346144\\
1.20976685026767	0.278030000030924\\
1.22504463566009	0.279710965158008\\
1.24075043971959	0.285223962054293\\
1.25645613439221	0.295475771332346\\
1.27216173703163	0.310466313158795\\
1.29474986135324	0.343125344354803\\
1.31927883837979	0.396013251202289\\
1.31927883840317	0.396013251261352\\
1.36639593972728	0.505414317474288\\
1.36639593975099	0.505414317524497\\
1.36639593977472	0.505414317574761\\
1.36639593982431	0.505414317679793\\
1.41351303578855	0.595589990827366\\
1.41351304114626	0.595589999988222\\
};
\addplot [color=mycolor6, line width=1.5pt, forget plot]
  table[row sep=crcr]{%
2.09574318576006e-23	0.4368\\
0.0120721169075488	0.43654890384474\\
0.0242714268381766	0.435785005208779\\
0.0367252812185313	0.434476174795048\\
0.0484913529856822	0.432387710118743\\
0.060694710620305	0.428955070127989\\
0.0734505624370627	0.423988452990606\\
0.0850560484753475	0.418245542182911\\
0.0972619247134961	0.410947066711071\\
0.110175843655594	0.401820588396924\\
0.127494317485398	0.387314242255636\\
0.14651087139629	0.368393658587461\\
0.146901124874125	0.367972581013788\\
0.146916280563363	0.367956202733305\\
0.146934801328469	0.367936188159279\\
0.183626406092657	0.328801224951765\\
0.183626406092751	0.328801224951666\\
0.183626406092845	0.328801224951566\\
0.220351687311188	0.296798165581469\\
0.223820549736117	0.294473132984907\\
0.240449292339064	0.285258544657746\\
0.257076968529719	0.27923958698437\\
0.284369973883784	0.279239586988854\\
0.289086114691192	0.280710362829409\\
0.293802249748251	0.282614536332887\\
0.306044308258645	0.289231197668554\\
0.318286070906833	0.298070292741926\\
0.330527530966782	0.309131653718038\\
0.342769583561538	0.321529072605478\\
0.355011339391295	0.334374655157125\\
0.367252812185313	0.347668383919201\\
0.379494800418983	0.360537979939878\\
0.391736550645601	0.372109521720659\\
0.403978093403845	0.382383108968004\\
0.41129820651838	0.388807768245506\\
0.418618300441398	0.396571084639909\\
0.440703374622376	0.428105485565942\\
0.449306930359902	0.442443141738574\\
0.457910466503541	0.456139744106358\\
0.477428655840907	0.484835513415171\\
0.477428655849572	0.484835513427178\\
0.479886911136005	0.488215733719049\\
0.514153937059439	0.529885398493478\\
0.526662220569708	0.542562430781267\\
0.538857500875048	0.553617751536726\\
0.55087921827797	0.563255140784586\\
0.562970746756373	0.571686014346681\\
0.575164952351746	0.57890617113389\\
0.587604499496501	0.584944740287142\\
0.599352675612022	0.589417275121133\\
0.611551143264307	0.592796389587607\\
0.624329780715033	0.594954197000783\\
0.628779313063295	0.595373621259829\\
0.646099422501005	0.595373621253466\\
0.661055061933564	0.593283510925403\\
0.697780295556283	0.579932706932289\\
0.697780342292722	0.579932682510039\\
0.697780343152095	0.579932682060971\\
0.697780343162478	0.579932682055545\\
0.697780343172887	0.579932682050105\\
0.734505624370627	0.562051331618632\\
0.746747385504919	0.555945348262125\\
0.758989145749661	0.548675266504093\\
0.771230905589158	0.540241086241606\\
0.783472666220637	0.531021605587011\\
0.795714426618693	0.521395624228082\\
0.807956186807689	0.511363142167701\\
0.820197947233686	0.502037137478051\\
0.832439707591771	0.494530588825157\\
0.844681468026221	0.488843496122432\\
0.856923228436784	0.484511003593965\\
0.869164988846335	0.481068255392948\\
0.881406749244752	0.478515251520903\\
0.893648509685202	0.476128916281944\\
0.905890270090455	0.473186174007456\\
0.918132030463284	0.469687024701385\\
0.930373790900825	0.465436875763512\\
0.942615551306088	0.46024113465556\\
0.954857311681815	0.454099801383529\\
0.967099072110391	0.446944322827422\\
0.97934083251491	0.438706145952999\\
0.991582592900346	0.429385270762639\\
1.00382435330832	0.418950918046622\\
1.01606611371088	0.407372308648923\\
1.02830787411888	0.394649442559279\\
1.04054963447237	0.380768082045683\\
1.05279139486729	0.365713989199924\\
1.06503315533741	0.349487163963719\\
1.07727491317159	0.332268282525479\\
1.089516674879	0.314238007911133\\
1.10175843655594	0.295396345488205\\
};
\addplot [color=white!15!black, dashed, line width=1.5pt, forget plot]
  table[row sep=crcr]{%
0	0.59559\\
2.44986470069434	0.59559\\
};
\addplot [color=white!15!black, dashed, line width=1.5pt, forget plot]
  table[row sep=crcr]{%
0	0.27803\\
2.44986470069434	0.27803\\
};
\end{axis}

\end{tikzpicture}%

%% file: Graphics/tikz/pump_u_over_t.tex
%
%
\definecolor{mycolor1}{rgb}{0.00000,0.44700,0.74100}%
\definecolor{mycolor2}{rgb}{0.85000,0.32500,0.09800}%
\definecolor{mycolor3}{rgb}{0.92900,0.69400,0.12500}%
\definecolor{mycolor4}{rgb}{0.49400,0.18400,0.55600}%
\definecolor{mycolor5}{rgb}{0.46600,0.67400,0.18800}%
\definecolor{mycolor6}{rgb}{0.30100,0.74500,0.93300}%
\begin{tikzpicture}

\begin{axis}[%
width=1.125in,
height=0.575in,
scale only axis,
xmin=0,
xmax=2.44986470069434,
xlabel style={font=\color{white!15!black}},
xlabel={$t\,$[s]},
ymin=-150,
ymax=500,
ylabel style={font=\color{white!15!black}},
ylabel={$u^\star\,$[N]},
axis background/.style={fill=white},
axis x line*=bottom,
axis y line*=left,
xmajorgrids,
ymajorgrids
]
\addplot[const plot, color=mycolor4, line width=1.5pt, forget plot] table[row sep=crcr] {%
1.03765445735025e-23	205.65604797951\\
0.0816621566898113	-108.249999999136\\
0.163324313379623	-108.249999997509\\
0.244986470069434	-83.5179367471736\\
0.326648626759245	-108.24999999756\\
0.408310783449057	-108.249999993166\\
0.489972940138868	346.891670564415\\
0.571635096828679	-56.53473435106\\
0.653297253518491	26.3492603060721\\
0.734959410208302	-10.6997265934364\\
0.816621566898114	12.6188408275456\\
0.898283723587925	24.8544496797824\\
0.979945880277736	-51.2395499767572\\
1.06160803696755	-1.63676431051969\\
1.14327019365736	20.258442536246\\
1.22493235034717	2.27429748835928e-07\\
1.30659450703698	-4.42805506143925e-07\\
1.38825666372679	309.067031080203\\
1.4699188204166	-108.249999992975\\
1.55158097710642	-108.249999996268\\
1.63324313379623	-108.249999972294\\
1.71490529048604	-108.249999991436\\
1.79656744717585	-85.40206058929\\
1.87822960386566	-44.1823604617863\\
1.95989176055547	253.551922301903\\
2.04155391724528	-0.0534498365542314\\
2.1232160739351	0.0226263890527613\\
2.20487823062491	324.692569073811\\
2.28654038731472	-108.249999922712\\
2.36820254400453	-108.249999961062\\
2.44986470069434	nan\\
};
\addplot[const plot, color=mycolor5, line width=1.5pt, forget plot] table[row sep=crcr] {%
1.88778178659998e-15	346.390479019027\\
0.0471171013715421	-108.249999993629\\
0.0942342027430839	-108.249999996324\\
0.141351304114626	-108.249999997023\\
0.188468405486168	-108.249999997126\\
0.235585506857708	-108.24999987439\\
0.282702608229251	-58.4905971068259\\
0.329819709600794	0.817111892619012\\
0.376936810972336	14.2291225188029\\
0.424053912343878	-35.0340554446358\\
0.47117101371542	72.9215722720811\\
0.518288115086962	38.775878537766\\
0.565405216458504	241.165759347033\\
0.612522317830046	376.874999752946\\
0.659639419201587	-108.249999947762\\
0.706756520573129	-108.249999986587\\
0.753873621944671	-108.249999992942\\
0.800990723316213	-108.249999995051\\
0.848107824687755	-108.249999995869\\
0.895224926059297	-108.249999996038\\
0.942342027430839	-108.249999990637\\
0.989459128802381	-6.48001441894481\\
1.03657623017392	107.328308517077\\
1.08369333154547	117.420305525902\\
1.13081043291701	-11.2967399982766\\
1.17792753428855	147.477462914455\\
1.22504463566009	240.141507990137\\
1.27216173703163	376.874999258328\\
1.31927883840317	-108.249999604624\\
1.36639593977472	-108.249999800785\\
1.41351304114626	nan\\
};
\addplot[const plot, color=mycolor6, line width=1.5pt, forget plot] table[row sep=crcr] {%
2.09574318576006e-23	-43.0738019153611\\
0.0367252812185313	-108.24999999306\\
0.0734505624370627	-108.249999996119\\
0.110175843655594	-108.249999997006\\
0.146901124874125	9.57783393558037\\
0.183626406092657	123.291794602745\\
0.220351687311188	144.447234248994\\
0.257076968529719	243.571035884981\\
0.293802249748251	185.385995727346\\
0.330527530966782	37.4063298025749\\
0.367252812185313	-108.249994163159\\
0.403978093403845	312.284876550384\\
0.440703374622376	-108.249999683936\\
0.477428655840907	-108.24999996567\\
0.514153937059439	-108.249999985661\\
0.55087921827797	-108.249999990756\\
0.587604499496501	-108.249999992359\\
0.624329780715033	-108.249999992737\\
0.661055061933564	-108.249999992961\\
0.697780343152095	24.2734648369834\\
0.734505624370627	-97.0984095885824\\
0.771230905589158	-33.9065614471663\\
0.807956186807689	151.762271601585\\
0.844681468026221	74.2142819444067\\
0.881406749244752	-46.410353889008\\
0.918132030463284	-78.8725958335416\\
0.954857311681815	-90.3087289725517\\
0.991582592900346	-95.4433603336355\\
1.02830787411888	-97.8185329950792\\
1.06503315533741	-67.678492066453\\
1.10175843655594	nan\\
};
\addplot [color=white!15!black, dashed, line width=1.5pt, forget plot]
  table[row sep=crcr]{%
0	376.875\\
2.44986470069434	376.875\\
};
\addplot [color=white!15!black, dashed, line width=1.5pt, forget plot]
  table[row sep=crcr]{%
0	-108.25\\
2.44986470069434	-108.25\\
};
\end{axis}

\end{tikzpicture}%

%% file: Content/6_conclusion.tex
\section{CONCLUSION}
\label{sec:conclusion}
This paper has proposed two novel problems for  simulation and optimal control of nonsmooth dynamics: a basic ski jumping model and optimal control of a bicycle on a pump track. Due to the fact that the underlying physics in both examples are straightforward to understand, these problems have potential for teaching purposes. Additionaly, the scalability of the bicycle problem makes it potentially useful for comparing and evaluating state-of-the-art methods in numerical nonsmooth optimal control.

Simulation results using a time-freezing approach for treating the state jumps of the systems have been presented. The solutions of the pump track model, which extend on the initial investigations in \cite{golembiewski2024dynamics} by incorporating jumps, closely match real-world observations on mountain bike tracks. Future work will extend to a multi-body system with two tires and incorporate friction.